\documentclass[11pt,a4paper,onecolumn]{amsart}
\usepackage{etex}
 
\usepackage[utf8]{inputenc}
\usepackage[left=2.9cm,right=2.9cm,top=2.8cm,bottom=2.8cm]{geometry}

\usepackage[foot]{amsaddr}
\usepackage[numbers]{natbib}

\usepackage{mathtools}
\usepackage{amssymb}
\usepackage{amsfonts}
\usepackage{amsthm}
\usepackage{stmaryrd}
\usepackage{multicol}
\usepackage{color}
\usepackage{graphicx}
\usepackage{setspace}

\usepackage{dblfloatfix}

\usepackage{relsize}
\usepackage{verbatim}
\usepackage{subfig}

\usepackage{caption}

\usepackage{booktabs}
\usepackage{multirow}
\usepackage{soul}

\usepackage{upgreek}

\usepackage[
colorlinks=true,
linkcolor=black, 
anchorcolor=black,
citecolor=blue, 
filecolor=magenta, 
menucolor=red, 
urlcolor=cyan,
plainpages=false,
pdfpagelabels,
hypertexnames=false,
linktocpage 
]{hyperref}

\usepackage{doi}

\usepackage{tensorNotation}

\makeatletter
  \newcommand{\srcsize}{\@setfontsize{\srcsize}{5pt}{5pt}}
\def\smalloverbrace#1{\mathop{\vbox{\m@th\ialign{##\crcr
  \noalign{\kern.5\fontdimen5\textfont2}%
  \tiny\downbracefill\crcr
  \noalign{\kern.7\fontdimen5\textfont2\nointerlineskip}%
  $\hfil\displaystyle{#1}\hfil$\crcr}}}\limits}
  \def\tinyoverbrace#1{\mathop{\vbox{\m@th\ialign{##\crcr
  \noalign{\kern.5\fontdimen5\textfont2}%
  \srcsize\downbracefill\crcr
  \noalign{\kern.7\fontdimen5\textfont2\nointerlineskip}%
  $\hfil\displaystyle{#1}\hfil$\crcr}}}\limits}
\makeatother

\title[Molecular mechanism behind the bubble rise velocity discontinuity]{On the molecular mechanism behind the bubble rise velocity jump discontinuity in viscoelastic liquids}

\author[Bothe et al.]{Dieter Bothe$^1$}
\address[]{$^1$Mathematical Modeling and Analysis, Technische Universit{\"a}t Darmstadt, Darmstadt, Germany}
\email[]{bothe@mma.tu-darmstadt.de}

\author[]{Matthias Niethammer$^1$}
\email[]{niethammer@mma.tu-darmstadt.de}

\author[]{Christian Pilz}

\author[]{G{\"u}nter Brenn$^2$}
\address[]{$^2$Institute of Fluid Mechanics and Heat Transfer, Graz University of Technology, Graz, Austria}
\email[]{guenter.brenn@tugraz.at}

\keywords{sub-/supercritical bubble state,
kinematic polymer orientation and stretching,
Lagrangian polymer transport,
self-amplifying acceleration mechanism,
conformation tensor analysis,
local stress distribution,
hoop stress,
negative wake,
PIV measurements,
extended Volume of Fluid method}

\begin{document}

\begin{abstract}
Bubbles rising in viscoelastic liquids may exhibit a jump discontinuity of the rise velocity as a critical bubble volume is exceeded. This phenomenon has been extensively investigated in the literature, both by means of experiments and via numerical simulations. The occurrence of the velocity jump has been associated with a change of the bubble shape, accompanied by the formation of a pointed tip at the rear end and to the appearance of a so-called negative wake, with the liquid velocity behind the bubble pointing in a direction opposite to that in Newtonian fluids.
We revisit this topic, starting with a review of the state of knowledge on the interrelations between the mentioned characteristic features. 
In search for a convincing explanation of the jump phenomenon, we performed detailed numerical simulations of the transient rise of single bubbles in 3D, allowing for
a local analysis of the polymer conformation tensor.
The latter shows that polymer molecules traveling along the
upper bubble hemisphere are stretched in the circumferential direction,
due to the flow kinematics.
Then, depending on the relaxation time scale of the polymer,
the stored elastic energy is either unloaded essentially above or below
the bubble's equator. In the former case, this slows down the bubble, while the bubble gets accelerated otherwise.
In this latter case, the velocity of motion of the polymer molecules along the bubble is increased, giving rise to a self-amplification of the effect and thus causing the
bubble rise velocity to jump to a higher level.
Detailed experimental velocity measurements in the liquid field around the bubble confirmed the conclusion that
the ratio of the time scale of the Lagrangian transport of polymer molecules along the bubble contour to the relaxation time scale of the polymer molecules determines the sub- or supercritical state of the bubble motion.
\end{abstract}

\maketitle



\section{Introduction and literature survey}
\label{sec:introduction}
Several applications in biotechnology, bio-process engineering, polymer processing and others involve
bubbly flows in rheologically complex liquids.
It is then of core relevance to understand the rise behavior of bubbles in viscoelastic liquids, since this determines in particular the residence (i.e., the contact) time. 
Since the pioneering paper by Astarita and Apuzzo \cite{Astarita1965} it has been known that for single bubbles rising in quiescent viscoelastic liquids, the terminal bubble rise velocity
as a function of the bubble volume 
may exhibit a jump discontinuity at a critical volume such that a supercritical bubble rises up to an order of magnitude faster than a subcritical one.
We call this \emph{the bubble rise velocity jump discontinuity}.
At this critical bubble size, the terminal rise velocity can experience a discontinuous increase by up to an order of magnitude, depending on the material properties of the liquid.
Astarita and Apuzzo reported this jump in the bubble rise velocity to be accompanied by a change in the bubble shape from a convex to a “teardrop”-shaped surface. As a possible
explanation, the authors of \cite{Astarita1965} suggested that the jump of the bubble rise velocity corresponds to a change of the interfacial mobility from rigid to free. 
This switch in the boundary conditions at the bubble surface
would then result in a transition similar to the one from the Stokes to the Hadamard-Rybczynski regime for a Newtonian ambient fluid. They also showed that, for a shear-thinning liquid, the ratio of the steady
rise velocities just above and below the critical volume can be much larger than the well-known factor of 1.5 in the Newtonian case.

Some time later, the suggested explanation was supported by
investigations of the mass transfer coefficients 
of single carbon dioxide bubbles in an aqueous polyethylene oxide solution by Calderbank et al.\ \cite{Calderbank1970}.
Moreover, experiments with glass spheres moving in viscoelastic liquids by Leal et al. \cite{Leal1971} showed no discontinuity of the settling velocity as a function of the sphere volume, hence also supporting the relevance of the surface mobility for the velocity jump to occur. Furthermore, a numerical analysis based on the creeping flow equations for shear-thinning liquids, but without elastic effects, indicated that the height of the experimentally observed velocity jump cannot be explained without accounting for elastic effects. 

Acharya et al.\ suggested in \cite{Acharya1977} that polymer molecules act as surface-active agents (surfactants), generating surface stresses which dampen the interfacial
motion, in particular in the downstream part of the bubble. In this case, a partial cleansing of the surface could be responsible for the rapid velocity change, and the authors claim that this is likely to occur much more abruptly in the case of a viscoelastic liquid rather than for Newtonian fluids. Analyzing the available data, they also derived a criterion for the critical bubble size, saying that a certain Bond number is of order of unity,
\begin{equation}
\textnormal{Bo} = \frac{\rho_l g R_c^2}{\sigma} \approx 1
\end{equation}
(liquid density $\rho_l$, gravitational acceleration $g$, critical bubble radius $R_c$, and surface tension $\sigma$ against the gas in the bubble). This turned out inappropriate for predicting the discontinuity; note that Bo can be unity for Newtonian fluids, while no rise velocity jump discontinuity has ever been observed in this case. Despite this, the study of  surfactant effects has been continued later by others, as will be briefly mentioned below.

Liu et al.\ \cite{Liu1995} studied the formation of a cusp at the rear end of bubbles rising in viscoelastic liquids. More precisely, the bubble shape displays a conical point at the rear end with a locally concave profile in the vertical section, i.e.\ a “teardrop”-like shape. They related the cusp formation and the associated increase in rise velocity to a critical capillary number, viz.\
\begin{equation}
\textnormal{Ca} = \frac{\eta_0 U_T}{\sigma} \approx 1
\label{eq:capnumber}
\end{equation}
(zero-shear viscosity $\eta_0$ and terminal bubble rise velocity $U_T$). Recall that the capillary number expresses the ratio between viscous forces and surface tension forces, while elastic forces are not included. The authors of \cite{Liu1995} have shown that data published previously by several authors also seem to correlate with a critical capillary number of order one. They also noted that the presence of a cusp is not sufficient for the appearance of a velocity jump. This has been confirmed, for instance, by De Kee et al.\ \cite{DeKee1990,Dekee1986}, Rodrigue and Blanchet \cite{Rodrigue2002} and Pilz and Brenn \cite{Pilz2007}, where, despite a clearly visible cusp at the rear end, in some cases no jump appeared. 

Extensive investigations of the effect of surfactants on the rise velocity jump discontinuity of bubbles in polymer solutions can be found in \cite{Rodrigue2002} - \cite{Rodrigue1998}. Rodrigue et al.\ \cite{Rodrigue1998} suggested that an appropriate jump criterion must represent elastic, viscous and capillary forces acting on the bubble. Their jump criterion states that a quantity $\alpha = \textnormal{Ca} \, \textnormal{De} / \textnormal{Ma}$ ($\textnormal{Ca}$ - capillary, $\textnormal{De}$ - Deborah, $\textnormal{Ma}$ - Marangoni number) changes its value from $< 1$ to $> 1$ upon transition from sub- to supercritical bubble state \cite{Rodrigue1998}. In the formulation of the quantity $\alpha$, the capillary number accounts for the shear-rate dependence of the dynamic viscosity of the polymer solution. The Deborah number accounts for the liquid elasticity by virtue of the first normal stress difference, which is also formulated as shear-rate dependent. In Rodrigue et al.\ \cite{Rodrigue1996} it is furthermore pointed out that polymer molecules act as surfactants, since the presence of a polymer alters the surface tension of the liquid against the gas phase. The authors proposed that the discontinuity is the result of an instability at the gas-liquid interface caused by an imbalance of forces, and that the origin of the instability should be related to normal forces which, under certain conditions, may extract surfactant as well as polymer molecules from the bubble surface, leaving a zone of different interfacial and rheological properties. Therefore, a Marangoni stress due to a surface tension gradient appears. According to this hypothesis, Rodrigue et al.\ \cite{Rodrigue1998} gave a physical interpretation of the discontinuity as follows. At the rear stagnation point, where the local strain is large, causing strong curvature and local deformation, high normal stresses are developed. Polymer and/or surfactant molecules are stretched along the liquid streamlines, and therefore induce a change of the fluid properties. The jump could be the result of the normal forces acting in the vicinity of the bubble, removing molecules from the bubble surface or causing a sudden change in the interfacial conditions. In the work of Rodrigue and Blanchet \cite{Rodrigue2002} it is shown that the jump can be eliminated by using surfactant concentrations above the critical association concentration (CAC). From their observations they concluded that the origin of the jump is most likely related to a change in interfacial conditions due to an imbalance in surface tension gradient and elastic forces at the gas-liquid interface. 
Since the presence of a surfactant introduces additional complexity in the nonlinear behavior underlying an already complicated phenomenon, and the rise velocity jump phenomenon appears in systems without surfactant, we do not go into further details on this topic.

Soto et al.\ \cite{Soto2006} investigated the rise behavior of single air bubbles in solutions of a hydrophobically modified alkali soluble polymer (HASE). These liquids exhibit a nearly constant shear viscosity over a wide range of shear rates, so that velocity enhancements due to shear thinning and due to elasticity can be distinguished. They compared the mean shear rate $\bar{{\dot \gamma}} = 2 U_T / d$ calculated from the terminal bubble rise velocity $U_T$ and the sphere-equivalent bubble diameter $d$ with data obtained from shear experiments. The mean shear rate, which can be attributed to the rise velocity jump, is approximately the same as the shear rate where the first normal stress difference $N_1$ became detectable in the shear experiments. The authors concluded that the jump can be directly related to elasticity, since the shear viscosity around the “jump shear rate” can be regarded as constant. In their opinion, the presence of significant normal stresses causes a change of the bubble shape, which finally leads to an abrupt drag reduction. They were able to classify the data of their rise velocity experiments by introducing the non-dimensional group 
\begin{equation}
\Pi = \frac{N_1 d}{\sigma},
\label{eq:group}
\end{equation}
which relates elastic forces to interfacial forces, the latter represented by $\sigma / d$. The jump appears at a critical value of $\Pi_{crit} \approx 0.25$ for their test liquids.

Flow visualization measurements by Funfschilling and Li \cite{Funfschilling2001} (Particle-Image Velocimetry (PIV) and birefringence) showed flow fields around bubbles rising in aqueous polyacrylamide solutions which were very different from those in a Newtonian glycerol solution. For polyacrylamide solutions, the flow field around the bubble can be divided into three distinct zones: an upward flow in front of the bubble, similar to that in the Newtonian case; a downward flow in the central wake, as described earlier as a negative wake (see, e.g., Hassager \cite{Hassager1979}); finally, a hollow cone of upward flow enclosing the region of the negative wake. The birefringence visualization qualitatively revealed a butterfly-like spatial distribution of shear stresses around the bubble within any plane going through the axis of symmetry.
Employing PIV measurements, Kemiha et al.\ \cite{Kemiha2006} observed a dependency of the opening angle of the hollow cone on the bubble Reynolds number, which converges to an asymptotic value for large bubble Reynolds numbers, whereas the evolution of the angle seems to depend on the nature of the fluid. Experiments carried out with glass spheres revealed similar flow patterns, thus providing evidence that the formation of the negative wake is not a consequence of the interface deformation (cusped bubble shape). Kemiha et al.\ \cite{Kemiha2006} concluded from their observations that the formation of a negative wake is governed by the liquid rheology (i.e., its viscoelasticity), which was supported by numerical simulations of a sedimenting sphere in a viscoelastic liquid with the Lattice-Boltzmann method. The numerically calculated flow field agreed qualitatively with the measured data.

Herrera-Velarde et al.\ \cite{Herrera-Velarde2003} found that the flow configuration changes drastically at the critical bubble volume, and the flow situation described as a negative wake appears for bubble volumes larger than the critical one. They noticed further that the size of the containment affects the magnitude of the jump, but not the critical bubble volume at which the jump occurs. This observation is in agreement with the PIV measurements by Soto and co-workers \cite{Soto2006}. 

Pilz and Brenn studied the rise of individual air bubbles in viscoelastic solutions of different polymers in various solvents \cite{Pilz2007}. Varying the bubble volume and measuring the steady bubble rise velocity, they showed that the jump discontinuity of the bubble rise velocity did not occur in all the liquids. The concentration of flexible polymers in the solutions needed to be high enough to enable the jump, and solutions of rigid, rod-like polymers  exhibited a bend rather than a jump in the curve of the rise velocity as a function of the bubble volume. These authors developed a criterion accounting for this fact. From their experiments they also derived an equation for the non-dimensional critical bubble volume at the jump discontinuity.

A breakthrough in the numerical description of the bubble motion may be seen in the paper by Pillapakkam et al.\ \cite{Pillapakkam2007}. In contrast to the simulations known from earlier literature, this paper presents a fully three-dimensional simulation of the flow field without any assumption, of axisymmetry for instance (see, e.g., Noh et al.\ \cite{Noh1993}, Frank and Li \cite{Frank2005}, Kemiha et al.\ \cite{Kemiha2006}, or Málaga and Rallison \cite{Malaga2007}). The constitutive equation in their calculations was the Oldroyd-B model with constant shear viscosity. The gas-liquid interface was captured by a level set method, using a constant value for the interfacial tension. The calculations revealed a rise velocity jump, showing increasing magnitude of the velocity enhancement with increasing “polymer concentration factor”, and the appearance of a negative wake for supercritical bubbles. Their calculations also showed an asymmetry in the cusp for bubbles with larger volumes, which was observed experimentally, e.g., by Liu et al.\ \cite{Liu1995} and Soto et al.\ \cite{Soto2006}. The paper \cite{Pillapakkam2007} demonstrates that the rise velocity jump is possible without any variations of interfacial tension or shear viscosity, but the only requirement seems to be three-dimensional flow. The authors conclude that the asymmetric cusped bubble shape, the additional vortex ring due to the negative wake, and the change in the velocity field due to the change in the bubble shape contribute to the velocity jump at a critical bubble volume in a certain parameter range. Theoretical investigations of the bubble shape revealed that the formation of a cusp- or corner-shaped trailing end can result in an upward force, since the integral of the surface tension force over the bubble surface is no longer zero. However, an estimation of the magnitude of that additional force with the para\-meters used in their simulations revealed that this force is too small to explain the strong increase of the rise velocity. On the other hand, it turned out that the additional upward force is approximately the same for different values of the polymer concentration parameter. Therefore, the authors of \cite{Pillapakkam2007} conclude that the global modification in the liquid velocity field around the bubble is presumably the reason for the steep increase of the bubble rise velocity. This is consistent with their calculations, since both the jump magnitude and the global modification of the velocity field in the ambient fluid are more substantial for higher values of the polymer concentration parameter.

Fraggedakis et al.\ \cite{Fraggedakis2016} numerically analyzed the velocity jump discontinuity by means of a dynamical systems approach. Using an axisymmetric arbitrary Lagrangian-Eulerian mixed finite element method, they computed quasi-steady solutions of the
governing equations for a single rising bubble in
a viscoelastic liquid. In particular, they studied the
setup of the experimental results from \cite{Pilz2007},
where the ambient liquid was approximated with an exponential Phan-Thien Tanner (EPTT) constitutive equation. The approach of \cite{Fraggedakis2016} is based on a continuation method, which allows to compute a manifold of quasi-steady state solutions for varying bubble radius. For two different polymer concentrations, very
good agreement with the experimental data from \cite{Pilz2007} has been obtained for the subcritical rise velocities, both concerning the rise velocity and the bubble shape. As the bubble radius approaches the critical value, the arc of simulated quasi-steady solutions in the diameter-velocity diagram exhibits a vertical tangent, i.e.\ infinite slope, and continues backward with increasing velocity until another turning point with infinite slope is reached and the arc of solutions continues further, with still increasing velocity but now also with increasing volume. This $S$-shaped arc of solutions is typical for dynamical systems which admit two stable as well as an unstable solution for a certain parameter range. Here, this parameter is the bubble diameter. Therefore, for increasing bubble diameter, the lower stable arc comes to an abrupt end at a certain critical value. For a slightly larger bubble, the only stable solution displays a significantly larger rise velocity.
The authors of \cite{Fraggedakis2016} reported  ``numerical difficulties when trying to continue on the upper branch of the solution family to higher velocities''. As possible reasons, the experimentally observed deviation from axisymmetry, also reported for the 3D numerical simulations in \cite{Pillapakkam2007},
and the increasing Weissenberg number which induces numerical problems, are mentioned.

In \cite{Niethammer2019}, an extended Volume-of-Fluid (VOF) method with a novel scheme for the second-order accurate finite-volume discretization of the interface stress term is employed to perform direct numerical simulations of the transient motion of a single bubble rising in a quiescent viscoelastic liquid. The rise velocity jump discontinuity and the negative wake are captured, with a quantitative match to the experimental data from \cite{Pilz2007}.
A similar VOF method has been used in \cite{Yuan2021} to simulate the same setup, i.e.\ single rising bubbles corresponding to the experiments from \cite{Pilz2007}, as well as different rise regimes concerning the Galilei number, including pulsating bubbles. The authors of \cite{Yuan2020} carried out three-dimensional numerical simulations with that same setup to study the stability of the bubbles during their rise in viscoelastic liquids characterized by the Weissenberg number and a viscosity ratio of the solvent to the solution. The influence of the fluid viscoelastic properties on the rise velocity jump discontinuity was pointed out. At high elasticity, for certain sets of the E\"otv\"os and Galilei numbers, vortex shedding in the wake region was found to yield a pulsating rise velocity. It was found that polymer stretching at the downstream end of the bubble may force the bubble to break up.

The investigation of the present paper identifies the molecular mechanism behind the rise velocity jump discontinuity for single bubbles rising in a viscoelastic liquid.
The present work combines three different approaches: experimental, numerical and theoretical. Put together, this--for the first time--gives a complete and clear picture of the underlying \emph{self-amplifying} mechanism leading not only to higher rise velocities, but explaining the abruptness of the change once the bubble volume exceeds a critical value.
Moreover, it sheds new light on the relation between the characteristic features mentioned above, i.e.\ the velocity jump, the negative wake and the cusped shape.
It also shows the intimate relationship of this phenomenon to the 3D flow kinematics. 

The paper is organized as follows:
Section~\ref{sec:experimentalmethod} presents the experimental methods used for the investigations of bubbles rising in viscoelastic liquids, including the flow field in the liquid. 
Section~\ref{sec:numericalmethod} details the numerical methods used for the simulations. In Section~\ref{sec:experimentalresults} we present the results from the experiments of this study, in particular the bubble rise velocities and the velocity and vorticity fields in the liquid. Section~\ref{sec:numericalresults} presents and validates the results from the numerical simulations in comparison with the experiments. The results from a theoretical analysis of the transport and conformation of polymer molecules in the flow around a bubble are presented and discussed in Section~\ref{sec:theory-results} in the context of the molecular processes behind the formation of the bubble rise velocity jump discontinuity. The findings are discussed and the conclusions are drawn in Section~\ref{sec:discussconclusions}.




\section{Experimental methods}
\label{sec:experimentalmethod}
\begin{figure}[ht!]    
  \centering
  \includegraphics[scale=0.7]{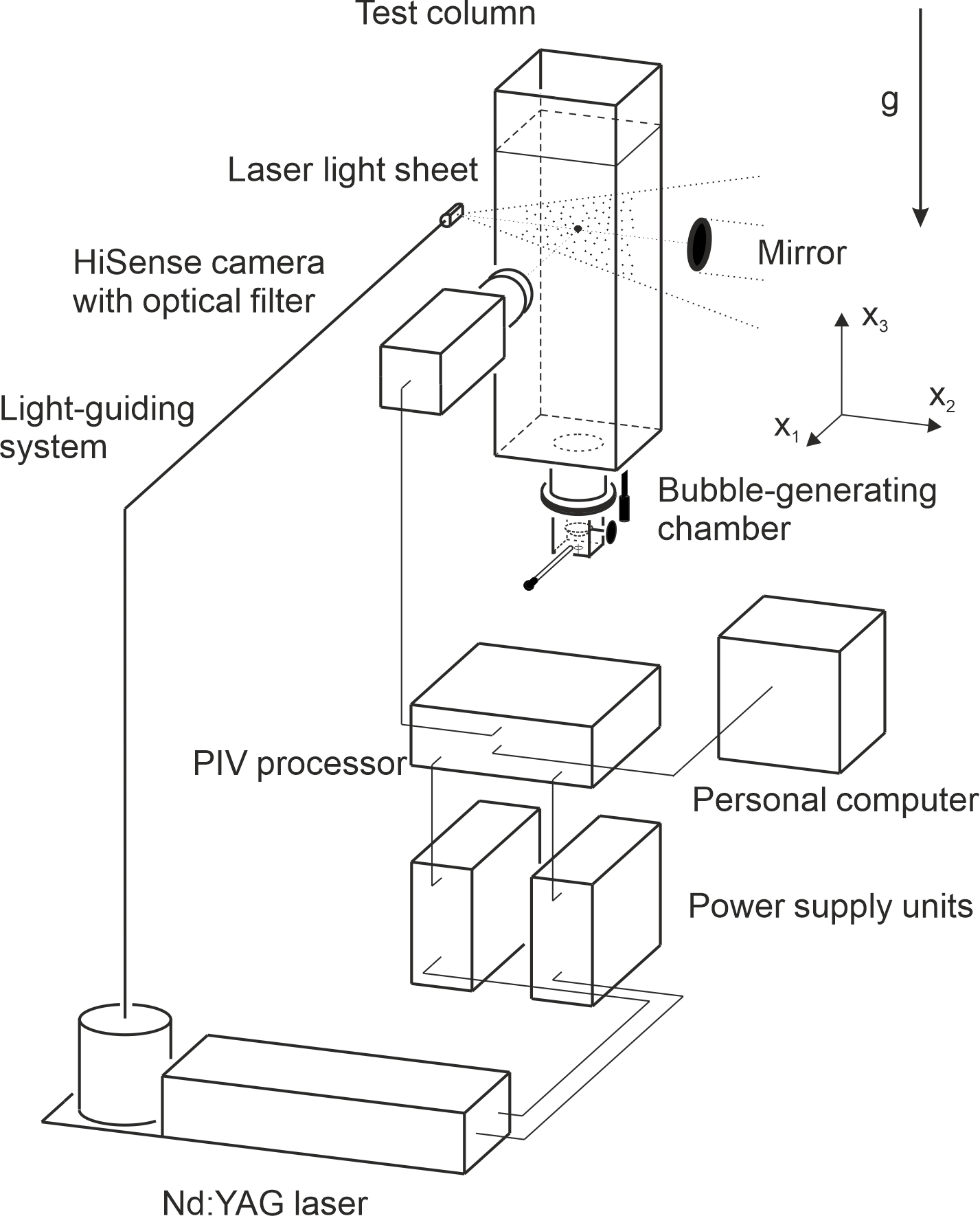}
  \caption{\small Experimental setup for investigating the flow field around bubbles in 
           viscoelastic liquids.}
     \label{fig:setup-PIV}        
\end{figure}

The experiments for measuring the terminal rise velocity of single bubbles in viscoelastic liquids
were carried out in an apparatus as shown in Fig.~\ref{fig:setup-PIV}. The apparatus and the experimental techniques were described in detail in \cite{Pilz2007}. Single air bubbles with controllable and well-defined volume in the range between $2$ and $1000~\textnormal{mm}^3$ were produced in the bubble-generating chamber by collecting individual bubbles detached from the end of a capillary. The resulting single bubbles were then allowed to rise in the liquid to a position half a meter above the location of their production to exclude any influence from the initial deformed shape of the bubble on the rise velocity. The viscoelastic liquids were solutions of poly(acrylamide)s and poly(ethyleneoxide) in water, in a $70\!:\!30$~wt.~\% mixture of glycerol with water, and in ethylene glycol. We limit the present study to the flexible polymers, excluding the hydrolyzed poly(acrylamide)s \cite{Pilz2009}. 

\subsection{Viscoelastic liquid characterization}
\label{sec:relax}
An important aspect in the dynamics of bubble motion in viscoelastic polymer solutions is the relaxation behavior of the polymer molecules in the solution, which is briefly reviewed in the following, since we will use characteristic time scales of the relaxation behavior in our explanation of the bubble rise velocity jump discontinuity phenomenon. 

Larson points out in his textbook \cite{Larson1999} that the stress in any polymeric system, dilute or concentrated, depends on the conformation of the polymer molecules. The contribution to the stress increases as the orientation and degree of stretch is increased. The total stress in a fluid element results from the individual conformations of all molecules in that element. A description of the stress arising from molecular theories must therefore include the distribution of conformations that results from deformation \cite{Larson1999}.

In order to describe the molecular process related to diffusion or relaxation, which changes the conformation for a flexible polymer, the polymer coil is usually modeled as a series of beads connected by springs \cite{Larson1999}. Although such a bead-and-spring chain does not represent the chemical structure of the polymer (Bird et al.\ \cite{Bird1987a,Bird1987b}), it is able to mirror the forces acting on the molecule, namely viscous and elastic forces \cite{Larson1999}, including a high number of degrees of freedom, as one expects from a flexible polymer \cite{Bird1987a,Bird1987b}. The relaxation of the entire molecule, i.e., the slowest relaxation, can be modeled by a single elastic dumbbell, which consists of only two spheres connected by a linear spring.

Yarin \cite{Yarin1993} constructed a system of constitutive equations for topological restrictions based on micromechanical polymer models by studying the mechanism of stress relaxation after a step strain according to Doi and Edwards \cite{DoiEdwards1986} and Doi \cite{Doi1980}. 

The analysis shows that the polymer relaxation time may be represented as 
\begin{equation}
\lambda \approx \frac{N_0^2 b^2 \zeta}{3 \pi^2 k_B T}.
\label{eq:timeconstant-sim}
\end{equation}
Herein, $N_0$ represents the number of Rouse segments in the entire chain, $b^2$ stands for the mean squared length of the Rouse segment at equilibrium, $\zeta$ is the friction factor (including solvent friction and interaction with the wall of a reptation tube \cite{Yarin1993}), $k_B$ the Boltzmann constant, and $T$ the temperature. Equation (\ref{eq:timeconstant-sim}) mirrors the statement given in Yarin (\cite{Yarin1993}, p.\ 363) that, under strong deformation (i.e., inside the liquid filament generated with a filament-stretching elongational rheometer, as used in our characterization of the polymer solutions), the relaxation processes which are faster than the reptational diffusion appear to be very important.

The paper \cite{Pilz2007} was -- to the best of our knowledge -- the first to account for liquid properties obtained from elongational rheometry in investigations of the critical bubble size at the rise velocity jump discontinuity. Our hypothesis is that the elongational behavior of the polymer solution plays a significant role in the rise of bubbles, since the liquid flow field around the bubbles exhibits considerable elongational components. Our rheometric characterization of the polymer solutions included both the shear viscosity at the first Newtonian plateau and the relaxation time obtained from filament stretching experiments carried out by means of the elongational rheometer described in Stelter et al.\ \cite{Stelter2000}. The device enables one to measure the diameter decrease of a self-thinning liquid filament with time after inducing a step strain within a liquid bridge between two plates.

In Stelter et al.\ \cite{Stelter2000} it was shown that the self-thinning of filaments of semi-dilute solutions of flexible polymers can be subdivided into two regimes. In the first regime, the filament diameter decreases exponentially with time, and in the second regime it decreases linearly with time. 
In the first regime, the polymer solution exhibits viscoelastic behavior. The filament diameter decreases according to the law
\begin{equation}
d_f (t) = d_{f0} ~\exp \left( - t / 3 \lambda_E \right) ~,
\label{eq:filthin}
\end{equation}
where $d_f$ is the filament diameter, $d_{f0}$ is its initial value (at time $t = 0$), and $\lambda_E$ is the relaxation time, which is equal to the one defined in Eq.\ \eqref{eq:timeconstant-sim}, i.e.\ $\lambda_E \equiv \lambda $.
In the second regime, a maximum of the polymer extension is reached, and the polymer solution exhibits Newtonian-like behavior, which allows the steady terminal elongational viscosity $\eta_{E}^\infty$ to be determined from the relation
\begin{equation}
d_f (t) = d_{f,0} - \frac{\sigma}{\eta_{E}^\infty } t~,
\label{eq:termvisc}
\end{equation}
where $\sigma$ denotes the surface tension.

It should be pointed out that, in the elasticity-dominated regime, the rate of stretching within the filament,
\begin{equation}
\dot \epsilon = \frac{\partial u_z}{\partial z} = - \frac{2}{d_f} \frac{{\rm d}\, d_f}{{\rm d}\, t}  
\label{eq:thinning}
\end{equation}
is constant during the self-thinning process. We show this by combining Eqs.\ \eqref{eq:filthin} and \eqref{eq:thinning} to obtain
\begin{equation}
\dot \epsilon = \frac{2}{3 \lambda_E}.
\label{eq:thinconst}
\end{equation}
The relaxation time $\lambda_E$ obtained from the viscoelastic regime of the filament self-thinning can be related to the longest time constant of the Zimm spectrum obtained from shear rheometric measurements \cite{Anna2001}.

It is therefore emphasized that, regardless of the mechanical properties of the polymer, the polymer concentration or the solvent quality, the dependency of the filament diameter on time (i.e., Eq.\ \eqref{eq:filthin}) does not change, and the relaxation time $\lambda_E$ calculated from the measured evolution of $d_f$ does not depend on any particular assumptions on the macromolecular model and polymer/polymer interactions \cite{Stelter2002}. This parameter $\lambda_E$ plays an essential role in the prediction of the critical bubble size at the rise velocity jump discontinuity \cite{Pilz2009}.

\subsection{PIV for flow field measurements}
\label{sec:PIV}
For investigating the flow field around the bubble, Particle-Image Velocimetry (PIV) was used as the measuring technique. The PIV system employed was the Dantec FlowMap 1500 system (Dantec Dynamics, Skovlunde, Denmark), together with a Dantec 80C60 HiSense camera ($1280 \times 1024$ pixel) and a NewWave Gemini double-cavity Nd:YAG laser ($120~\textnormal{mJ}/\textnormal{pulse}$). 
A special light-sheet optics consisting of a spherical lens ($600~\textnormal{mm}$ focal length), a cylindrical lens ($-10~\textnormal{mm}$ focal length), and a prism was used to illuminate the seeding particles in plane sections of the flow. The light sheet with a thickness of approximately $5~\textnormal{mm}$ was placed in a symmetry plane of the test column, which contains the trajectory of the bubble center. 

The system was calibrated using a Dantec calibration target, a $100 \times 100~\textnormal{mm}$ white plate with a square grid of black dots of spacing $2.5~\textnormal{mm}$ and size $1.5~\textnormal{mm}$.

\begin{figure}[bt]    
  \centering
  \includegraphics[scale=0.5]{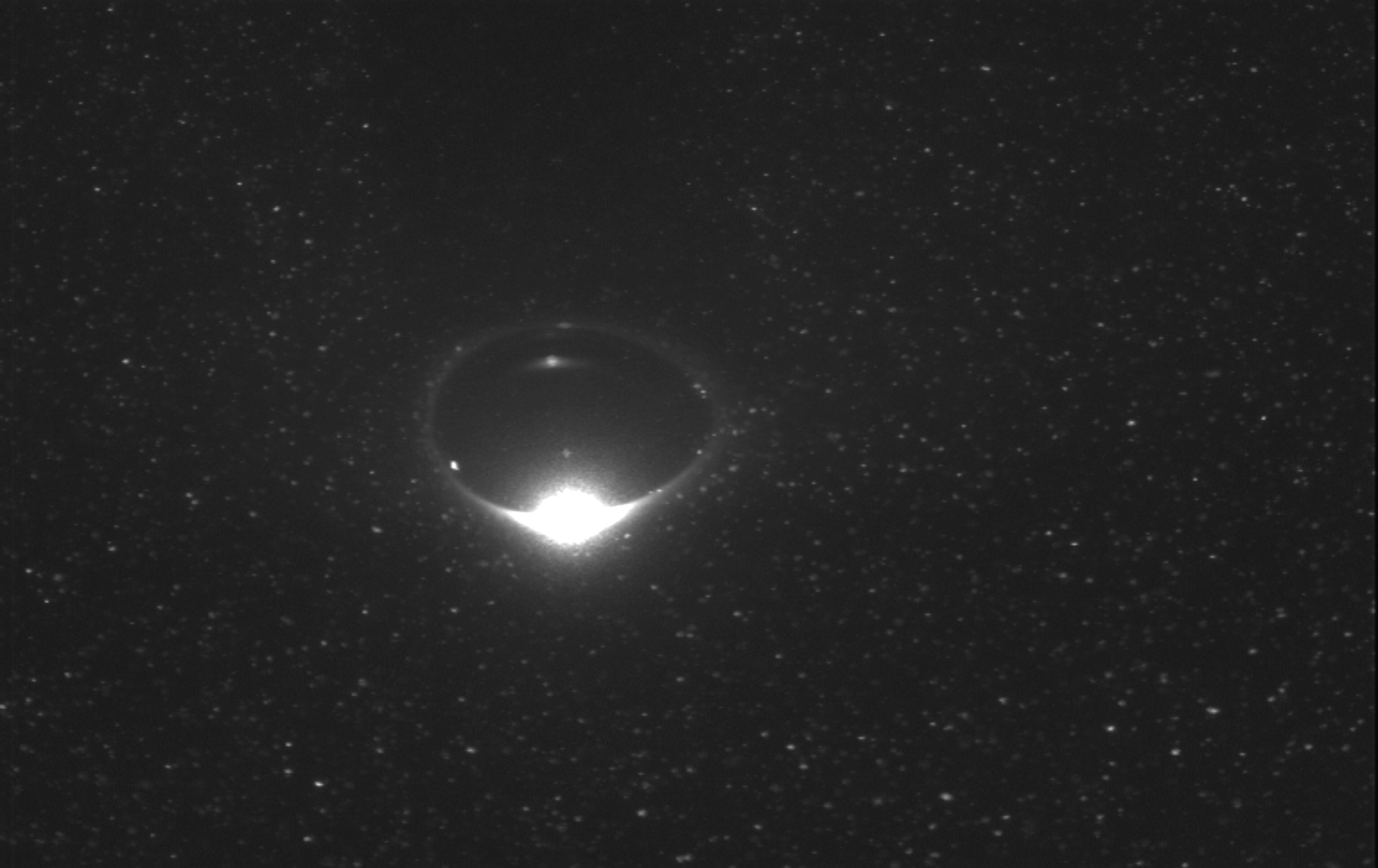}
  \caption{\small Scattered light from the bubble surface.}
   \label{fig:PIV-reflexion-bubblesurface}
\end{figure}

The essential difficulty when measuring with PIV in a liquid field with bubbles is the strong reflection of the irradiated laser light from the surfaces of the bubbles, as shown in Fig.~\ref{fig:PIV-reflexion-bubblesurface}. 
The bright zone on the images hinders an accurate velocity measurement in the vicinity of a part of the bubble surface, which must be avoided. 
A remedy is the use of fluorescent particles as tracers in the liquid phase, and of an optical filter, which lets only the fluorescent light pass and removes the Mie-scattered laser light reflected from the bubble surface. 
Fluorescent particles were produced for the present investigations by saturating polyamide particles ($20~\mathrm{\upmu m}$ diameter, PSP20, Dantec Dynamics, Skovlunde; Denmark) with the dye Eosin 
$(\mathrm{C}_{20}\mathrm{H}_{6}\mathrm{Br}_{4}\mathrm{Na}_{2}\mathrm{O}_{5})$ 
and drying them such that the dye became insoluble in the polymer solutions. For this purpose, the polyamide particles were soaked in an alcoholic Eosin solution for $24$ hours 
and then washed with distilled water, filtered, and dried in air. 
Washing and filtering were repeated seven times. 

A quantity of $0.4~\textnormal{g}$ particles was added to $40~\textnormal{ml}$ of the polymer solution and dispersed using an ultrasonic bath. 
This concentrated solution was then stirred into $7.5$ liters of the aqueous polymer solution. 
An additional mirror opposite from the light sheet probe helped to reflect light from the light sheet into the shadow from the bubbles.

As a result, pictures were obtained with clearly visible particle images and sufficiently visible bubble contours to determine the bubble volume by the image processing method described in section~\ref{sec:imageprocessing-PIV} below; cf.~\cite{Pilz2009}.

The double images from one PIV recording were evaluated using a cross-correlation 
technique and an interrogation area size of $32 \times 32$ pixel with 50~\% overlap.
The time between the two pulses was 1--100 ms, depending on flow situation and setup. 
Finally, a range validation, a correlation peak ratio validation, 
and a moving average validation were applied to reject invalid vectors 
(Dantec Flow Manager software, v.\ 4.60.28). 
Maps of $79 \times 63$ vectors were obtained. 
The results from these investigations are presented below in section~\ref{sec:PIV-velocityfields}.

\subsection{Image Processing}
\label{sec:imageprocessing-PIV}
\begin{figure}[ht!]    
  \centering
  \includegraphics[scale=1]{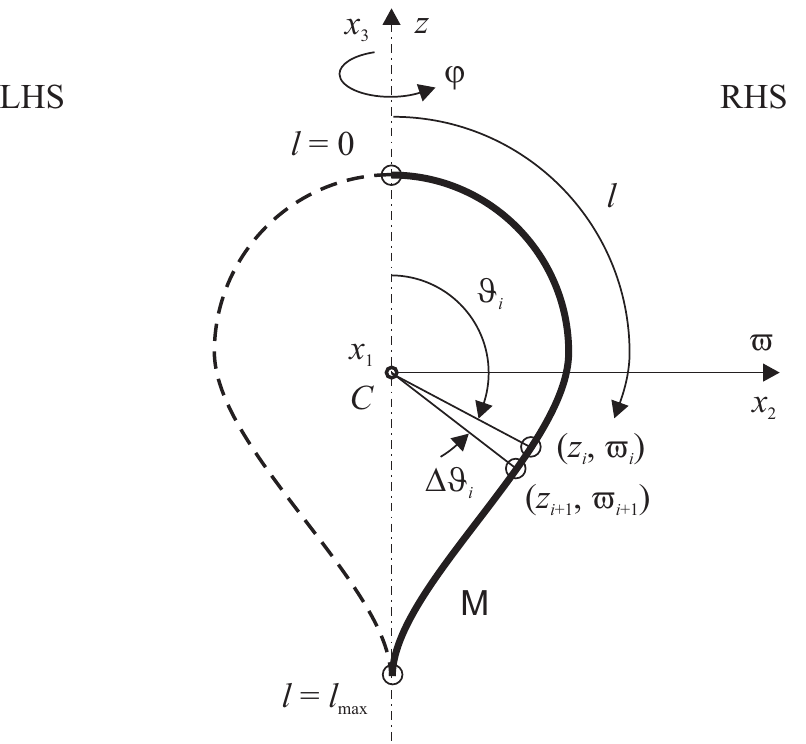}
  \caption{\small Extraction of the boundary curve.}
    \label{fig:image-processing-PIV}
\end{figure}

The shape of the bubble and its position relative to the picture frame were extracted from each picture using image processing tools provided by Adobe Photoshop and Matlab. 
In order to preserve the contour of the bubble, images of the seeding particles had to be 
removed from the frame. For this purpose, filter options for contrast enhancement and masking tools provided by Adobe Photoshop were applied first. 
Furthermore, the centroid $C$ of the meridional area was determined, and the bubble contour points were filtered out. The coordinates of the bubble contour points were transformed into a local coordinate system $(x_{1}, x_{2}, x_{3})$ with right-handed orientation and the $x_{1}$ axis pointing towards the camera. 
Furthermore, the origin of the coordinate system coincides with the centroid $C$ of the bubble's meridional area.
The two in-plane axes $(x_{2}, x_{3})$ represent the principal axes of the bubble's 
meridional area computed via image processing tools with Matlab. 
Since, for the volume range of interest, axial symmetry around the direction of motion of the bubbles can be assumed 
and, furthermore, the bubbles rise along a straight vertical path, 
the axis of symmetry (denoted $x_{3}$ in Fig.~\ref{fig:image-processing-PIV}) 
points in the direction of motion. 
The volume of the air bubble and the arc length of the contour 
are obtained from an approximation of the meridional curve $M$,
which represents one half of the previously extracted boundary curve by means of piecewise 
linear functions through two adjacent points $(z_{i}, \varpi_{i})$ and $(z_{i + 1}, \varpi_{i + 1})$ 
of the boundary curve. 
The separation of the boundary curve into left-hand and right-hand sides was carried out 
after calculation of the corresponding polar angles $\vartheta_{i}$ for each 
data point of the curve. 
The data points were sorted in ascending order, based on the angles $\vartheta_{i}$, 
where negative values of $\vartheta_{i}$ refer to the left-hand side (LHS) of the bubble contour, 
and positive values refer to the right-hand side (RHS) (see Fig. \ref{fig:image-processing-PIV}). 
A partial volume  $\mathrm{\Delta} V_{i}$ of the air bubble is then computed 
form data of one half of the boundary curve according to 
\begin{equation}
\label{eq:partial-volume}
\mathrm{\Delta} V_{i} =
\frac{\pi}{3} \left( \varpi_{i}^{2} + \varpi_{i} \varpi_{i + 1} 
+ \varpi_{i + 1}^{2} \right)(z_{i + 1} - z_{i}) \; \text{.}
\end{equation}
These partial bubble volumes were then added up to obtain the total volume. 

Similarly, the arc lengths $l_{i}$ corresponding to the discrete points $(z_{i}, \varpi_{i})$ 
of the meridian curve (i.e.\ one half of the boundary curve) 
were calculated in accordance with 
the approximation of the boundary curve by
\begin{equation}
\label{eq:arc-length-PIV}
  l_{i} = l_{i - 1} + \sqrt{\left(z_{i} - z_{i - 1} \right)^{2}%
   + \left(\varpi_{i} - \varpi_{i - 1} \right)^{2}}~.
\end{equation}

The bubble rise velocity was obtained from the relative displacement of the bubble  
contour curve within two consecutive images and checked by comparison with the maximal velocity measured at the boundary curve by PIV. For depicting liquid-phase velocity fields, the bubble rise velocity was subtracted from the instantaneous velocities measured by PIV.
Before doing this, the PIV data had to be transformed 
into the local coordinate system $(x_{1}, x_{2}, x_{3})$, 
since the centroid of the bubble meridian plane did not necessarily coincide 
with the reference mark (i.e.\ the ``zero marker'') of the calibration target \cite{Pilz2009}.
%

\section{Numerical method}
\label{sec:numericalmethod}
Direct numerical simulations (DNS) of bubbles rising in a viscoelastic fluid are carried out using an extended volume of fluid (VOF) method, introduced in Niethammer et al. \cite{Niethammer2019, Niethammer2019b}. Numerical experiments in \cite{Niethammer2019, Niethammer2019b, Niethammer2019c} demonstrate that the extended VOF method captures characteristic flow phenomena of single bubbles rising in a quiescent viscoelastic fluid, such as the rise velocity jump discontinuity and the negative wake. It is noteworthy that good quantitative agreement of the steady-state rise velocities was achieved between experimental measurements \cite{Pilz2007} and three-dimensional DNS \cite{Niethammer2019}. Besides the velocity jump discontinuity and the negative wake, the present study provides a detailed insight into the local stress and conformation tensor distributions in the neighborhood of the rising bubble. Moreover, the spectrum of the conformation tensor, especially the leading eigenvalues and corresponding eigenvectors, is investigated to analyze the orientation of the polymer molecules; cf.\ Appendix~A about the meaning of the eigenvalues of the conformation tensor for the orientation of the polymer molecules.

The extended VOF method employs a one-field formulation based on the volume-averaged two-phase flow equations. The reader is referred to \cite{Niethammer2019} for the derivation of the volume-averaged local instantaneous bulk equations and interface jump conditions. In section
\ref{sec:vofmodel}, we provide a summary of the extended VOF model.

\subsection{Governing equations}
\label{subsec:governing-equations}
Assume the two-phase fluid system to fill the domain ${\CCV \subset \mathbb{R}^3}$ under consideration, such that the two bulk phases $\CCV^{\phaseOne}(t)$ and $\CCV^{\phaseTwo}(t)$ are separated by the two-dimensional surface ${\Sigma(t) \in \mathbb{R}^3}$, i.e.\ $\CCV = {\CCV^{\phaseOne}(t) \cup \CCV^{\phaseTwo}(t) \cup \Sigma(t)}$. Let us further denote the interface unit normal pointing towards the phase $\CCV^{\phaseOne}$ as $\vec{n}_{\Sigma}$.

Assuming isothermal flow with constant mass densities in the bulk phases, the balance equations for mass and momentum in their local formulation in the bulk read
\begin{align}
\label{contieq00}
\nabla \dprod \velocity &= 0 \quad \textnormal{in} \ \CCV \setminus \Sigma, \\
\label{monentumeq00}
\rho {\partial_t \velocity} + \rho \left(\velocity \dprod \nabla\right) \velocity &= \nabla \dprod \ST + \rho \vec{b} \quad \textnormal{in} \ \CCV \setminus \Sigma,
\end{align}
where $\velocity$ is the velocity, $\rho$ is the density and the term $\rho \vec{b}$ represents body forces with the mass specific density $\vec{b}$. The term $\nabla \dprod \ST$ represents contributions from contact forces, where the second rank tensor $\ST$ is the Cauchy stress.

At the interface $\Sigma$, additional jump conditions are needed in order to obtain a complete system. We define the jump $\jump{ \PhiT }$ of a quantity $\PhiT$ at $\Sigma$ as
\begin{align}
\label{defjump}
\jump{ \PhiT } (\vec{x}) = \lim\limits_{h \rightarrow 0^{+}} \left(\PhiT (\vec{x} + h\vec{n}_{\Sigma}) - \PhiT (\vec{x} - h\vec{n}_{\Sigma})\right), \quad \vec{x} \in \Sigma.
\end{align}
Then, assuming no-slip at $\Sigma$,
the interfacial jump conditions are 
\begin{align}
\label{jcontieqVV00}
\jump{ \velocity }  &= 0 \quad \textnormal{on} \ \Sigma, \\
\label{jmomentumeqVV00}
{\jump{- \ST}} \dprod \vec{n}_{\Sigma} &= \sigma \kappa \vec{n}_{\Sigma} 
\quad \textnormal{on} \ \Sigma, \\
\label{snd00}
\velocity \dprod \vec{n}_\Sigma &= V_\Sigma \quad \textnormal{on} \ \Sigma,
\end{align}
where $\sigma$ denotes the surface tension and the local curvature (more precisely, twice the mean curvature) is defined as ${\kappa = \nabla_{\Sigma} \dprod (-\vec{n}_{\Sigma})}$ with the surface gradient operator $\nabla_\Sigma$.
The kinematic condition (\ref{snd00}) for the interface normal velocity  is satisfied in absence of phase change, with $V_\Sigma$ denoting the speed of normal displacement of $\Sigma$.
Note that in the momentum transmission equation (\ref{jmomentumeqVV00}), we assume a constant surface tension $\sigma$ and neglect the effect of any intrinsic interfacial rheology, i.e.\ assume zero surface viscosities. This is a good assumption as long as local microstructures along the interface, e.g.\ interfacial polymer monolayers, have a minor impact compared to the bulk-fluid stresses acting across the interface.

We split the stress $\ST$ into an isotropic part, a viscous solvent component and an elastic polymer contribution, i.e.\
\begin{align}
\label{stressdef00}
\ST = -p \, \IT + \stressS + \stressP,
\end{align}
where $p$ is the pressure. The viscous stress obeys Newton's law, i.e.\
\begin{align}
\label{stressdef01}
\stressS = \svisc \big( \grad{\velocity} + \trans{\grad{\velocity}} \big),
\end{align}
where $\svisc$ is the solvent viscosity. 
The polymer stress is directly related to a conformation tensor $\CT$, representing the average configuration of the polymer molecules in the macroscopic fluid element \cite{Bird1987a, Larson1988}. In particular, the conformation tensor is related to the second moment of the distribution function of the molecular configuration \cite{Larson1988}. By its definition, this implies that $\CT$ is a symmetric and positive definite (spd, for short) tensor of second rank; cf.\ also Appendix~\ref{asec:conftens}.
Throughout this paper, we exploit the fact that every spd tensor is diagonalizable with real, positive eigenvalues.

The polymer stress $\stressP$ is a linear function of the  polymer conformation tensor $\CT$, viz.\
\begin{align}
\label{stressdef02}
\stressP = \frac{\pvisc}{\lambda (1-\zeta)} \left( h_{0} \IT + h_{1} \CT\right),
\end{align}
where $\pvisc$ is the polymer viscosity, $\lambda$ is the relaxation time and $\zeta$ is a material parameter (sometimes called ``slip parameter"). Note that (\ref{stressdef02}) can be derived from a closed form of the Kramers' expression for the polymer stress tensor \cite{Bird1987b, Larson1988}. The coefficients $h_{j}$ depend on the molecular model and the closure approximation. For most rheological models, the scalar coefficients $h_{j}$ are either functions of the first invariant of $\CT$, i.e.\ $h_{j} = \hat{h}_{j}(\tr{\CT})$, or constants (cf.\ Table \ref{tab:constitutiveEx}).

To obtain a closed system, additional constitutive equations for $\CT$ are required. In this work, we restrict our considerations to partial differential conformation tensor constitutive equations of the form
\begin{align}
\label{constCEq00}
\partial_t \CT + \left( \velocity \dprod \nabla \right)  \CT  - \LT \dprod \CT - \CT \dprod \trans{\LT}
=
\frac{1}{\lambda} \cvec{P}\left(\CT \right) \quad \textnormal{in} \ \CCV \setminus \Sigma
\end{align}
with $\LT = \trans{\grad \velocity} - \frac{\zeta}{2} \left({\grad \velocity} + \trans{\grad \velocity}\right)$ and the material parameter ${\zeta} \in [0, 2]$ characterizing the degree of non-affine response of the polymer chains to an imposed deformation. For ${\zeta} = 0$, the motion becomes affine, and in this case the left-hand side of (\ref{constCEq00}) reduces to the upper-convected derivative, first proposed by Oldroyd~\cite{Oldroyd1950}. The relaxation is introduced by a specification of the tensor function $\cvec{P} \left(\CT \right)$. We impose the restriction that $\cvec{P} \left(\CT \right)$ is a real analytic tensor function of $\CT$, such that
\begin{equation}
\label{defisotropicf}
\OT \dprod \cvec{P} \left(\CT \right) \dprod \trans{\OT} = \cvec{P} \big( \OT \dprod \CT \dprod \trans{\OT} \big)
\end{equation}
for every orthogonal tensor $\OT$. In this case, $\cvec{P} \left(\CT \right)$ is referred to as an isotropic tensor-valued function of the second-rank spd tensor $\CT$. It is well-known \cite{RivlinEricksen1955, Rivlin1955} that every real analytic and isotropic tensor-valued function $\cvec{P} \left(\CT \right)$ of a second-rank tensor $\CT$ has a representation of the form
\begin{equation}
\label{Pfunc}
\cvec{P}\left(\CT \right) = g_{0} \IT + g_{1} \CT + g_{2} {\CT}^2,
\end{equation}
where $g_{0}$, $g_{1}$ and $g_{2}$ are isotropic invariants (isotropic scalar functions) of $\CT$ and hence can be expressed as functions of its three scalar principal invariants, i.e.\ $g_{i} = \hat{g}_{i}(I_{1}, I_{2}, I_{3}), \ i = 0,1,2$. The principal invariants $I_{1}, I_{2}, I_{3}$ are
\begin{equation}
I_{1} = \textnormal{tr} \: \CT, \ I_{2} = \frac{1}{2} 
\left[ (\textnormal{tr} \: \CT)^2 - \tr{{\CT}^2} \right]\!, \ I_{3} = \det \CT.
\end{equation}

Here, we only use models for which $g_{2} = 0$, i.e.\ the quadratic term in (\ref{Pfunc}) vanishes. Table \ref{tab:constitutiveEx} summarizes the model-specific expressions for the constitutive equations used in the present work.
\begin{table}[h!]
\small
\caption{\small Model-dependent scalar-valued tensor functions for the generic constitutive equation (\ref{constCEq00}) and (\ref{Pfunc}).}
\label{tab:constitutiveEx}
\begin{center}
\begin{tabular}{@{}lcccccc@{}}
\toprule
constitutive model & \ ${\zeta}$ \ & \ $g_{0}$ \ & \ $g_{1}$ \ & \ $g_{2}$ \ & \ $h_{0}$ \ & \ $h_{1}$ \ \\ 
\midrule
Maxwell/Oldroyd-B & $0$ & $1$ & $-1$ & $0$ & $-1$ & $1$ \\
LPTT & $\in \left[0, 2 \right]$ & $1 + \frac{\varepsilon}{1 - {\zeta}} \left( \tr \CT - 3 \right) $ & $ - g_{0} $ & $0$ & $-1$ & $1$  \\  
EPTT & $\in \left[0, 2 \right]$ & $ \exp \left[  \frac{\varepsilon}{1 - {\zeta}} \left( \tr \CT - 3 \right) \right] $ & $ - g_{0} $ & $ 0 $ & $-1$ & $1$ \\
\bottomrule
\end{tabular}
\end{center}
\end{table}

\subsection{Stabilization}
\label{subsec:stab}
Because of the high Weissenberg number problem (HWNP)~\cite{Joseph1985, Keunings1986}, a numerical stabilization is required to increase the robustness of the numerical method at moderate and high degrees of fluid elasticity. Fattal and Kupferman~\cite{Fattal2004} showed that a logarithmic change of variables in the conformation tensor equation circumvents the HWNP. Balci et al.~\cite{Balci2011} proposed a square root conformation tensor representation that does not require any diagonalization of the conformation tensor. Afonso et al.~\cite{Afonso2011} generalized the conformation representations with generic kernel transformation functions. In the present work, we use the unified mathematical and numerical stabilization framework, proposed by Niethammer et al.~\cite{Niethammer2018} for the general conformation tensor constitutive equations (\ref{constCEq00}) and (\ref{Pfunc}). This generic framework has been validated~\cite{Niethammer2018, Niethammer2019c} and extended to non-isothermal~\cite{Meburger2020} and two-phase flows~\cite{Niethammer2019}. For a detailed description of the generic stabilization framework and its validation in computational benchmarks, we refer to~\cite{Niethammer2018, Niethammer2019b}. Below, we give a brief summary of the stabilization framework used in the present work.

The generic constitutive equation is obtained by introducing the real analytic tensor function $\cvec{F}(\CT)$ of the spd second-rank tensor $\CT$, such that
\begin{equation}
\label{diagonalizef0}
\OT \dprod \cvec{F}(\CT) \dprod \trans{\OT}
=
\cvec{F}(\OT \dprod \CT \dprod \trans{\OT})
\end{equation}
for every orthogonal tensor ${\OT}$, i.e.\ $\cvec{F}(\CT)$ is an \textit{isotropic} tensor-valued function of $\CT$. If $\CT$ is governed by the generic constitutive equation (\ref{constCEq00}), then $\cvec{F}(\CT)$ satisfies
\begin{equation}
\label{evolutionConf3}
\DDt {\cvec{F}(\CT)}
=
{2 \BT \dprod {\UpsilonT} \dprod \CT}
+
\OmegaT \dprod {\cvec{F}(\CT)} - {\cvec{F}(\CT)} \dprod \OmegaT
+
\frac{1}{\lambda} \UpsilonT \dprod \cvec{P} \left(\CT \right),
\end{equation}
where we exploit the representation $\CT = \QT \dprod \DT \dprod \trans{\QT}$ with the diagonal tensor $\DT$ containing the three real eigenvalues $d_1, d_2, d_3$ and the orthogonal tensor $\QT$ containing the corresponding set of eigenvectors. Then, $\cvec{F}(\CT) = \cvec{F}(\QT \dprod \DT \dprod \trans{\QT}) = \QT \dprod \cvec{F}(\DT) \dprod \trans{\QT}$. The deformation terms in the convective derivative are decomposed into the first three terms on the right-hand side of (\ref{evolutionConf3}), containing the tensors $\BT$ and $\OmegaT$, which are functions of the velocity gradient. This local decomposition of the velocity gradient was first proposed in~\cite{Fattal2004} and is in general necessary to transform the conformation tensor equation (\ref{constCEq00}) into an evolution equation for the analytic tensor function $\cvec{F}(\CT)$. The functional relationship of $\BT$ and $\OmegaT$ with $\LT$ is given in Appendix \ref{asec:lddt}.


The generic constitutive equation (\ref{evolutionConf3}) falls back to the classical logarithm or root conformation representations (LCR or RCR), depending on the choice of $\cvec{F}(\CT)$ and $\UpsilonT$.
Table~\ref{tab:JJ}
\begin{table}[htbp]
\caption{\small Function $\cvec{F}(\CT)$ and $\UpsilonT$ terms in (\ref{evolutionConf3}) for the conformation tensor representation, the $k$-th root of $\CT$ representation, and the logarithm of $\CT$ to base $a$ representation.}
\renewcommand{\arraystretch}{1.35}
\begin{center}
\linespread{1.15}
\begin{tabular}{rlll}
\toprule
\ & $\cvec{F}(\CT)$ & \ $\UpsilonT$ \ \\
\midrule
conformation tensor \ & $\CT$ & \ $\IT$ \ \\
$k$-th root of $\CT$ \ & $\RT = \QT \dprod \DT^{\frac{1}{k}} \dprod \trans{\QT}$ & \ $\frac{1}{k} \RT^{1-k}$ \ \\
logarithm of $\CT$ to base $a$ \ & $\ST = \QT \dprod \loga{(\DT)} \dprod \trans{\QT}$ & \ $\frac{1}{\operatorname{ln}(a)} a^{-\ST}$ \ \\
\bottomrule
\end{tabular}
\end{center}
\renewcommand{\arraystretch}{1.0}
\label{tab:JJ}
\end{table}
shows the corresponding terms for specific conformation representations. In the present work, we choose the RCR, given by
\begin{align}
\label{evolutionConfRTb}
\DDt \RT
&=
\frac{2}{k} \BT \dprod \RT
+
\OmegaT \dprod {\RT} - {\RT} \dprod \OmegaT
+
\frac{1}{k \lambda} \left(g_0 \RT^{1-k} + g_1 \RT + g_2 \RT^{1+k}\right).
\end{align}
In preliminary studies we tried root stabilization with $k=2,3, 4$, from which $k=4$ turned out to be the optimal choice concerning stability and convergence. 
We also tried the logarithmic transformation
but did not observe significant differences.
Therefore, all results shown below have been obtained using the $4^{\rm th}$ order root.
\subsection{The Volume of Fluid (VOF) model}
\label{sec:vofmodel}
The VOF method \cite{Hirt1981} is an interface capturing method, widely used for sharp interface direct numerical simulations (DNS). Due to its inherent mass conservation and robustness in handling strong morphological changes, the VOF method is an appropriate choice for the simulation of rising bubbles in a viscoelastic fluid. It was first shown in Niethammer et al.~\cite{Niethammer2019} that the VOF method is capable to capture the bubble rise velocity jump discontinuity. Recently, the VOF results for the velocity jump have been confirmed by Yuan et al.~\cite{Yuan2021}.

In the VOF method, the fluid interface is captured by a phase indicator function. We employ the approach of an algebraic VOF, where the propagation of the fluid interface is achieved implicitly by solving a scalar transport equation for the volume fraction field. The volumetric phase fraction $\alpha$ is defined over the entire domain as the volume average of the phase indicator function over a mesh cell and is constrained to satisfy $0 \leq \alpha \leq 1$. We let $\alpha = 1$ in the viscoelastic phase, hence $\alpha = 0$ in the gas phase of the bubble, and the interface is located in mesh cells, where the volume fraction has values $0 < \alpha < 1$.

The extended VOF equations for viscoelastic fluids are derived from the local instantaneous volume equations and the interfacial jump conditions from section~\ref{subsec:governing-equations} by successively applying a volume averaging procedure and the homogeneous mixture model as a closure. The definitions of volume-averaged quantities are found in the appendix \ref{asec:vadefinitions} and for a detailed derivation of the extended VOF equations we refer to \cite{Niethammer2019}. As a result, we obtain a single set of equations, valid for both phases throughout the computational domain, which is shown below. Introducing the definition for the mixture density
\begin{align}
\label{mixdensity00}
\rho_m = \alpha {\rho}^{\phaseOne} + \left(1 - \alpha \right) {\rho}^{\phaseTwo},
\end{align}
the mixture velocity
\begin{align}
\label{mixvelocity00}
{\velocity}_m = \frac{\alpha \overline{\velocity}^{\phaseOne} + \left(1 - \alpha \right) \overline{\velocity}^{\phaseTwo}}{\rho_m}
\end{align}
and the mixture stress
\begin{align}
\label{mixstress00}
{\ST}_m = \alpha \overline{\ST}^{\phaseOne} + \left(1 - \alpha \right) \overline{\ST}^{\phaseTwo},
\end{align}
the mass and momentum equations can be written as
\begin{align}
\label{mixContieq00}
\nabla \dprod {\velocity}_m &= 0, \\
\label{mixUeq00}
\partial_t {\rho}_m {\velocity}_m 
+ ({\velocity}_m \dprod \nabla) {\rho}_m {\velocity}_m
&= {\nabla \dprod {\ST}_m}
+ {\rho}_m \vec{b}
+ \vec{f}_\sigma,
\end{align}
where $\vec{f}_\sigma$ represents the surface tension force.
We employ the widely used continuous surface force (CSF) model proposed by Brackbill et al.~\cite{Brackbill1992} as a closure for the surface tension force.
The scalar transport equation for the volume fraction field reads
\begin{align}
\label{vofraceq02b}
\partial_t \alpha + \left({\velocity}_m \dprod \nabla\right)  \alpha  = 0. \end{align}
Note that the discretization of the advection term in (\ref{vofraceq02b}) requires special numerical schemes that simultaneously keep the interface sharp and produce monotonic profiles of the volume fraction between its bounds $0 \leq \alpha \leq 1$. In the present work, we use the multi-dimensional flux corrected transport (FCT) method, introduced by Boris and Book \cite{Boris1973} and extended to multiple dimensions by Zalesak \cite{Zalesak1979}. The FCT method is a multi-dimensional flux limiting strategy that applies a local weighting procedure between an unconditionally bounded low-order upwind differencing scheme and an anti-diffusive higher-order correction scheme, such that the higher-order flux is used to the greatest extent possible, while keeping the solution bounded between some local extrema. For the higher-order corrections, we use the inter-Gamma scheme \cite{Jasak1995} which introduces a certain amount of downwind differencing to maintain the sharpness of the interface. 

The mixture stress tensor ${\ST}_m$ in the momentum balance (\ref{mixUeq00}) is split according to
\begin{align}
\label{mixstress01}
{\ST}_m = -\grad p + {\nabla \dprod \stressSm} + {\nabla \dprod \stressPm},
\end{align}
where the mixture solvent stress is defined as
\begin{align}
\label{closestress04a}
{\stressSm} &= {\smvisc} \big(\grad{{\velocity}_m} + \trans{\grad{{\velocity}_m}}\big)
\end{align}
with mixture solvent viscosity
\begin{align}
\label{mixviscosity00}
{\smvisc} = \alpha {\eta}_{s}^{\phaseOne} + \left(1 - \alpha \right) {\eta}_{s}^{\phaseTwo}.
\end{align}
Using the reformulation 
\begin{align}
\label{closestress04b}
{\nabla \dprod \stressSm} = \nabla \dprod \left({\smvisc} \grad{{\velocity}_m}\right) + \grad{{\velocity}_m} \dprod \nabla {\smvisc}
\end{align}
for the solvent stress divergence in (\ref{mixstress01}) is beneficial due to the elliptic (Laplacian-type) term, which is discretized implicitly in this work.
For the closure of the polymer stress term in (\ref{mixstress01}), we introduce the definitions
\begin{align}
\label{mixpstress00}
 {\stressPm} &= \alpha \overline{\stressE}_{p}^{\phaseOne} + \left(1 - \alpha \right) \overline{\stressE}_{p}^{\phaseTwo},\\
 \label{mixG00}
 G_m &=  \alpha G^{\phaseOne} + \left(1 - \alpha \right) G^{\phaseTwo}, \\
  \label{mixC00}
 {\CT}_m  &= \frac{\alpha G^{\phaseOne} \overline{\CT}^{\phaseOne} + \left(1 - \alpha \right) G^{\phaseTwo} \overline{\CT}^{\phaseTwo}}{G_m},
\end{align}
where $G^{\pm} = \frac{\eta_{p}^{\pm}}{\lambda^{\pm} \left(1 - \zeta^{\pm} \right)}$. 
Inserting (\ref{mixpstress00})--(\ref{mixC00}) into the volume averaged form of (\ref{stressdef02}), we obtain the relation between the mixture polymer stress and the mixture conformation tensor as
\begin{align}
\label{mixstresstoC00}
 {\stressPm} = G_m \left({\CT}_m - \IT\right).
\end{align}
Equation (\ref{mixstresstoC00}) is valid for two viscoelastic phases. However, we know a priori that the polymer stress is zero in the gas phase of the bubble. For the special case that only one phase is viscoelastic, eq.\ (\ref{mixstresstoC00}) reduces to
\begin{align}
\label{mixstresstoC01}
 {\stressPm} =
 \alpha \overline{\stressE}_{p}^{\phaseOne}
 = \alpha G^{\phaseOne} \big(\overline{\CT}^{\phaseOne} - \IT\big).
\end{align}
Furthermore, the conformation tensor equation can be written as
\begin{align}
\label{constCEqva02}
&\partial_t \overline{\CT}^\phaseOne + \left( {\velocity}_m \dprod \nabla \right)  \overline{\CT}^\phaseOne  - \alpha {\LT}_m \dprod \overline{\CT}^\phaseOne - \alpha \overline{\CT}^\phaseOne \dprod \trans{({\LT}_m)} =
{\frac{\alpha}{\lambda^{\phaseOne}} \cvec{P}\big(\overline{\CT}^\phaseOne \big)}.
\end{align}
The derivation of (\ref{constCEqva02}) from (\ref{constCEq00}) is found in \cite{Niethammer2019}. To arrive at a more robust form, we transform (\ref{constCEq00}) using the change-of-variable representation (\ref{evolutionConf3}) to obtain
\begin{align}
\nonumber
&\partial_t {\cvec{F}(\overline{\CT}^\phaseOne)} + \left( {\velocity}_m \dprod \nabla \right)  {\cvec{F}(\overline{\CT}^\phaseOne)} \\
= \; &
\label{constCEqva03}
\alpha \Big(
{2 {\BT}_m \dprod \overline{\UpsilonT}^\phaseOne \dprod \overline{\CT}^\phaseOne }
+
{\OmegaT}_m \dprod {\cvec{F}(\overline{\CT}^\phaseOne)} - {\cvec{F}(\overline{\CT}^\phaseOne)} \dprod {\OmegaT}_m
+
\frac{1\;}{\lambda^\phaseOne} \overline{\UpsilonT}^\phaseOne \dprod \cvec{P} (\overline{\CT}^\phaseOne ) \Big).
\end{align}
The change-of-variable representation (\ref{constCEqva03}) is the volume-averaged counterpart for the mixture model of the generic change-of-variable representation (\ref{evolutionConf3}) for a single-phase system. It is valid only if the first phase is viscoelastic, while the second is not. The application of a certain tensor function $\cvec{F}(\overline{\CT}^\phaseOne)$ can affect the numerical stability, as described in section~\ref{subsec:stab}. According to the benchmark results \cite{Niethammer2018}, an appropriate choice is the 4th-root tensor function, which is also used in the present work.
Equations (\ref{mixstresstoC01}) and (\ref{constCEqva03}) are used as a closure for the polymer stress term in (\ref{mixstress01}). Note that the polymer stress divergence in (\ref{mixstress01}) can be decomposed using the product rule as
\begin{align}
\label{closestress07}
\nabla \dprod \left(\alpha \overline{\stressE}_{p}^\phaseOne\right) =  \alpha \nabla \dprod \overline{\stressE}_{p}^\phaseOne + \overline{\stressE}_{p}^\phaseOne \dprod \nabla \alpha.
\end{align}
This decomposition is used to separate pure interface contributions in the second term on the right-hand side of (\ref{closestress07}) from the remainder before the numerical discretization is applied. A discretization scheme for the interface contribution has been proposed in \cite{Niethammer2019}.

\subsection{Numerical setup}
\label{subsec:numericalsetup}
\begin{figure}[htbp!]
\centering{
\includegraphics[width=440pt]{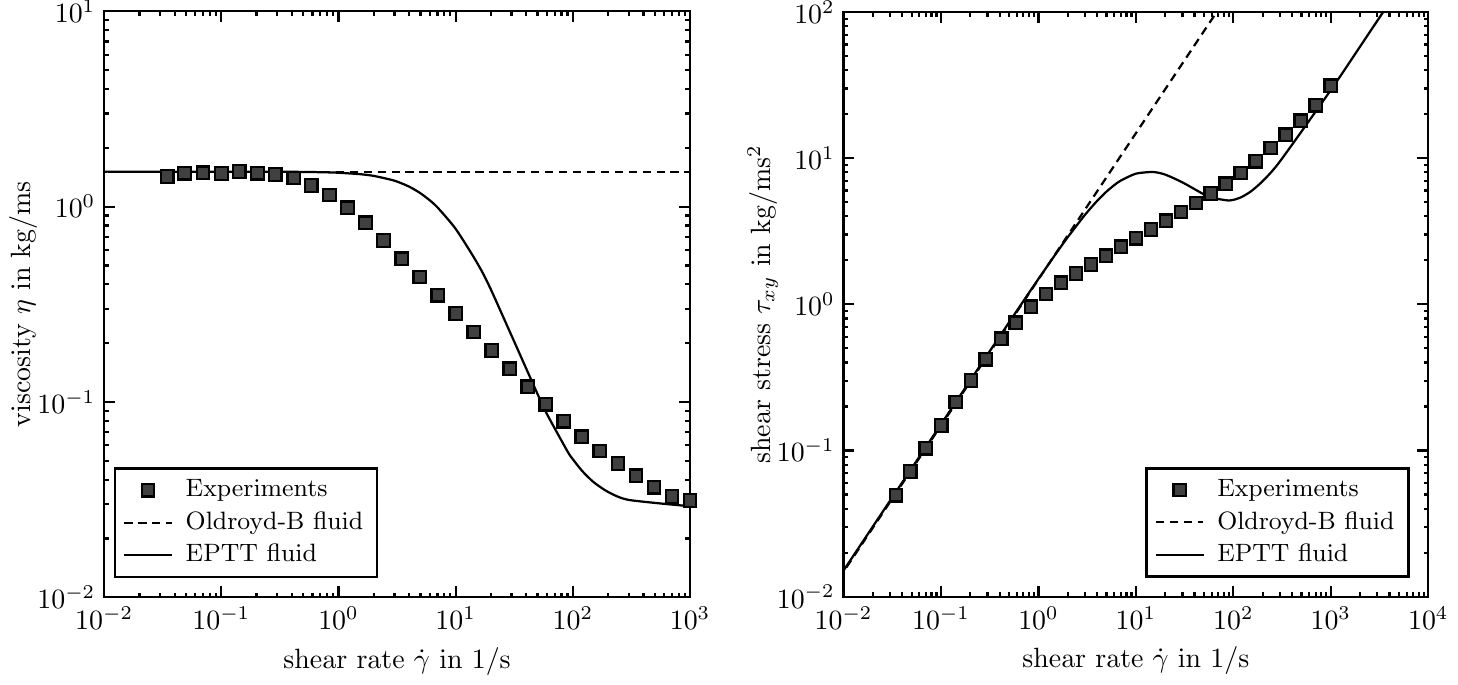}
}
\caption{\small Flow curves for the aqueous $0.8$ wt.~\% P2500 solution. The symbols show the measurements in simple shear flow \cite{Pilz2007}. The curves show the viscosity and the shear stress, using the Oldroyd-B and the EPTT model with the parameters given in Table \ref{tab:parameters}; Figure reproduced from \cite{Niethammer2019} with permission.
}
\label{fig:viscosityfit}
\end{figure}

The flow curve and the shear stress as a function of the shear rate for the aqueous $0.8$~wt.~\% Praestol $2500$ (P2500) solution in simple shear flow are depicted in Fig.~\ref{fig:viscosityfit}. The experimental data show the shear-thinning behavior at shear rates above 1 $s^{-1}$. This behavior cannot be captured with simple two-parameter constitutive equations like the Oldroyd-B model. The Oldroyd-B model does not account for shear thinning, but uses a constant shear viscosity over $\dot{\gamma}$. The four-parameter exponential Phan-Thien Tanner (EPTT) model, in contrast, shows a shear-thinning behavior, which corresponds better to the measurements, as can be seen from the solid line in Fig.~\ref{fig:viscosityfit}. The EPTT parameter fit is identical to our previous work \cite{Niethammer2019}. Note that the fluid relaxation time $\lambda$ is estimated from measurements in elongational flow \cite{Pilz2007}. Therefore, only the remaining EPTT parameters, $\pvisc$, $\svisc$, $\epsilon$ and $\zeta$ are adjusted to the data from shear rheometry in Fig.~\ref{fig:viscosityfit}. In simple shear flow, a change of $\epsilon$ and $\zeta$ results in a different slope of the flow curve, but does not shift the curve along the shear-rate coordinate. Hence, when keeping $\lambda$ fixed, the viscosity curve must have a steeper slope to achieve good agreement with the experiments. The parameters are listed in Table \ref{tab:parameters}.\!\!\!\!
\begin{table}[h!]
\small
\begin{center}
\caption{\small Parameters for the polymer solution used for the investigation.}
\label{tab:parameters}
\begin{tabular}{@{}llllllll@{}}
\toprule
$w$ $[\%]$ \ & $\rho^\phaseOne$  $[\textnormal{kg}/\textnormal{m}^3]$ \ & $\eta_p^\phaseOne$  $[\textnormal{Ns}/\textnormal{m}^2]$ \ & $\eta_s^\phaseOne$  $[\textnormal{Ns}/\textnormal{m}^2]$ \ & $\lambda^\phaseOne$  $[\textnormal{s}]$ \ & $\sigma$  $[\textnormal{N}/\textnormal{m}]$ \ & $\epsilon^\phaseOne$ \ & $\zeta^\phaseOne$ \ \\ 
\midrule
$0.8$ & $1000.9$ & $1.483$ & $0.03$ & $0.203$ & $0.07555$ & $0.05$ & $0.12$ $\!$ \\
\bottomrule
\end{tabular}
\end{center}
\end{table}\
\    

Regarding the flow and shear stress curves in Fig.~\ref{fig:viscosityfit}, an important question is how shear thinning behavior affects the velocity jump discontinuity. Results from other researchers \cite{Pillapakkam2007} suggest that the velocity jump occurs also for rheological models with shear rate-independent viscosities. This means that shear thinning cannot be a necessary condition for the appearance of the velocity jump discontinuity. In the preparation of this work we made similar observations, employing the Oldroyd-B model. However, to obtain quantitative agreement of the bubble rise velocity with the experiments, account for shear-thinning was shown to be necessary \cite{Niethammer2019}. Therefore, the EPTT equations are used for all the simulations of the present study. The potential of this material model to exhibit shear-banding (cf.~\cite{Divoux2016}) for shear stresses between 5 Pa and 8 Pa (Fig. \ref{fig:viscosityfit} right) does not affect the bubble dynamics, since the shear stress $\tau_{r\theta}$ with effect for the $z$-momentum of the bubble, occurring along the lateral parts of the bubble, is well below 5 Pa. This was shown in Fig. 14 of our earlier paper \cite{Niethammer2019}.

\begin{figure}[b!t]    
\centering{
\includegraphics[width=0.99\textwidth]{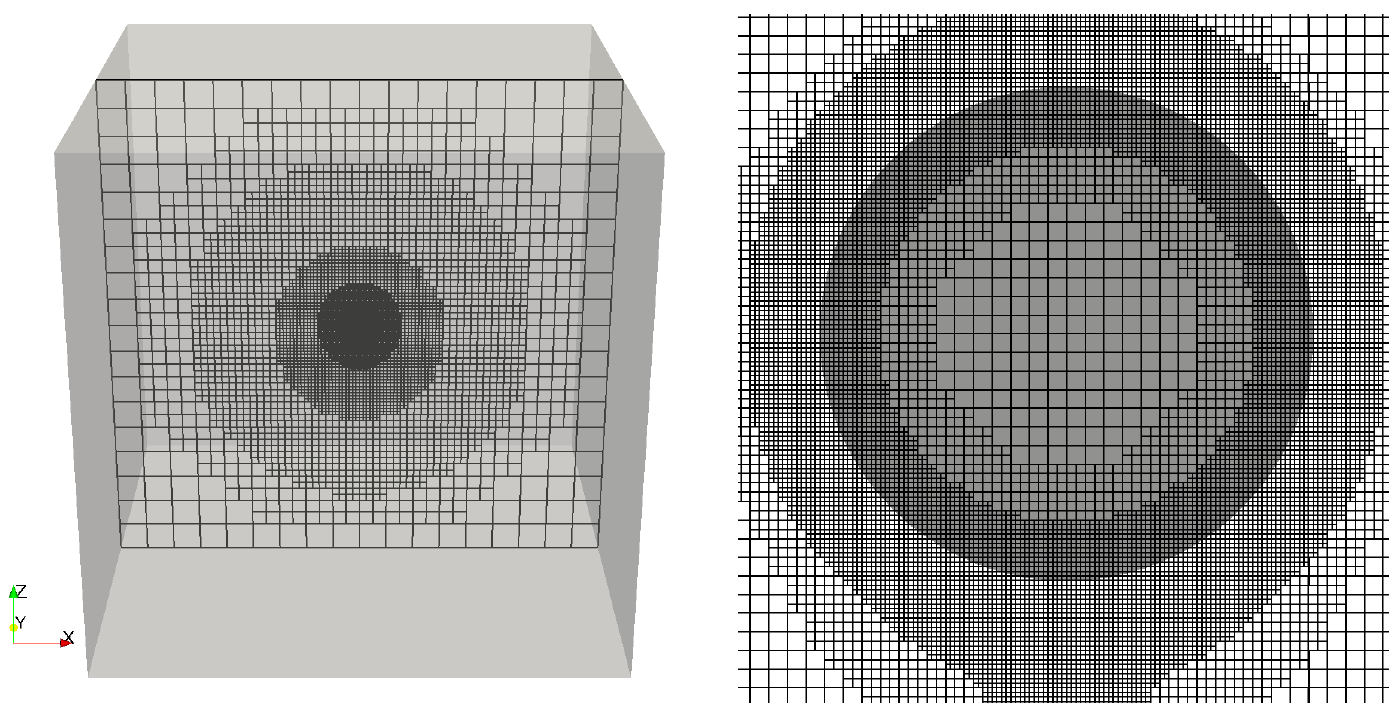}
}
\caption{\small Visualization of the three-dimensional computational mesh M1. Left: Cutting planar mesh slice through the center of the computational domain. Right: Detail of the mesh in the vicinity of the bubble. The initial bubble shape is represented in gray.}
\label{fig:meshdetail}
\end{figure}

The simulation is initiated by placing a spherical bubble in the center of a cubic computational domain. The size of the three-dimensional computational domain is defined with respect to the initial bubble diameter. The domain length $L$ equals 21.7 times the initial bubble diameter $d_b$. This ratio of $L/d_b$  is sufficient to eliminate wall effects in the flow field around the bubble for the mesh shown in Fig.~\ref{fig:meshdetail} \cite{Niethammer2019}. Local adaptive mesh refinement is used to increase
spatial resolution in the vicinity of the bubble interface, while less resolution is provided far away from the bubble to save computational resources. In this way, the best mesh quality is provided in the most important region near the bubble interface, where the mesh cells are uniform and well-spaced around the bubble, such that throughout the simulation the deforming fluid interface is always located inside the region of finest resolution. The resolution on the finest refinement level equals 108 cells per initial bubble diameter. The total number of mesh cells is $1.58$ million.

The system is solved in a moving reference frame, co-moving with the bubble's center of mass, such that the bubble remains in the center of the computational domain throughout the simulation. The use of a moving reference frame on a static mesh saves computational costs that otherwise would be caused by a re-meshing procedure. Due to the moving reference frame, the velocity boundary conditions of the computational domain must be adjusted relative to the motion of the reference frame at each time step. This is achieved by defining a Dirichlet boundary condition for the velocity on each side of the domain, except on the bottom. The Dirichlet boundary condition is set with the time-dependent frame velocity $\velocity_F$. The boundary values are updated in every time step of the simulation. Zero normal gradient boundary conditions are applied for the pressure, the polymer stress and the volumetric phase fraction. On the bottom face of the cube, different boundary conditions are employed. Since we expect an outflow there, zero normal gradient boundary conditions are applied for the velocity, the polymer stress and the volumetric phase fraction. For the pressure, a Dirichlet boundary condition is used with a constant value.

Due to the moving reference frame, an additional force term must be considered in the momentum balance, which results from the different acceleration of the bubble in the two reference frames. Denoting the frame acceleration with $\vec{a}_F$, the body force in the non-inertial frame becomes
\begin{align}
{\rho} \vec{b} = {\rho} \left(\vec{g} - \vec{a}_F\right),
\end{align}
where the $\vec{g}$ is the gravitational acceleration and the frame acceleration $\vec{a}_F$ is given by
\begin{align}
\vec{a}_F = \frac{d {\velocity}_F}{d t}.
\end{align}
The velocity ${\velocity}_F$ is computed in every time step of the simulation based on the relative movement of the center of mass of the bubble. To dampen small oscillations in the velocity field, a PD controller is applied for the computation of ${\velocity}_F$.

\section{Experimental results}
\label{sec:experimentalresults}
\subsection{Velocity fields around the bubbles}
\label{sec:PIV-velocityfields}
Velocity fields in the liquid phase around the rising bubbles were measured with PIV. For doing this, the liquid phase was seeded with fluorescent particles, as detailed in Sec. \ref{sec:PIV}. Upon formation of the bubbles from individual small bubbles, as detailed in Sec. \ref{sec:experimentalmethod}, it was noted that, for producing bubbles of a certain volume, more individual small bubbles were needed in the setup for the PIV measurements than in the bubble rise experiments (BRE) of \cite{Pilz2007}. For the BRE in \cite{Pilz2007}, 12 single bubbles were needed to form a bubble of $42.4~\textnormal{mm}^3$. 
In the experiments for the PIV measurements, 12 single bubbles created a bubble of only $37.7~\textnormal{mm}^3$. According to the equation \cite{Brauer1971}
\begin{equation}
d = \left( \frac{3 \Phi \sigma d_{cap}}{\rho_l g} \right)^{1/3},
\label{eq:brauer1971}
\end{equation}
which predicts the size $d$ of gas bubbles formed by a quasi-static process at the end of a capillary, as used in the present experiments, this indicates that the surface tensions in the two sets of experiments were not the same. Equation (\ref{eq:brauer1971}) states that the ratio of the single-bubble volume in the PIV measurements (subscript PIV) to the one from the bubble rise experiments (subscript BRE) \cite{Pilz2007} is related to the ratio of the respective surface tensions according to \cite{Pilz2009}
\begin{equation}
  \label{eq:diam-ratio-it}
  \left(\frac{d_{\mathrm{PIV}}}{d_{\mathrm{BRE}}}\right)^{3} = 
  \frac{V_{\mathrm{PIV}}}{V_{\mathrm{BRE}}} = 
  \frac{\sigma_{\mathrm{PIV}}}{\sigma_{\mathrm{BRE}}} \; \text{,}
\end{equation}
provided that all the other parameters are identical. 
Thus, when inserting the bubble volumes and the surface tension given in 
\cite{Pilz2007} into (\ref{eq:diam-ratio-it}), 
it follows that the interfacial tension in the PIV measurements amounted to only 
$0.067~\textnormal{J}/{\textnormal{m}^2}$. The reduced surface tension may have been due to some Eosin that may have escaped from the tracer particles into the liquid. Inserting this value for the surface 
tension into the correlation for the critical bubble volume at the jump given in \cite{Pilz2007}, 
a critical bubble volume of $38.8~\textnormal{mm}^3$ results.

\begin{figure}[t]    
  \centering
    \includegraphics[scale=1.0]{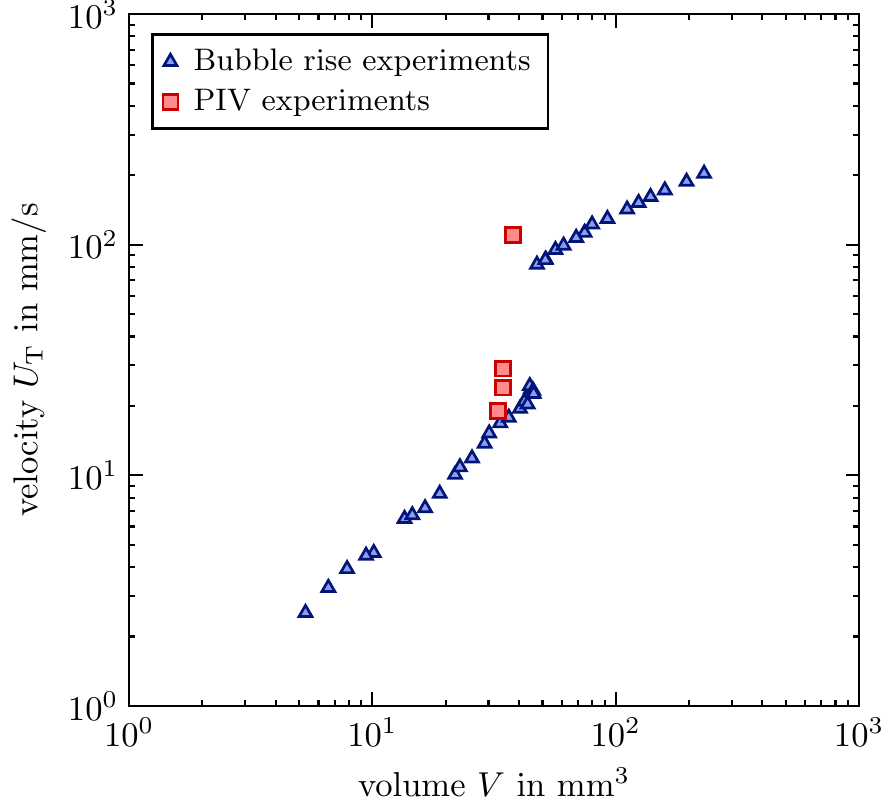}
    \caption{\small Terminal bubble rise velocity vs.\ bubble volume in an aqueous
           $0.8$~wt.~\% P2500 solution obtained from the BRE and the PIV experiments.}
  \label{fig:JumpPIV-P2500}
\end{figure} 

Terminal bubble rise velocities measured in the BRE of \cite{Pilz2007} and in the PIV experiments are depicted in Fig. \ref{fig:JumpPIV-P2500}.  Comparison of the data shows that the jump is clearly visible in both data sets, and that the rise velocities from both sets of experiments are in the same order of magnitude. The critical bubble volumes are $46~\textnormal{mm}^3$ and $37~\textnormal{mm}^3$ in the BRE and the PIV experiments, respectively. The latter is close to the value of $38.8~\textnormal{mm}^3$ predicted due to the reduced surface tension. The lower surface tension may also be the reason that for the bubble rise velocities from the PIV experiments are slightly higher than in the BRE in \cite{Pilz2007}.

\begin{figure}[t!]    
  \centering{
  \begin{tabular}{@{}cc@{}}
  \includegraphics[width=0.49\textwidth]{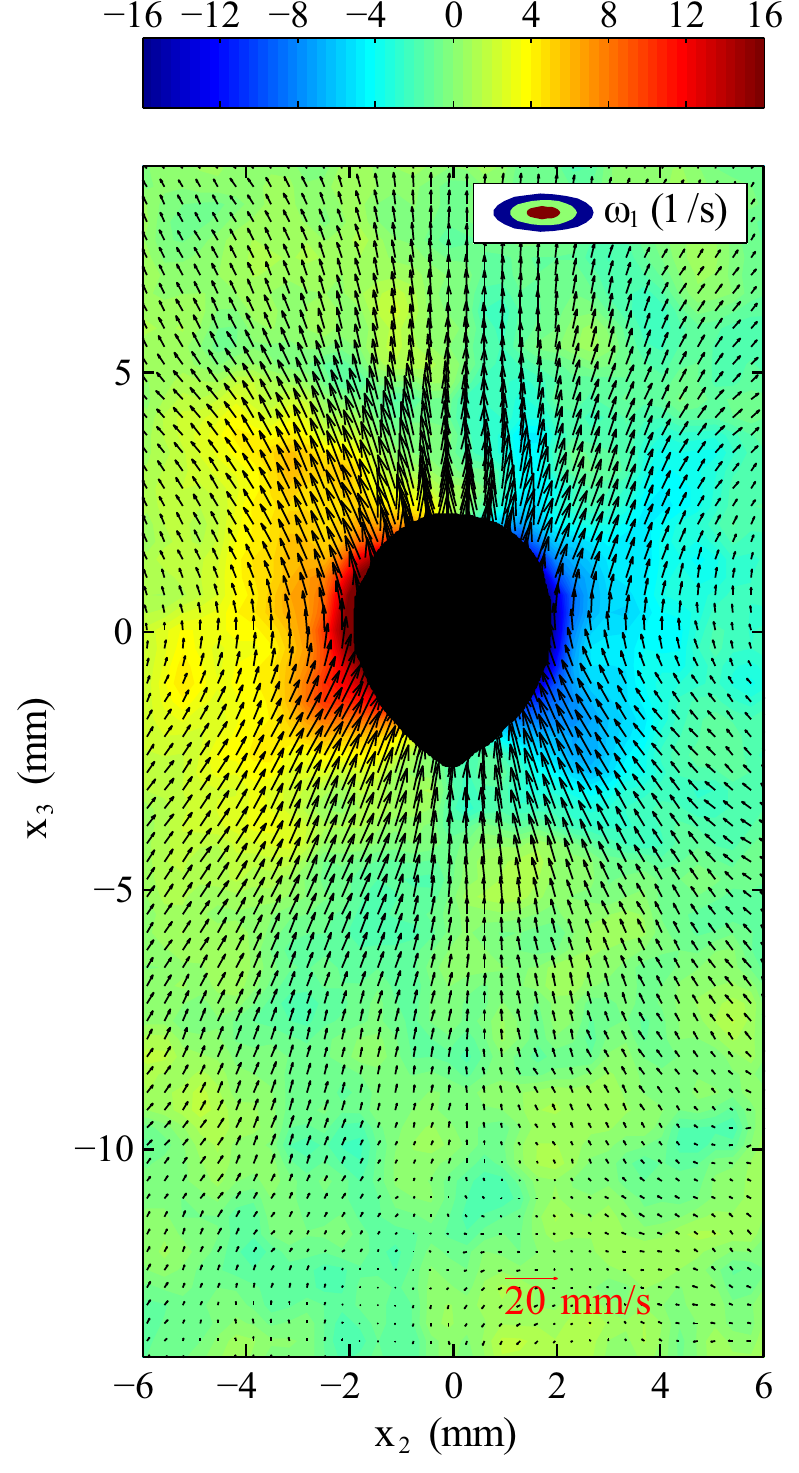} &
  \includegraphics[width=0.49\textwidth]{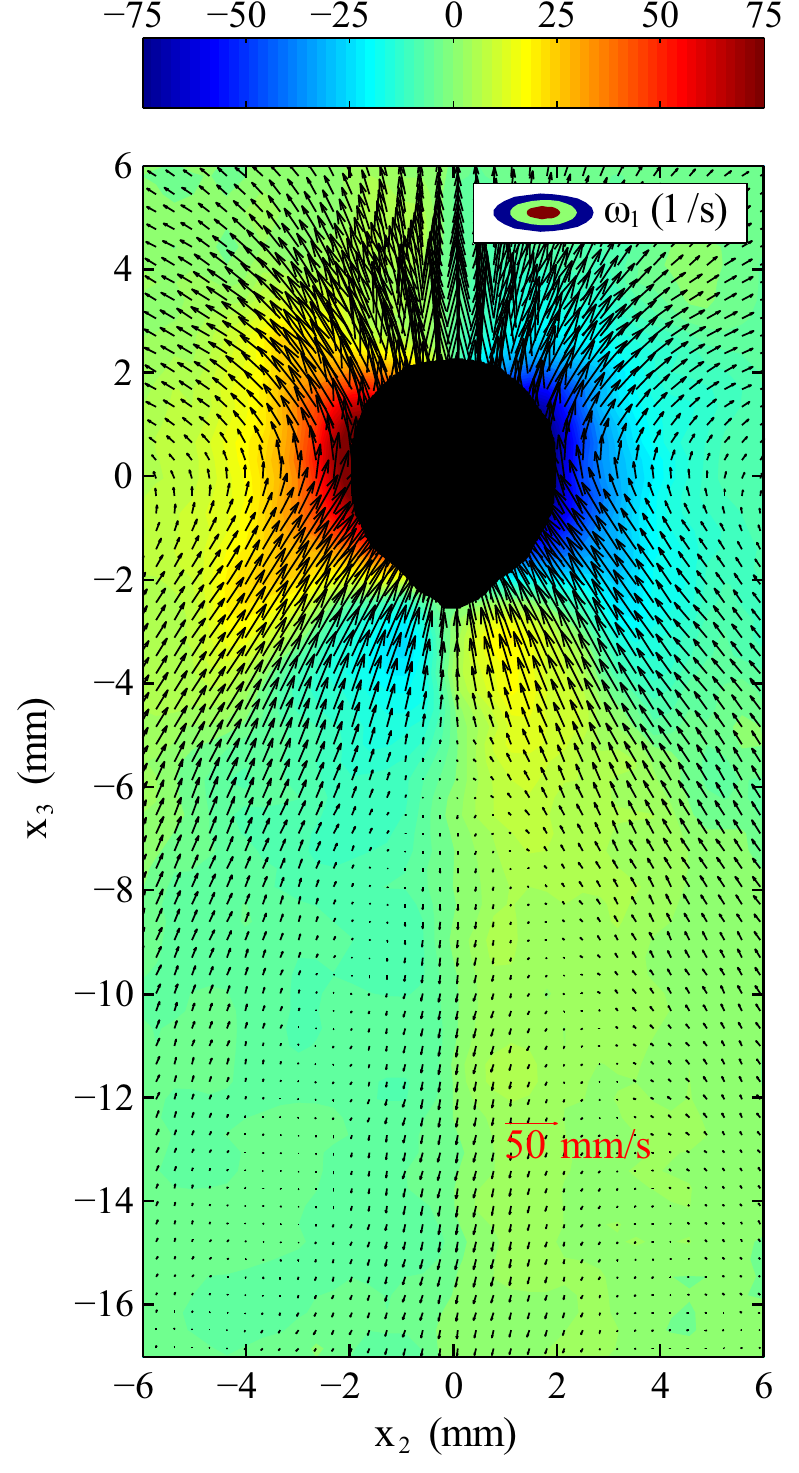} \\
  (a) & (b) \\
  \end{tabular}
  }
  \caption{\small Instantaneous velocity field around a bubble with (a) subcritical volume
  ($U_{\mathrm{T}} = 24~\textnormal{mm}/\textnormal{s}$, $V = 34.3~\textnormal{mm}^3$) and (b) supercritical volume
  ($U_{\mathrm{T}} = 110~\textnormal{mm}/\textnormal{s}$, $V = 37.7~\textnormal{mm}^3$), rising in an aqueous $0.8$~wt.~\% solution of Praestol $2500$. The colors depict the vorticity $\omega_{1}$ defined in (\ref{eq:vorticity-outofplane-PIV}).}
  \label{fig:vorticity-P2500-subsupcritical}
\end{figure}

The velocity fields in the liquid phase around two different bubbles rising in an aqueous $0.8$~wt.~\% Praestol $2500$ solution measured with PIV are depicted in Fig.~\ref{fig:vorticity-P2500-subsupcritical}. 
 Fig.~\ref{fig:vorticity-P2500-subsupcritical}(a) shows a plot 
of the instantaneous velocity vectors around a rising 
bubble with a volume of $34.3~\textnormal{mm}^3$, which is just below the critical value for this setup. 
The terminal bubble rise velocity is $24~\textnormal{mm}/\textnormal{s}$. In  Fig.~\ref{fig:vorticity-P2500-subsupcritical}(b), the velocity field around a bubble with the volume $V = 37.7~\textnormal{mm}^3$ is shown. This bubble volume is above the critical value, and the terminal rise velocity is $110~\textnormal{mm}/\textnormal{s}$. The velocity fields correspond to the sub- and supercritical states, respectively, the supercritical field showing the negative wake. 
The contours show the vorticity component $\omega_{1}$ corresponding to the measured velocity vectors, defined as
\begin{equation}
  \label{eq:vorticity-outofplane-PIV}
  \omega_{1} = \frac{\partial u_{3}}{\partial x_{2}} - \frac{\partial u_{2}}{\partial x_{3}}\; \text{.}
\end{equation}
The spatial derivatives in (\ref{eq:vorticity-outofplane-PIV}) were obtained using 
the numerical gradient operation provided by the Matlab package.
 
\subsection{Transport of polymer molecules along the bubble contour}
\label{sec:transport-time}
\begin{figure}[t]    
  \centering{
  \includegraphics[width=0.8\textwidth]{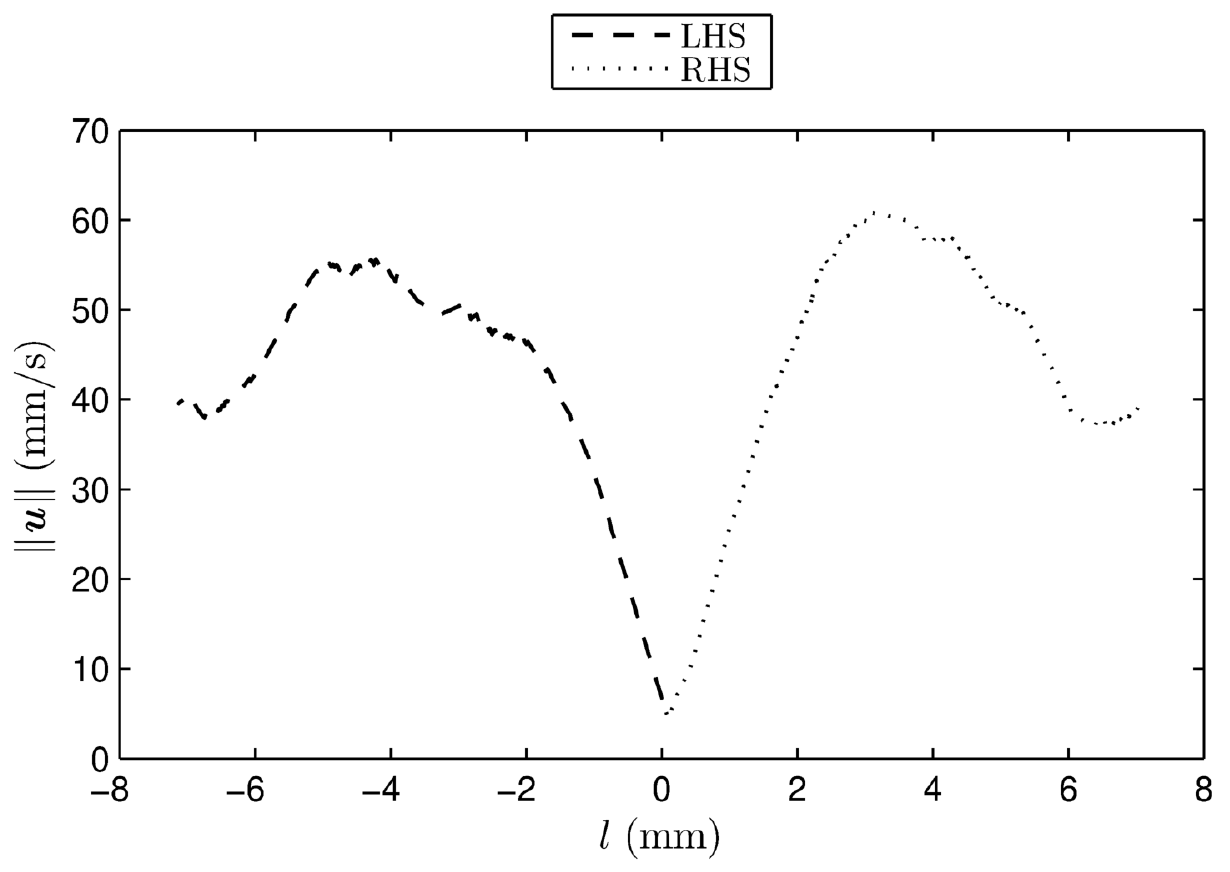}
  }
  \caption{\small Norm of the velocity vectors versus arc length evaluated from the measured velocity field of an air bubble rising in an aqueous $0.8$~wt.~\% solution of Praestol $2500$ ($U_{\mathrm{T}} = 110~\textnormal{mm}/\textnormal{s}$, $V = 37.7 ~\textnormal{mm}^3$). The corresponding velocity field is shown in Fig.~\ref{fig:vorticity-P2500-subsupcritical}.}
  \label{fig:velocity-arclength}
\end{figure}

In the literature discussed in the introduction, a possible explanation for the occurrence of the velocity jump phenomenon was proposed, 
which is based on the assumption that the phenomenon is due to a change of 
dominance between capillary and elastic forces. 
If a given polymer solution has the potential for the discontinuity to occur, 
the time scale of elastic stress relaxation in the liquid, 
as compared to the time it takes to 
transport the relaxing liquid portions downstream along the bubble contour, 
decides about the sub- or supercritical state of the bubble: 
the (hoop) stress set free due to the polymer relaxation generates 
an additional elastic force on the bubble surface. 
The vertical component of this force represents either a push upwards if the relaxation takes very long, or, if the relaxation is fast, an additional resistance, 
depending on the position along the 
bubble contour where the polymers have substantially relaxed and set free the hoop stress. 
In order to evaluate the times of liquid transport down the bubble contour, 
the transport time $\xi$ of a material element was estimated
by numerical evaluation of the integral
\begin{equation}
  \label{eq:time-scale-bubblecontour}
  \xi(l) = \int_{0}^{l} \frac{\mathrm{d} \tilde{l}}{\| \vec{u} \|}\; \text{,}
\end{equation}
where $\vec{u}$ represents the vector of the fluid velocity relative to the bubble, evaluated 
along the bubble contour in the meridional plane, and $l$ is the arc length measured 
from the north pole of the bubble (see Fig.~\ref{fig:image-processing-PIV}). 
The front stagnation point is identified by $l = 0$, while the maximum value of $l$ 
is reached at the rear stagnation point and depends on the bubble volume and shape.
Fig.~\ref{fig:velocity-arclength} shows the distribution of $\| \vec{u} \|$ over
the arc length $l$ as measured by PIV along the supercritical bubble in Fig.~\ref{fig:vorticity-P2500-subsupcritical}b. The symmetry is reasonable. The velocity maxima are located in the equatorial plane of the bubble.
\begin{figure}[h!t]    
  \centering{
  \begin{tabular}{@{}cc@{}}
  \includegraphics[width=0.49\textwidth]{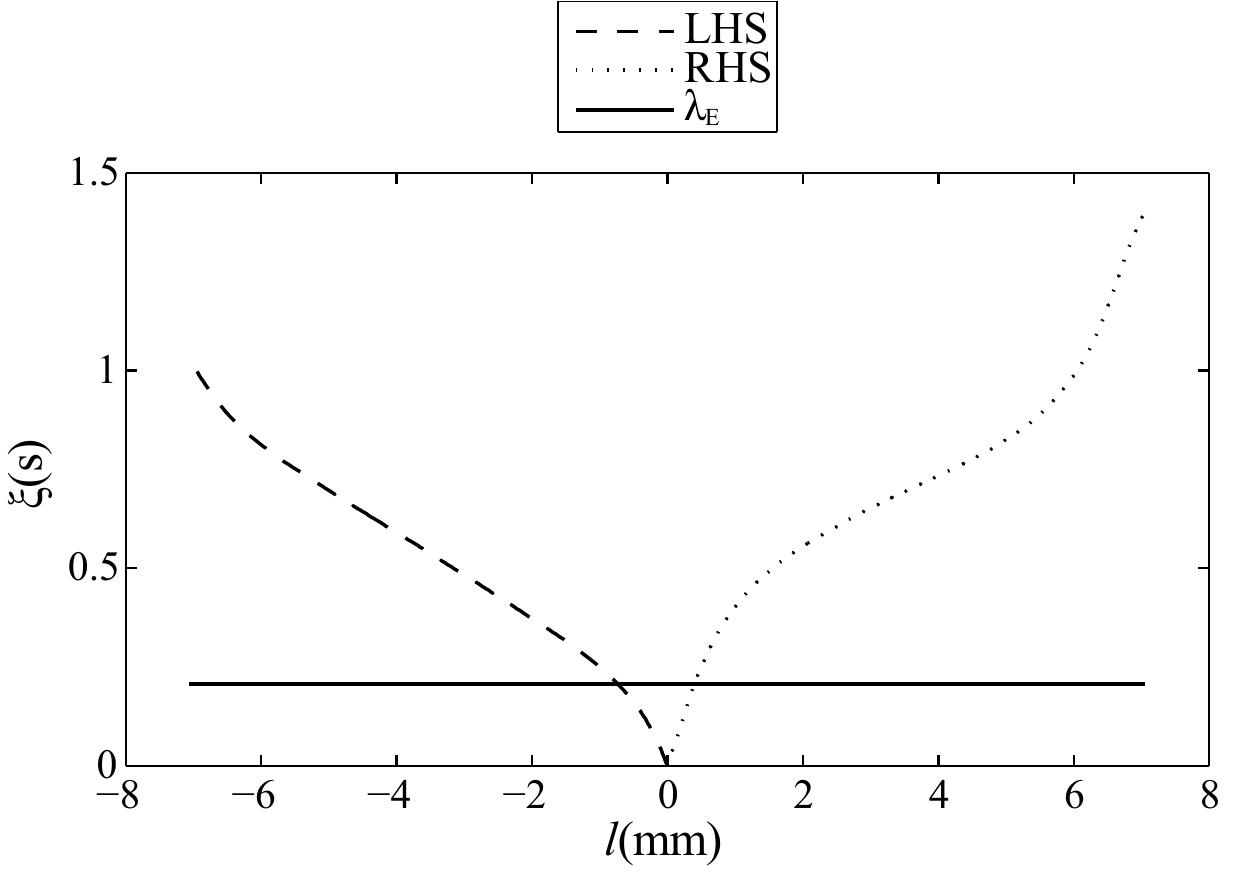} &
  \includegraphics[width=0.49\textwidth]{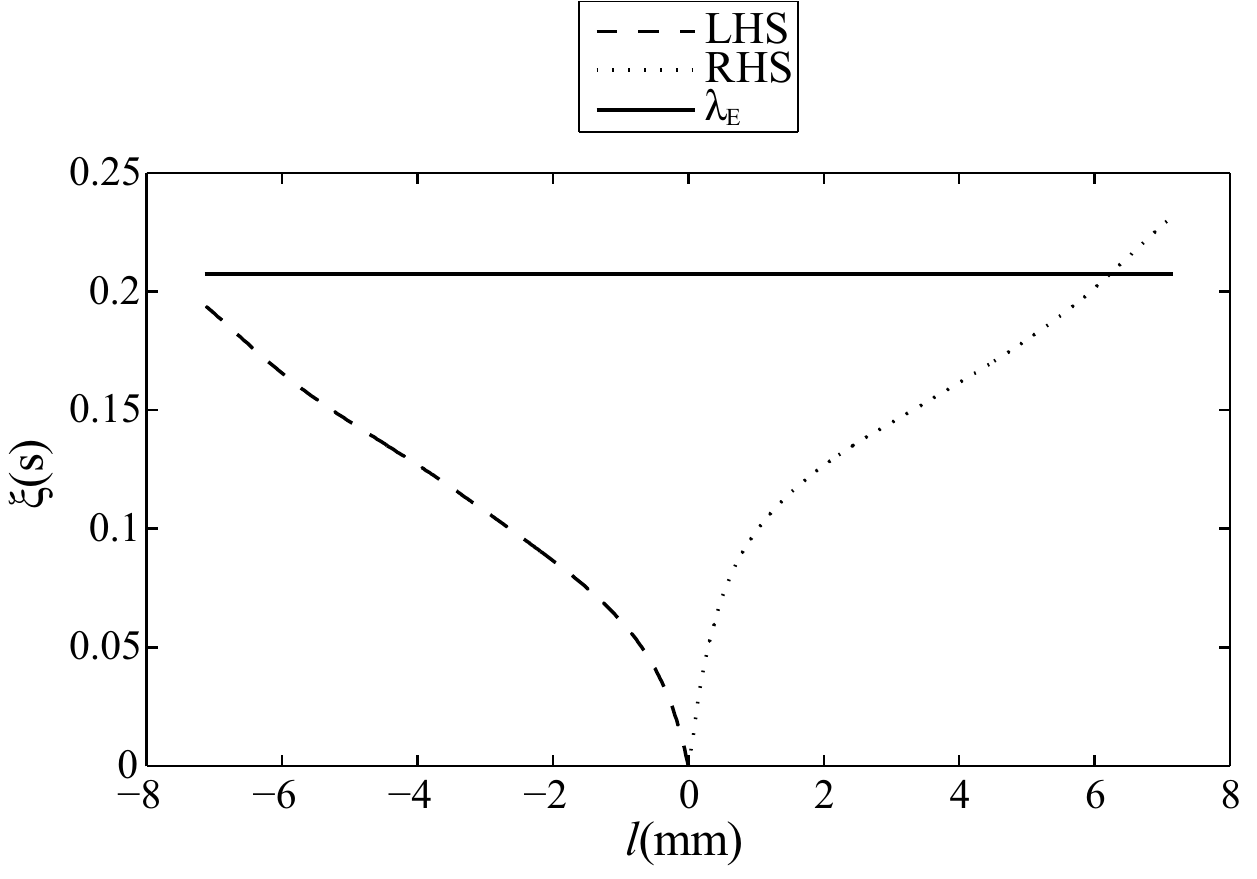} \\
  (a) & (b) \\
  \end{tabular}
  }
  \caption{\small Time $\xi$ elapsed for a fluid particle to travel the arc length $l$ along the contour of (a) a subcritical bubble ($U_{\mathrm{T}} = 24~\textnormal{mm}/\textnormal{s}$, $V = 34.3~\textnormal{mm}^3$) and (b) a supercritical bubble ($U_{\mathrm{T}} = 110~\textnormal{mm}/\textnormal{s}$, $V = 37.7~\textnormal{mm}^3$) in an aqueous $0.8$~wt.~\% solution of Praestol $2500$, compared to the relaxation time $\lambda_E = 0.207~\textnormal{s}$ of the solution.}
  \label{fig:timescale-P2500-subcritical}
\end{figure}
The values of $\xi(l)$ defined by (\ref{eq:time-scale-bubblecontour}) are determined from the PIV velocity data along the half contours in the bubble's meridional plane, identified as the left-hand side (LHS) and right-hand side (RHS) of the contour, for check of symmetry. 
The transport times for the flow along the bubble contour are compared to the 
corresponding relaxation times $\lambda_{\mathrm{E}}$ obtained from elongational rheometry of the test liquids \cite{Pilz2009}. 

Figure \ref{fig:timescale-P2500-subcritical}(a) shows results obtained for the 
bubble with the subcritical volume of $34.3~\textnormal{mm}^3$ in Fig. \ref{fig:vorticity-P2500-subsupcritical}a, rising in an aqueous solution of $0.8$~wt.~\% Praestol $2500$. It is clearly visible that the relaxation process is already finished at the upper part of the bubble. 
For the bubble with the supercritical volume of $37.7~\textnormal{mm}^3$ in Fig. \ref{fig:vorticity-P2500-subsupcritical}b, however, the transport time down the bubble and the relaxation time of the polymer solution appear to be of the same order of magnitude, as shown in Fig.~\ref{fig:timescale-P2500-subcritical}(b). In the supercritical case, therefore, the relaxation of polymeric stress in the liquid persists to the downstream part of the bubble \cite{Pilz2009}, providing a push and leading to the higher rise velocity.

\section{Numerical results}
\label{sec:numericalresults}
Numerical results of bubbles rising in an aqueous solution of $0.8$~wt.~\% Praestol $2500$ are presented in sections \ref{subsec:steadyjump} to \ref{subsec-conformation}. Transient, three-dimensional simulations were performed for various bubble volumes in, both, the subcritical and the supercritical state. The numerical setup described in section \ref{subsec:numericalsetup} was used for all simulations. In section \ref{subsec:steadyjump}, we first show a comparison of the steady-state rise velocities between experiment and simulation. In section \ref{subsec:negativewake}, the velocity field around the bubble is presented to examine the negative wake effect. 
While similar results obtained with the same numerical method have already been reported in \cite{Niethammer2019}, we include these short sections for providing a complete picture.
The main numerical results originate from a local polymer stress and conformation tensor analysis, which is presented in sections \ref{subsec:stressdistribution} and \ref{subsec-conformation}, providing new insights into the molecular orientation and extension for the subcritical and the supercritical states of bubble motion.

\subsection{Terminal bubble rise velocity}
\label{subsec:steadyjump}
\begin{figure}[b!]
\centering{
\includegraphics[scale=1.1]{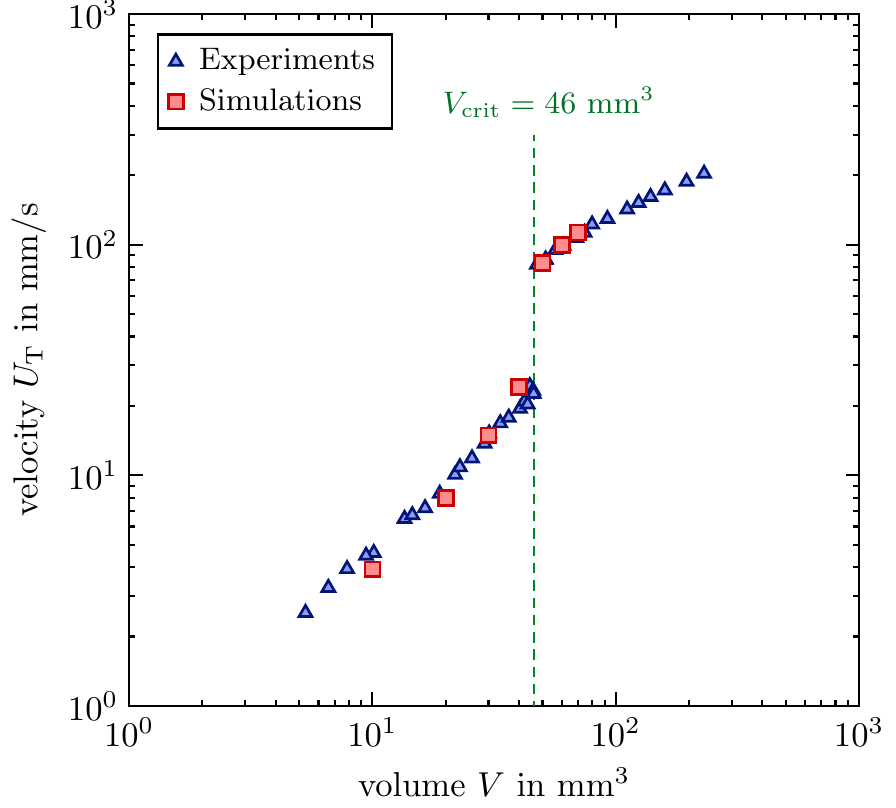}
}
\caption{\small Terminal bubble rise velocity $U_{\operatorname{T}}$ as a function of the bubble volume $V$ for bubbles rising in an aqueous 0.8 wt.\ \% P2500 solution. Comparison between experiments \cite{Pilz2007} and numerical simulations \cite{Niethammer2019}; figure adapted from  \cite{Niethammer2019}.}
\label{fig:velvol}
\end{figure}

The validity and accuracy of the numerical method is verified by comparing the terminal rise velocities of bubbles of different volumes. A comparison of the terminal rise velocities as a function of the bubble volume is shown in Fig.~\ref{fig:velvol}. 
The simulation results obtained with the VOF method show good quantitative agreement with the experiments of Pilz and Brenn \cite{Pilz2007}. The terminal rise velocities of the supercritical bubble volumes are nearly identical to the experimental measurements. The subcritical bubble rise velocities show a slightly steeper increase with bubble volume than the experiments. It is worth to note that the critical bubble volume at about $V_{crit} = 46~\textnormal{mm}^3$ is quantitatively correctly captured by the simulations.

\subsection{Negative wake}
\label{subsec:negativewake}
\begin{figure}[t!]
\centering{
\begin{tabular}{@{}ccc@{}}
\small subcritical $V = 30$ $\textnormal{mm}^3$ &
\small supercritical $V = 50$ $\textnormal{mm}^3$ &
\small supercritical $V = 70$ $\textnormal{mm}^3$ \\
\includegraphics[width=140pt]{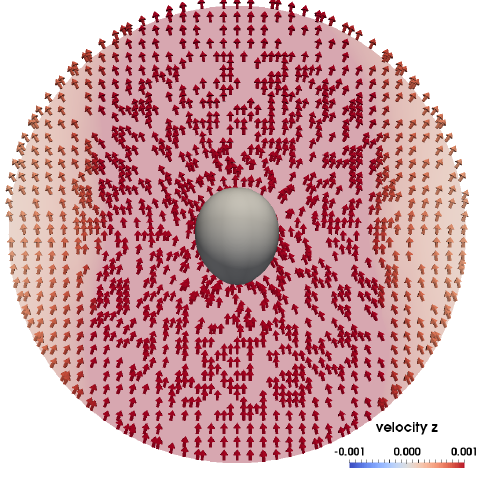} &
\includegraphics[width=140pt]{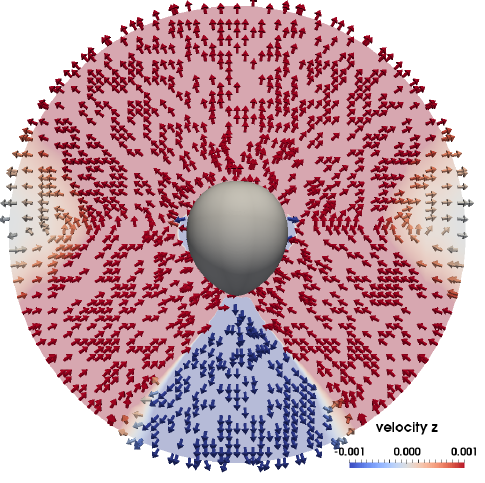} &
\includegraphics[width=140pt]{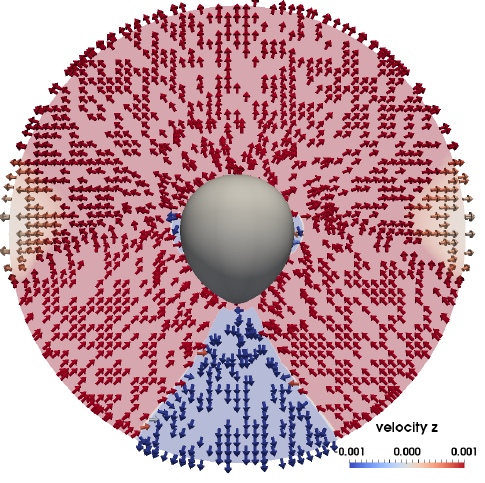}
\end{tabular}
}
\caption{\small Velocity vector field relative to the laboratory frame in a vertical cutting plane through the rising bubble. The color scale is related to the vertical velocity component (in z direction), where blue represents negative and red positive values. The negative wake behind the bubble is visible for the two supercritical states. Figure reproduced from \cite{Niethammer2019} with permission.}
\label{fig:negwake}
\end{figure}

Figure \ref{fig:negwake} shows the velocity vector field in a vertical cutting plane through the center of the rising bubble for three different bubble volumes when the terminal rise velocity is reached.
The direction of the fluid velocity vectors of the subcritical and supercritical volumes differ significantly in the bubble wake. The flow field of the smallest bubble with $V = 30~\textnormal{mm}^3$ is directed upwards into the direction of motion of the rising bubble. This agrees with the flow behavior known from rising bubbles in Newtonian fluids. However, for the two supercritical bubbles with $V = 50~\textnormal{mm}^3$ and $V = 70~\textnormal{mm}^3$, the liquid velocity behind the bubble is pointing in the opposite direction to the bubble motion. The spatial extension of the negative velocities in the bubble wake can be determined from the color scheme, where blue represents negative velocities with respect to the vertical coordinate direction. The negative velocity is observed in an almost cone-shaped region below the bubble's trailing end. This is consistent with experimental findings of the negative wake in non-Newtonian fluids. For example, the PIV measurements in Fig.~\ref{fig:vorticity-P2500-subsupcritical}~(b) show negative fluid velocities in the bubble wake, too. Comparing the two supercritical bubbles with $V = 50~\textnormal{mm}^3$ and $V = 70~\textnormal{mm}^3$ suggests that, above the supercritical volume, the flow field is only subject to minor changes. In particular, the spatial extension of the negative wake is very similar for both bubble volumes.

\subsection{Stress distribution}
\label{subsec:stressdistribution}
The components of the stress tensor are transformed into spherical coordinates $\left(r, \theta, \phi \right)$, which is convenient for the analysis in the vicinity of the bubble interface. The origin of the spherical coordinate system is placed in the bubble’s center of mass. We use standard notation: $r$ is the radial distance from the origin, $\theta$ is the polar angle, with zero value in the vertical upward direction, and $\phi$ is the azimuthal angle. Since the bubble interface is approximately spherical in the upper half, the components $\stressE_{\theta \theta}$ and $\stressE_{\phi \phi}$ represent the tangential stress components, with the hoop stress $\stressE_{\phi \phi}$ acting in the circumferential direction (i.e., the $\phi$-direction) in a horizontal cutting plane through the bubble interface.
\begin{figure}[t!]    
\centering{
\includegraphics[width=\textwidth]{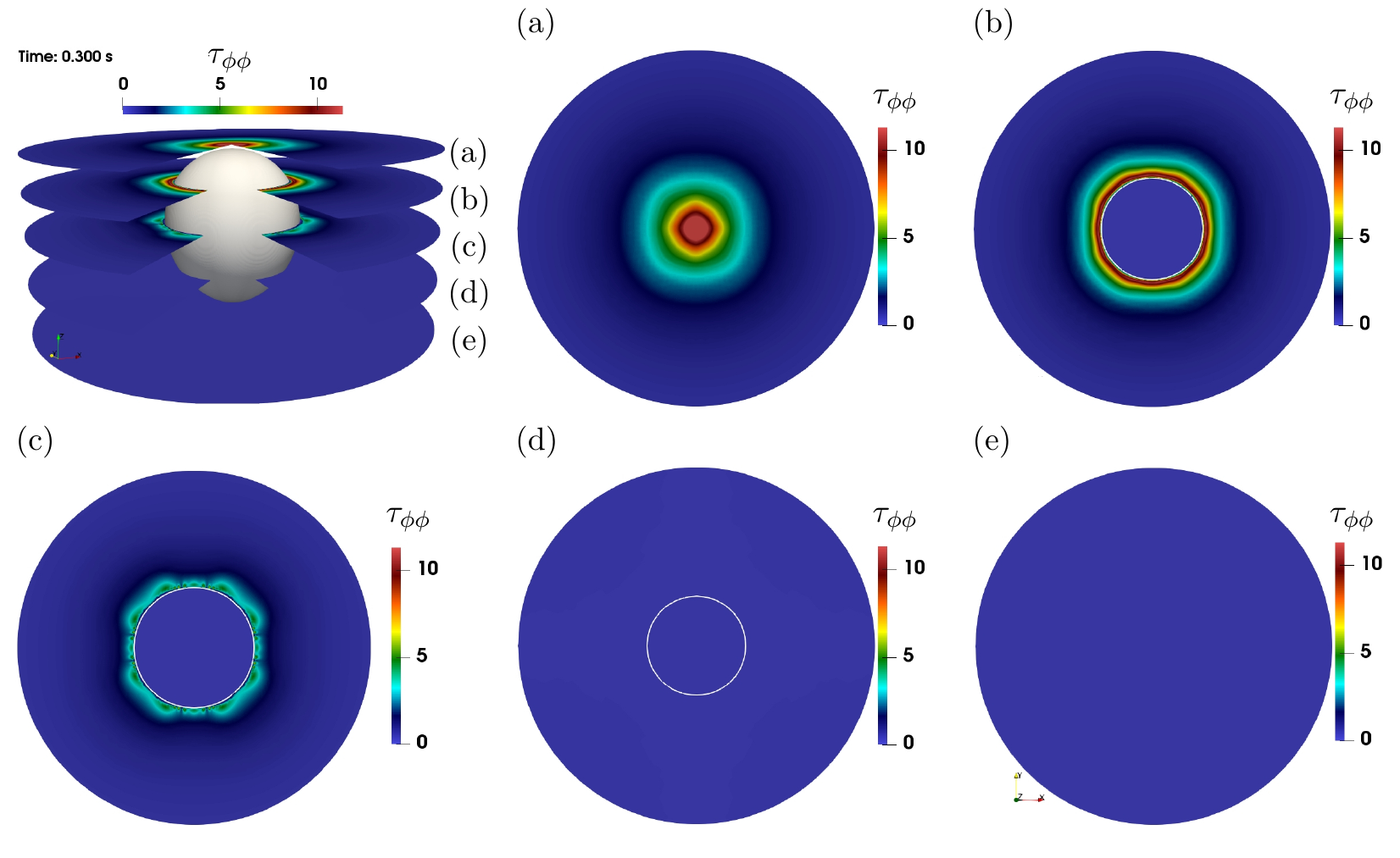}\vspace{-22pt}
}
\caption{\small Stress tensor component ${\tau}_{\phi\phi}$ (in Pa) in horizontal cutting planes at different heights (a) - (e). Subcritical bubble volume $V = 30~\textnormal{mm}^3$ in an aqueous $0.8$~wt.~\% Praestol $2500$ solution. The bubble interface is represented by the $0.5$ iso-surface of the volume fraction field (white) in (b) - (d). The color scale is adapted to the maximum stress in (b).}
\label{fig:tau_PP_sub}
\end{figure}
\begin{figure}[h!]    
\centering{
\includegraphics[width=\textwidth]{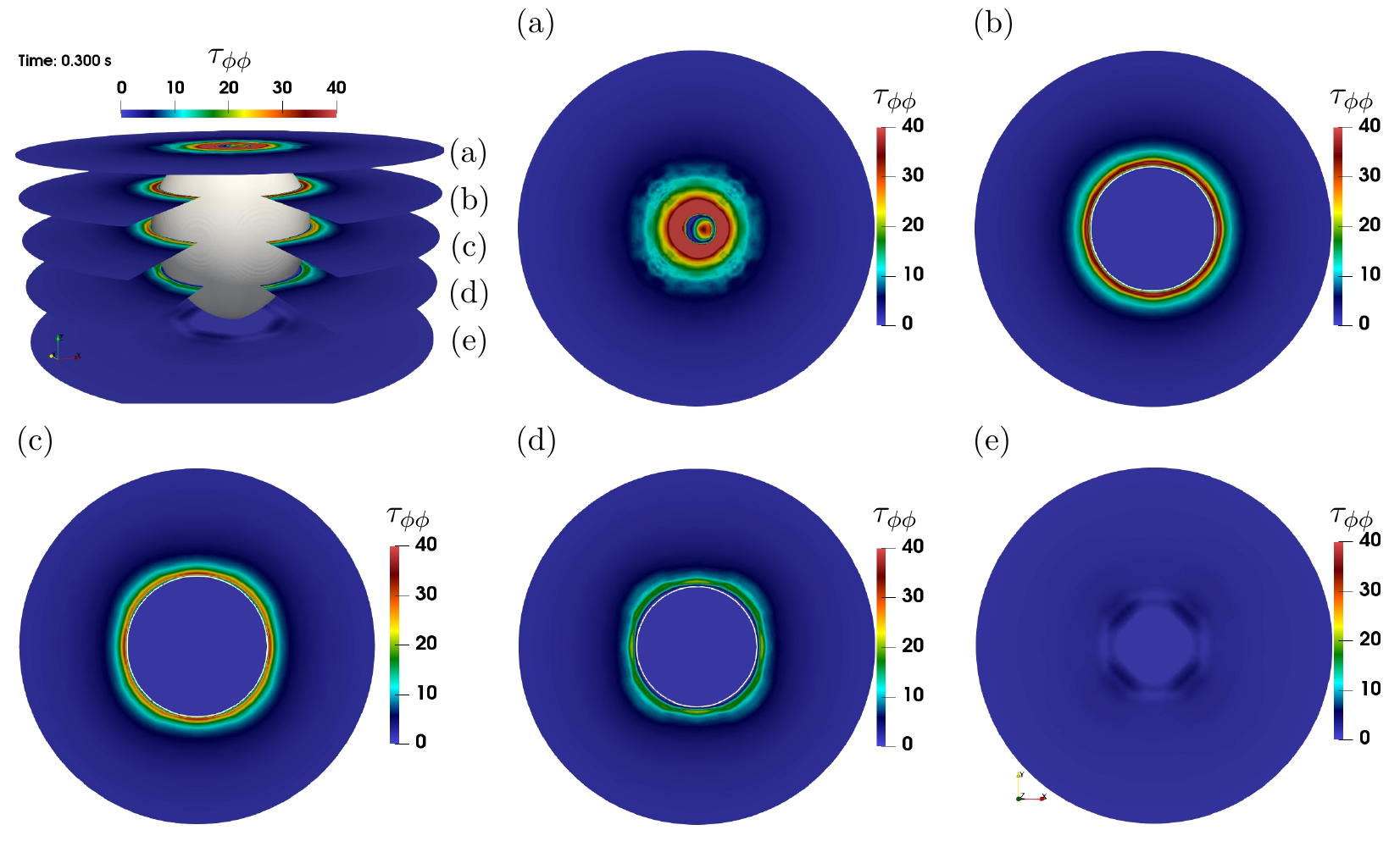}\vspace{-22pt}
}
\caption{\small Stress tensor component ${\tau}_{\phi\phi}$ (in Pa) in horizontal cutting planes at different heights (a) - (e). Supercritical bubble volume $V = 60~\textnormal{mm}^3$ in an aqueous $0.8$~wt.~\% Praestol $2500$ solution. The bubble interface is represented by the $0.5$ iso-surface of the volume fraction field (white) in (b) - (d). The color scale is adapted to the maximum stress in (b).}
\label{fig:tau_PP_super}
\end{figure}

The stress component $\stressE_{\phi \phi}$ is displayed in Figs.~\ref{fig:tau_PP_sub} and \ref{fig:tau_PP_super} for a subcritical and a supercritical bubble volume, respectively. The stress field is shown in five horizontal cutting planes (a) to (e) at different heights. The planes (a) and (e) are located slightly above and below the bubble, respectively. The planes (b) to (d) intersect the bubble, where (c) represents the equatorial cutting plane\footnote{Note that the plane passing through this center point is not precisely the same as the plane in which the intersection with the bubble has maximum diameter. But the distance between these planes, measured by the angle between the lines passing through the center and the intersection point between the bubble surface and the respective plane, is less than $3^\circ$ in all cases.}. In the horizontal cutting planes (a), we see the formation of a hoop stress $\stressE_{\phi \phi}$ near the north pole. As the polar angle increases, the hoop stress decreases, which implies a relaxation of the polymer in the circumferential direction as it flows around the bubble; cf.~(b)~to~(d). Comparing the subcritical and the supercritical states, the hoop stress is significantly larger in the super- than in the subcritical state of the bubble, which can be seen from the color scales. The regions in the liquid affected by the stress are furthermore significantly larger in the super- than in the subcritical state. Moreover, the stress persists further downstream along the bubble, strongly present down to plane (d) in the supercritical state, while it has relaxed almost completely in the plane (c) for the subcritical state. The hoop stress for further bubble volumes and the stress components $\tau_{rr}$ and $\tau_{\theta\theta}$ are presented in Appendix \ref{asec:stresstensor}.

\subsection{Conformation tensor analysis}
\label{subsec-conformation}
The conformation tensor is a key element in a group of material models for polymeric substances. It is employed for the modeling of polymeric flows by formulating the relationship between the molecular polymer conformation and the bulk stress. The conformation is defined as the ensemble-averaged dyadic product of the end-to-end vector of the polymer molecule,
resulting from the respective placement of the rotatable Rouse segments along the macromolecule, with itself. This tensor, often denoted as $\CT$, can be interpreted as a tensorial measure of the molecular orientation \cite{Guenette1992}; see Appendix~\ref{asec:conftens} for more details. The tensor is symmetric and positive definite (spd). Its trace is the ensemble-average of the squared length of the molecules. 
Values of the trace increasing in time therefore represent a stretching of the macromolecules. But note that the trace of $\CT$ does not describe the spatial orientation of the molecules.

\begin{figure}[b!]    
\centering{
\includegraphics[width=\textwidth]{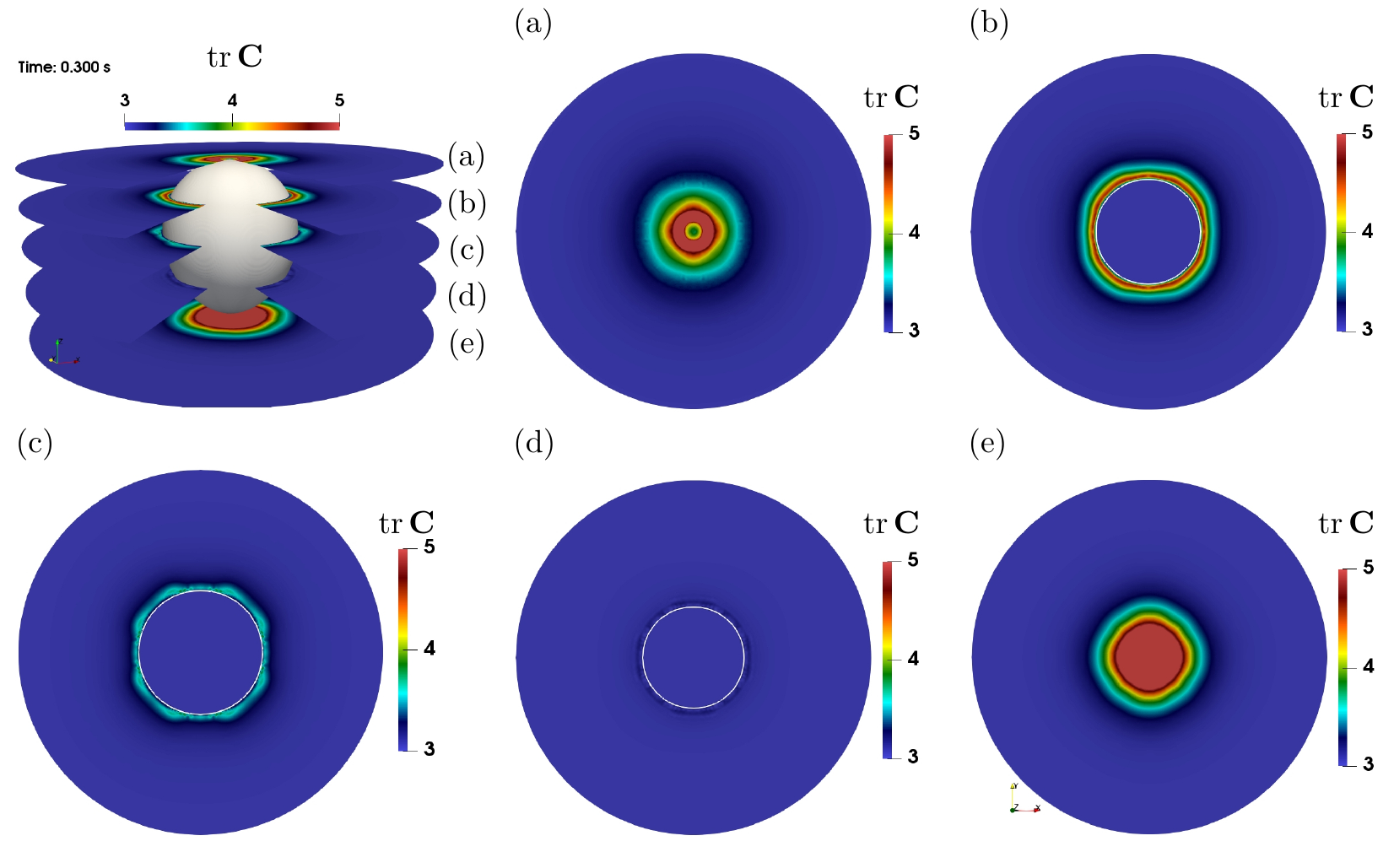}\vspace{-18pt}
}
\caption{\small Trace of the conformation tensor $\CT$ in horizontal cutting planes at different heights (a) - (e). Subcritical bubble volume $V = 30~\textnormal{mm}^3$ in an aqueous $0.8$~wt.~\% Praestol $2500$ solution. The bubble interface is represented by the $0.5$ iso-surface of the volume fraction field (white) in (b) - (d). The color scale is adapted to the maximum $\tr{\CT}$ values in (b).}
\label{fig:tr_C_sub}
\end{figure}
\begin{figure}[htbp!]    
\centering{
\includegraphics[width=\textwidth]{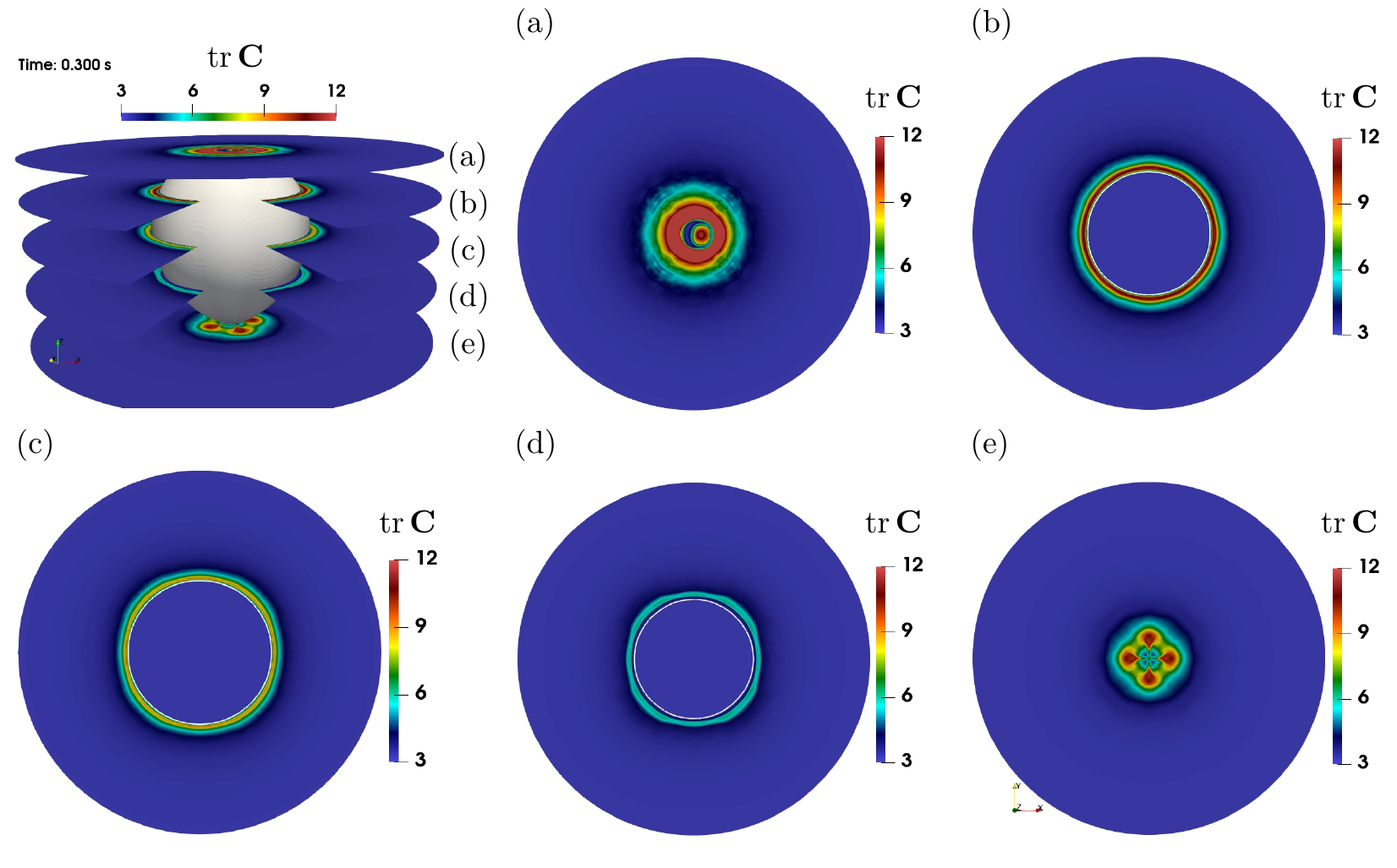}\vspace{-18pt}
}
\caption{\small Trace of the conformation tensor $\CT$ in horizontal cutting planes at different heights (a) - (e). Supercritical bubble volume $V = 60~\textnormal{mm}^3$ in an aqueous $0.8$~wt.~\% Praestol $2500$ solution. The bubble interface is represented by the $0.5$ iso-surface of the volume fraction field (white) in (b) - (d). The color scale is adapted to the maximum $\tr{\CT}$ values in (b).}
\label{fig:tr_C_super}
\end{figure}
Figures~\ref{fig:tr_C_sub} and \ref{fig:tr_C_super} depict the spatial distribution of the trace of $\CT$ for a sub- and a supercritical bubble, respectively.
The stretching of the molecules in the biaxial elongational flow around the upper bubble pole and in the uniaxial elongational flow around the downstream end of the bubble is clearly seen in the images. For the subcritical bubble, the flow causes molecular elongation in the biaxial flow along the upper hemisphere of the bubble
(Fig.~\ref{fig:tr_C_sub}~(a) and (b)), which relaxes before the molecules pass the equatorial plane (Fig.~\ref{fig:tr_C_sub}~(c))
and persists in this relaxed state (Fig.~\ref{fig:tr_C_sub}~(d)). A new elongation appears in the flow field around the lower bubble pole and in the wake flow downstream of it (Fig.~\ref{fig:tr_C_sub}~(e)).

In contrast, the flow around the supercritical bubble leads to a stronger stretching of the polymer molecules which also persists below the bubble equator (Fig.~\ref{fig:tr_C_super}~(a) to (e)) and down to the lower pole. 
Consequently, the polymer molecules get strongly elongated near the upper pole and never fully relax during their traveling along the bubble surface.

Around the downstream pole, the trace of the conformation tensor does not show strong differences between the sub- and the supercritical states. This indicates that the elongation state in the lower pole region is caused by a different mechanism than the elongation in the upper bubble hemisphere.
\begin{figure}[t!]    
\centering{
\includegraphics[width=\textwidth]{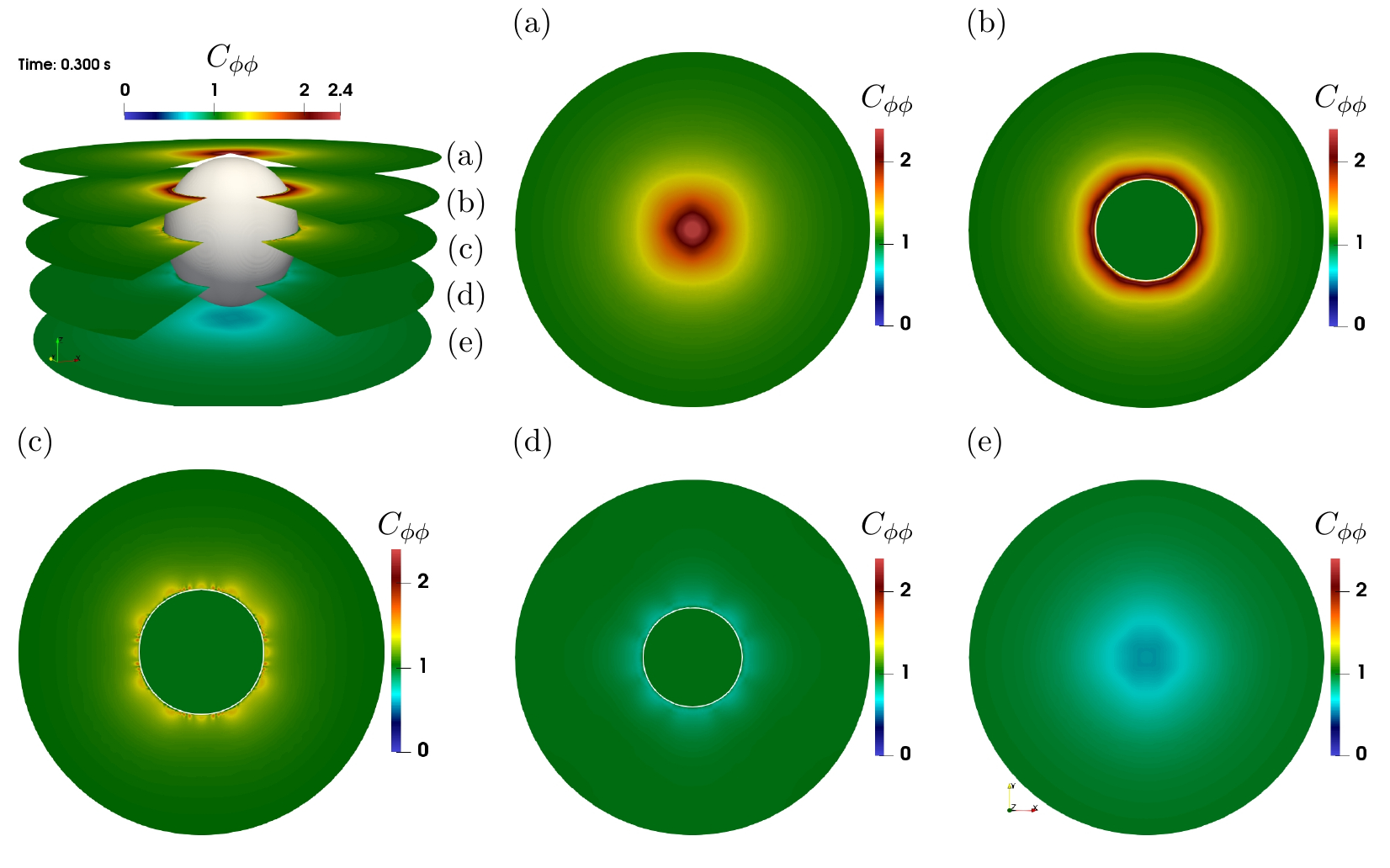}\vspace{-28pt}
}
\caption{\small Conformation tensor component ${C}_{\phi\phi}$ in horizontal cutting planes at different heights (a) - (e). Subcritical bubble volume $V = 30~\textnormal{mm}^3$ in an aqueous $0.8$~wt.~\% Praestol $2500$ solution. The bubble interface is represented by the $0.5$ iso-surface of the volume fraction field (white) in (b) - (d). The color scale is adapted to the maximum ${C}_{\phi\phi}$ values in (b).}
\label{fig:C_PP_sub}
\end{figure}
\begin{figure}[h!]    
\centering{
\includegraphics[width=\textwidth]{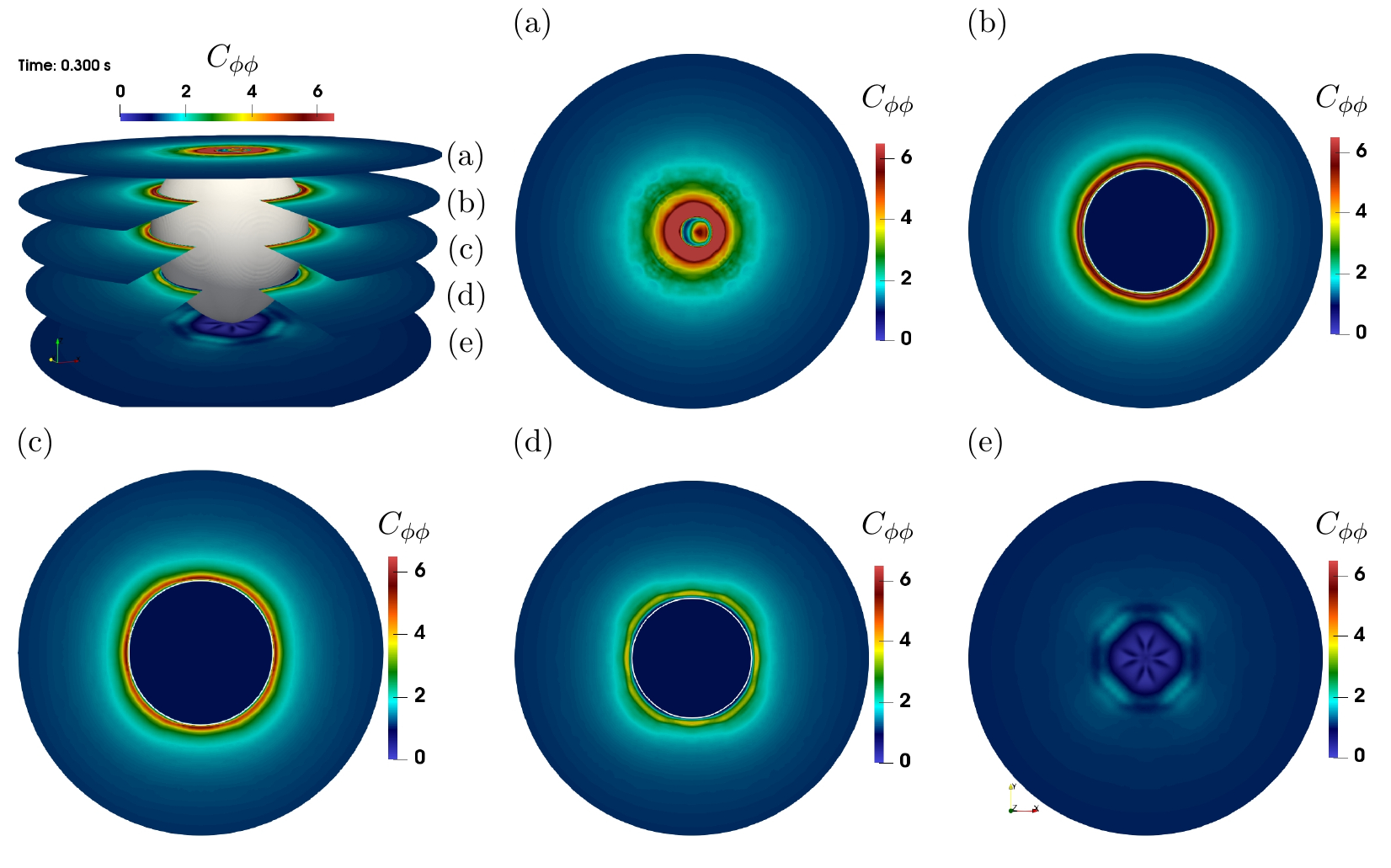}\vspace{-28pt}
}
\caption{\small Conformation tensor component ${C}_{\phi\phi}$ in horizontal cutting planes at different heights (a) - (e). Supercritical bubble volume $V = 60~\textnormal{mm}^3$ in an aqueous $0.8$~wt.~\% Praestol $2500$ solution. The bubble interface is represented by the $0.5$ iso-surface of the volume fraction field (white) in (b) - (d). The color scale is adapted to the maximum ${C}_{\phi\phi}$ values in (b).}
\label{fig:C_PP_super}
\end{figure}
Inspection of the conformation tensor component ${C}_{\phi\phi}$ in Figs. \ref{fig:C_PP_sub} and \ref{fig:C_PP_super} shows that the elongation in the azimuthal angular (i.e., circumferential) direction is dominant around the upper bubble hemisphere, and, again, that this elongation survives far down the supercritical bubble (Fig.~\ref{fig:C_PP_super}). The data downstream from the  bubble's rear pole show that the molecular elongation there is not due to the azimuthal component. It is even smaller there than in the surrounding liquid bulk, where the azimuthal orientation occurs randomly. The very small values of the tensor component ${C}_{\phi\phi}$ in this region shows that the elongation which is present there (cf.\ the results on the conformation tensor trace discussed above) must result from stretching in another direction,
which turns out to be the radial one; the latter conclusion is supported from the supplementary results on ${C}_{rr}$ given in Appendix~\ref{asec:conformationtensor}.

The conformation and orientation of the polymer molecules in the flow around the bubble can be quantified by a spectral analysis of the conformation tensor. For this purpose, the eigenvectors of the tensor with their corresponding eigenvalues are studied, as depicted in Figs.~\ref{fig:ev_sub} and \ref{fig:ev_super} for the sub- and supercritical bubble, respectively.
{\begin{figure}[t!]    
\centering{
\includegraphics[width=0.993\textwidth]{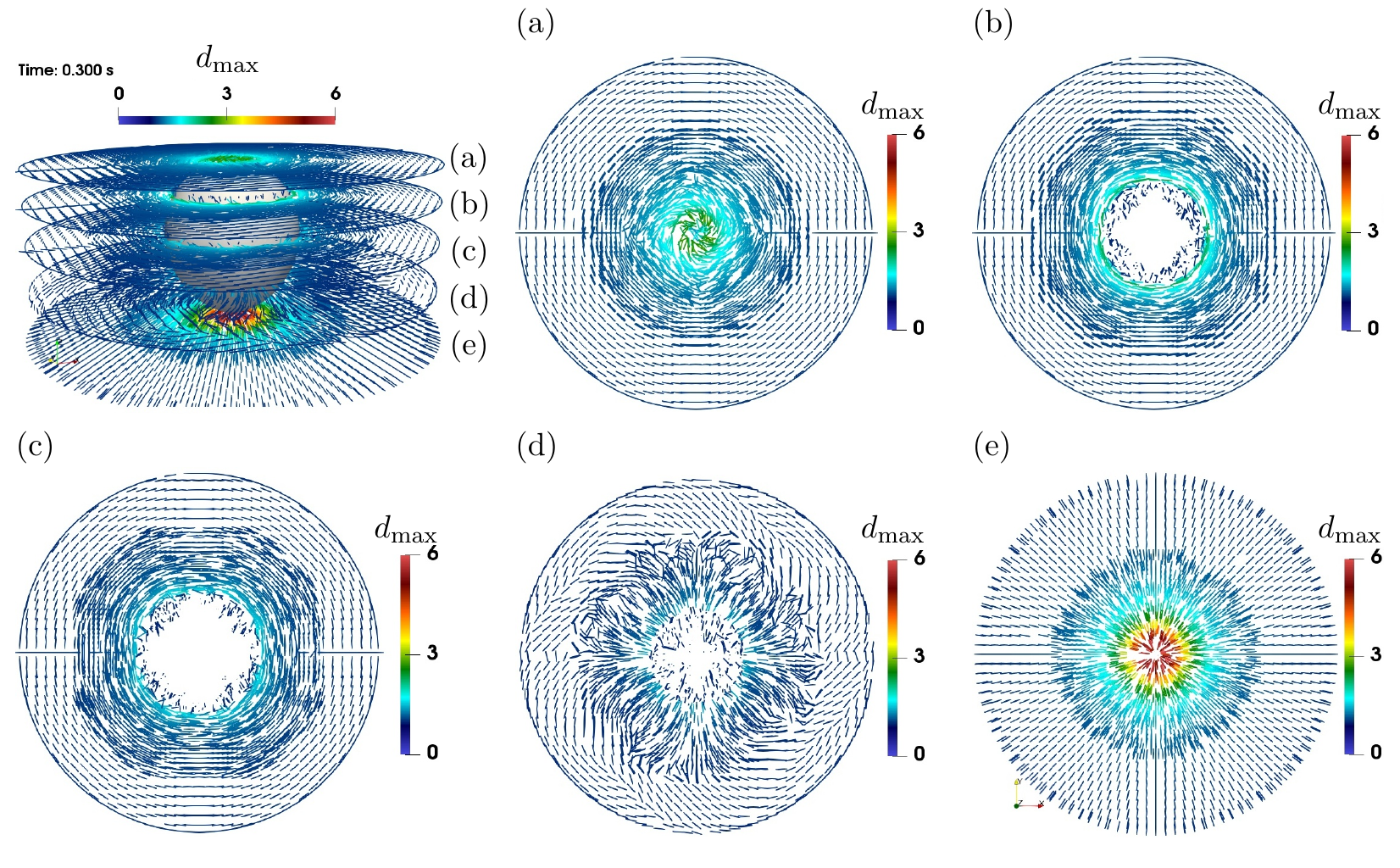}\vspace{-21.5pt}
}
\caption{\small Line elements visualizing the direction of the eigenvectors which belong to the largest eigenvalues ${d_{\max}}$ of the conformation tensor in horizontal cutting planes at different heights (a) - (e), colored corresponding to the value of ${d_{\max}}$. Subcritical bubble volume $V\,=\,30~\textnormal{mm}^3$ in an aqueous $0.8$~wt.~\% Praestol $2500$ solution.}
\label{fig:ev_sub}
\end{figure}
\begin{figure}[h!]    
\centering{
\includegraphics[width=0.993\textwidth]{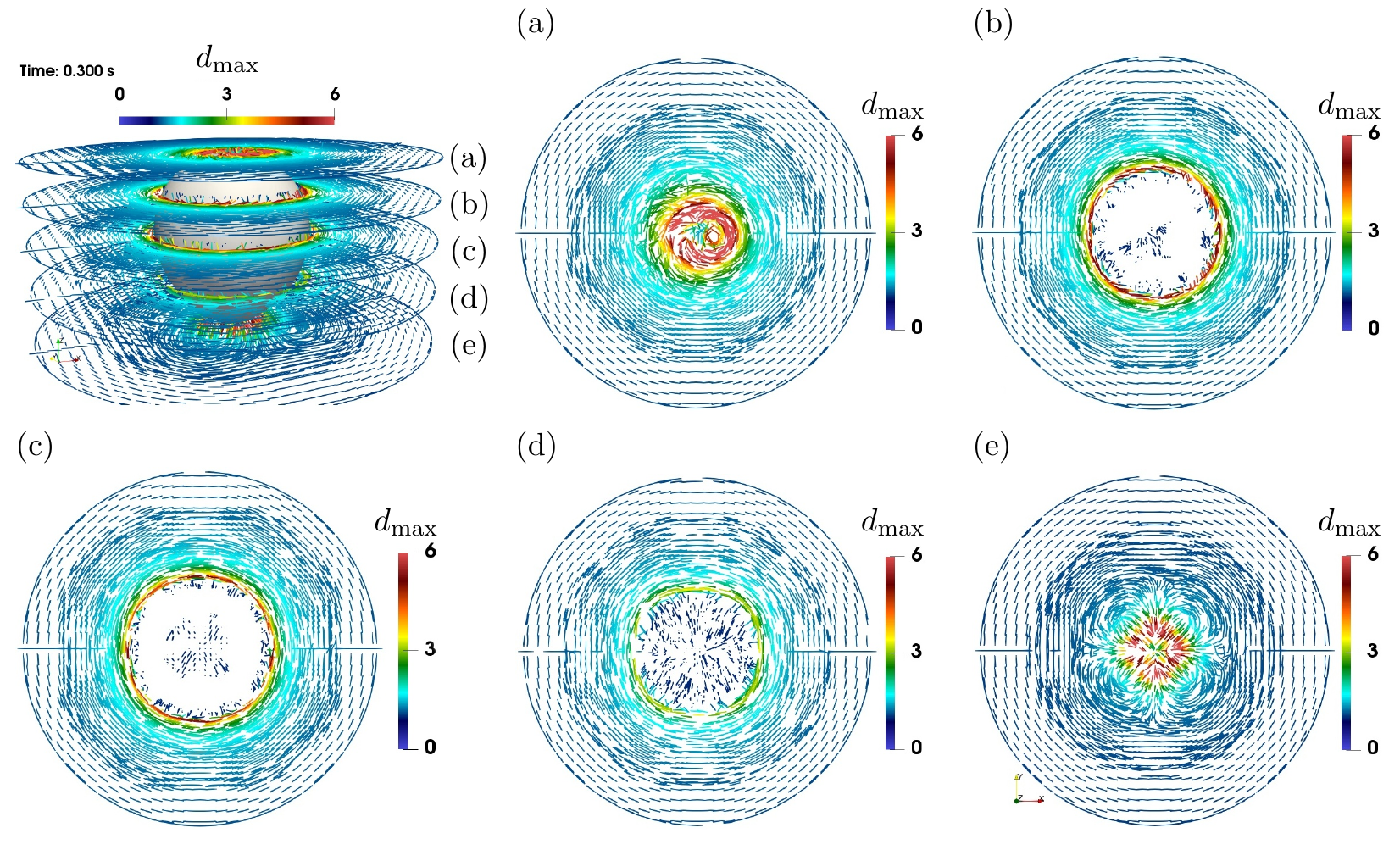}\vspace{-21.5pt}
}
\caption{\small Line elements visualizing the direction of the eigenvectors which belong to the largest eigenvalues ${d_{\max}}$ of the conformation tensor in horizontal cutting planes at different heights (a) - (e), colored corresponding to the value of ${d_{\max}}$. Supercritical bubble volume $V~=~60~\textnormal{mm}^3$ in an aqueous $0.8$~wt.~\% Praestol $2500$ solution.}
\label{fig:ev_super}
\end{figure}}
The figures display, locally in every mesh cell, the direction of the eigenvector which corresponds to the largest eigenvalue in this position. This is indicative for the mean polymer molecular orientation, since the square root of the eigenvalues is a measure for the length of the polymer end-to-end vector projected on the direction given by the corresponding eigenvector; cf.\ Appendix~\ref{asec:conftens}. From this analysis, it is seen that the formation of large eigenvalues is more pronounced in the supercritical than in the subcritical case, and that the large eigenvalues persist further downstream along the supercritical bubble. While the predominantly circumferential orientation of the eigenvectors, which is characteristic of planes (a) - (c) in both cases, is lost in plane (d) for the subcritical bubble, it persists for the supercritical one. Even in plane (e), where the eigenvectors are re-oriented in a predominantly radial direction for the subcritical bubble, the supercritical case still exhibits a marked circumferential component. This different behavior of the polymer molecular conformation in the sub- and the supercritical cases will be further analyzed in the following section by theoretical means.

\section{Theoretical results}
\label{sec:theory-results}

The analysis of the local conformation tensor from the DNS results as given in subsection~\ref{subsec-conformation} shows that, predominantly, the polymer molecules get oriented and stretched in a circumferential direction as they move along the bubble's upper hemisphere. This is a purely kinematic effect, since the distance on neighboring longitudinal curves grows with the polar angle $\theta$ as $\sin \theta $. The stretching of the polymer molecules is mediated via friction between the solvent and the polymer molecules. A part of the kinetic energy is stored in the polymer molecules due to their entropic elasticity. 
The release of the stored energy takes place with a time delay which is intimately related to the relaxation time of the polymer molecules.
The retraction of the polymer particles generates a hoop stress, which acts on the bubble surface.
Depending on the ratio of the polymer relaxation time to the time scale of advection, this hoop stress is either applied above or below the bubble's equator. In the former case this leads to a force against the bubble rise, while the hoop stress pushes the bubble upwards in the latter case.
In order to understand whether this perception is correct and explains the abruptness of the change in rise velocity at a critical bubble state, we further investigate the transport of the polymer molecules around the bubble. Since the numerical simulations revealed the relevance of the circumferential component of the conformation tensor, we study the evolution of this component in more detail by theoretical means.

\subsection{Polymer orientation-deformation evolution} 

The key process in the flow of the polymeric liquid around the rising bubble is the transport and deformation of the polymer molecules. The ensemble-average state of conformation of individual polymer 
molecules is described by the orientation-deformation tensor $\pmb L$, which satisfies the transport equation \cite{Yarin1993}
\begin{equation}
\frac{{\rm d} \pmb{L}}{{\rm d}t} = \pmb{\nabla} \pmb{u} \cdot \pmb{L} + \pmb{L} \cdot \pmb{\nabla} \pmb{u}^{\sf T} - \frac{2}{\lambda} \left( \pmb{L} - \frac{1}{3} N_{01} b_1^2  \pmb{I}\right),
\label{eq:conform}
\end{equation}
where $\lambda$ is the stress relaxation time, and $N_{01}$ and $b_1$ are the number and length of the Kuhn segments constituting the polymer molecule. The flow around the upstream hemisphere of the rising bubble is a biaxial straining flow with the dominant straining component oriented in the direction of the azimuthal angle $\phi$ of the spherical coordinate system. Consequently, as already underlined by the numerical results shown above, it is advisable to analyze the transport of the orientation-deformation tensor component $L_{\phi \phi}$. Under the reasonable assumption of an axisymmetric, non-swirling flow (both w.r.\ to the vertical axis through the bubble center), the differential equation for the azimuthal conformation tensor component decouples from the other components and reads 
\begin{equation}
\frac{{\rm d}L_{\phi\phi}}{{\rm d}t}=
2 \, \left( \frac{u_r}{r} + \frac{u_\theta}{r} \cot \theta \right) L_{\phi\phi} - \frac{2}{\lambda} \left(L_{\phi\phi} -\frac{1}{3} N_{01} b_1^2 \right)
\label{eq:conform_phi}
\end{equation}
where $u_r$ and $u_\theta$ denote the radial and polar angular velocity components in spherical coordinates, respectively.

\begin{figure}[htbp!]    
\centering{
\includegraphics[width=0.49\textwidth]{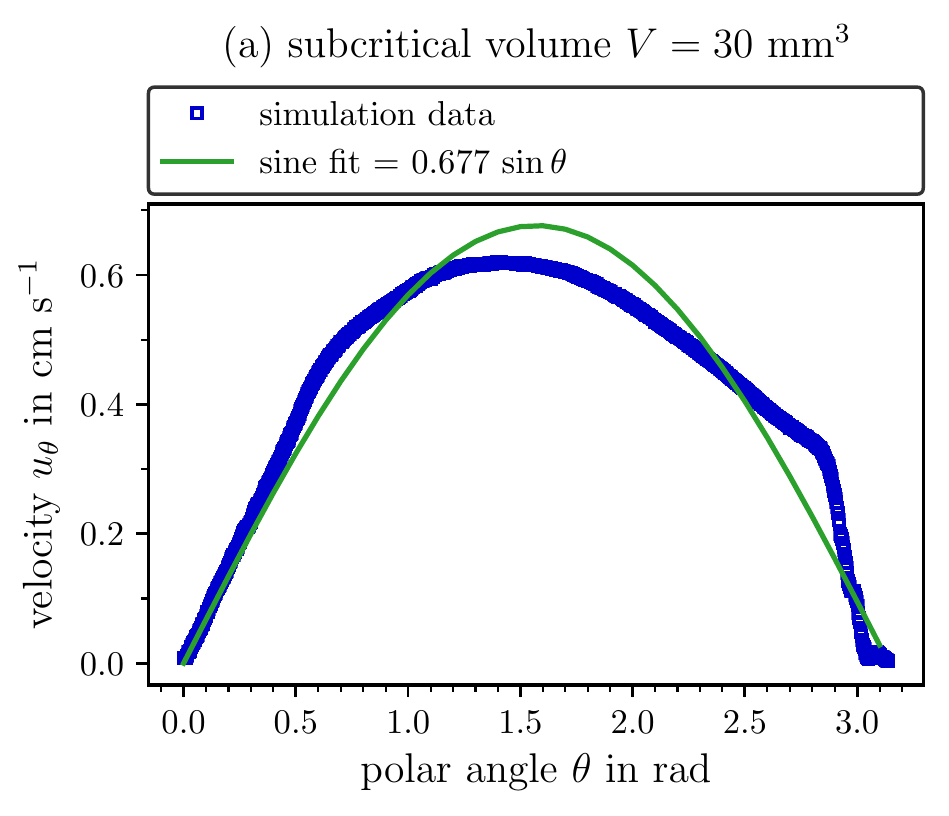}
\includegraphics[width=0.49\textwidth]{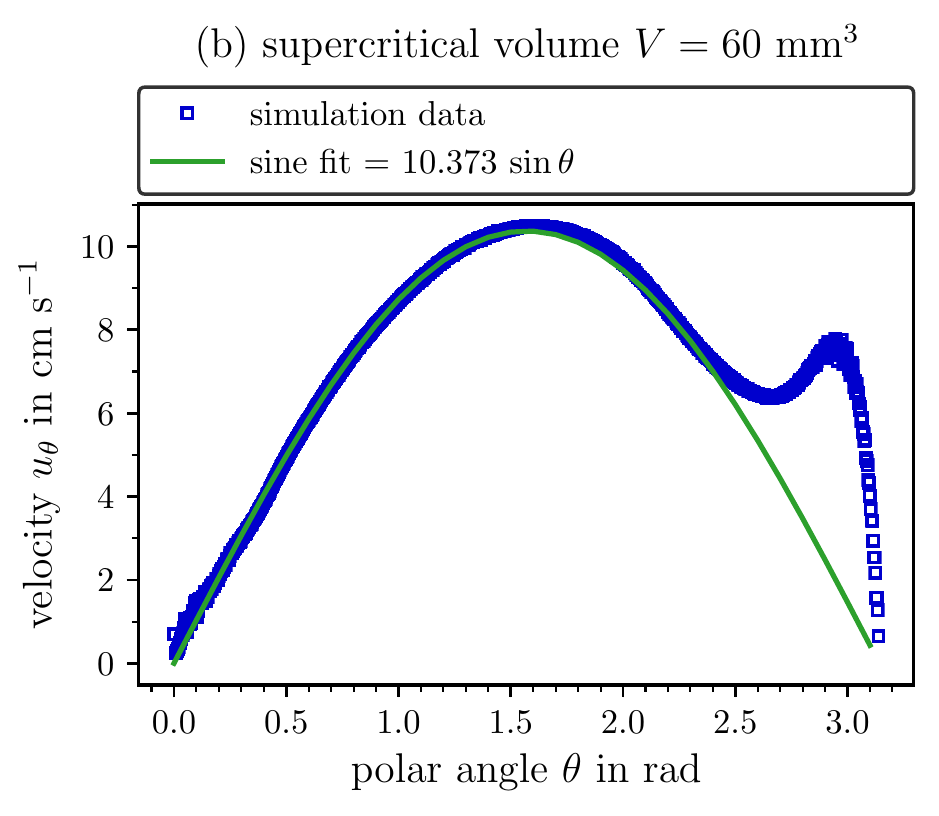}
}
\caption{\small Velocity component $u_{\theta}$ near the bubble interface as a function of the polar angle. Note that the interface is represented as an iso-surface of the volume fraction field and therefore does not correspond to the exact position of the interface, which cannot be determined precisely in the algebraic VOF method.}
\label{fig:velocity_at_surface}
\end{figure}

We add the assumption of a solenoidal flow field, which is appropriate for polymer solutions with small variations in the mixture density. Since we are interested in the quasi-steady situation of a bubble rising at fixed terminal velocity and with constant shape, we switch to the inertial frame with the bubble center of mass as its origin.
Then the velocity is given via a Stokes stream function $\psi = \psi(r,\theta)$ according to
\begin{equation}
u_r = - \, \frac{1}{r^2 \sin \theta}\, \partial_\theta \psi, \qquad
u_\theta = \frac{1}{r \sin \theta}\, \partial_r \psi.
\label{eq:Velocity_from_streamfunction}
\end{equation}
In order to arrive at a closed form of the ordinary differential equation (ODE) (\ref{eq:conform_phi}), we further assume that the bubble shape resembles a half-sphere in the upper half space (i.e., for $0\leq \theta \leq \pi / 2$) and that the Hadamard-Rybczynski flow provides a reasonable approximation of the velocity field at least in the upper half space. More precisely, we assume the
functional dependence of the flow (in the upper half space) to be governed by 
that of the Hadamard-Rybczynski solution, with the rise velocity $U_{\mathrm{T}}$ taken as the actual terminal velocity of the bubble (and not determined as a part of the Hadamard-Rybczynski solution). This is an admissible flow field, since the quasi-steady two-phase Stokes problem in a co-moving frame with fixed interface is a linear problem. As visible in Fig. \ref{fig:velocity_at_surface}, these assumptions are indeed sensible, since the numerically computed flow field resembles this analytical solution even quantitatively. This corresponds to the Reynolds numbers $\leq O(0.1)$ of the bubble rise in the experiments; cf.\ also
Figure~\ref{fig:velocity-arclength}.

With these assumptions, the stream function in the form normalized by the bubble rise velocity $U_{\mathrm{T}}$ and radius $R$, i.e.\ $\psi^* := \psi / (U_{\mathrm{T}} R^2 / 2)$, and with the normalized radial coordinate $r^* := r/R$ reads
\begin{equation}
\psi^* (r^* ,\theta ) = r^* (r^* -1)   {\sin}^2 \theta.
\label{eq:HR-streamfunction}
\end{equation}
With these notations, the  velocity components normalized by $U_{\mathrm{T}}$ read
\begin{equation}
u^*_r (r^*, \theta) = \Big( \frac{1}{r^*} -1 \Big) \cos \theta, \qquad u^*_\theta (r^*, \theta) = \Big( 1 - \frac{1}{2 r^*} \Big) \sin \theta.
\label{eq:hadaryb}
\end{equation}
Around the upstream bubble hemisphere, the radial velocity component in the liquid is negative, and the polar angular component is positive, indicating a flow towards the bubble directed in the negative $z$ direction at large distances from the bubble. Insertion of \eqref{eq:hadaryb}
into \eqref{eq:conform_phi} and normalizing the time variable by $R/U_{\mathrm{T}}$ and the circumferential conformation tensor component by $N_{01} b^2_1 / 3$,
yields the dimensionless form of the circumferential conformation evolution as
\begin{equation}\label{eq:Lphiphi_norm}
\frac{{\rm d}L^*_{\phi\phi}}{{\rm d}t^*}=
\frac{\cos \theta}{(r^*)^2} L^*_{\phi\phi} 
- \frac{1}{{\rm De}_c} \big( L^*_{\phi\phi} -1 \big),
\end{equation}
where the dimensionless tensor $\pmb{L}^*$ is identical to the tensor $\pmb{C}$ in Section~\ref{sec:numericalmethod}. 
In eq.\ \eqref{eq:Lphiphi_norm} we have used the convective Deborah number, which we define as
\begin{equation}
{\rm De}_c = \frac{\lambda U_{\mathrm{T}}}{2 R}.
\label{eq:def_Deborah}
\end{equation}
To finally obtain a closed form of the ODE, one replaces the time variable by either $\theta$ or $r^*$ and then eliminates the other remaining variable
using \eqref{eq:HR-streamfunction} along the streamlines $\psi^*  = {\rm constant}$. We choose $\theta$ as the new independent variable, let $y(\theta):=L^*_{\phi\phi}(t^* (\theta))$
and obtain, employing
\begin{equation}
\frac{{\rm d}\theta}{{\rm d} t^*} = \frac{u^*_\theta}{r^*}
=\frac{1}{r^*} \Big( 1-\frac{1}{2 r^*} \Big) \sin \theta,
\end{equation}
the ODE
\begin{equation}
y'(\theta)=
\frac{2}{\sqrt{4 \psi^* + {\sin}^2 \theta}} \,
\Big( \cos \theta \, y(\theta)- \frac{r^* (\theta)^2}{{\rm De}_c} (y(\theta) -1) \Big),
\label{eq:conform_phi_theta}
\end{equation}
where, along the streamline, the normalized radial coordinate $r^*$ depends
on $\theta$ according to
\begin{equation}
r^* (\theta) = \frac{1}{2} + \sqrt{\frac{1}{4} + \frac{\psi^*}{\sin^2 \theta}}.
\label{eq:streamline}
\end{equation}
Equation \eqref{eq:conform_phi_theta} is a linear, inhomogeneous ordinary
differential equation of first order and with variable coefficients.
Since the coefficient functions are continuously differentiable (hence, in particular, locally Lipschitz continuous), the initial value problems associated to \eqref{eq:conform_phi_theta} are uniquely solvable.
Since $r^* (\theta)$ is bounded away from $\theta = 0$ and $\theta = \pi$, it is also clear that solutions exist globally, i.e.\ on the full interval $(0,\pi)$.
Moreover, any solution starting in some $\theta_0 \in (0, \pi)$ with an initial value $y_0 > 0$,
stays positive on $(\theta_0, \pi)$, which follows from
the fact that the right-hand side in \eqref{eq:conform_phi_theta}
is positive if a solution reaches zero, i.e.\ $y(\theta)=0$ implies $y' (\theta)>0$.

An analytical representation of the general solution can -- in principle -- be obtained by the variation-of-constants-formula. But since the  primitives which appear in the integration are not given by elementary functions, this does not help to analyze the behavior w.r.\ to the Deborah number. 
\begin{figure}[b!]    
\centering{
\includegraphics[width=\textwidth]{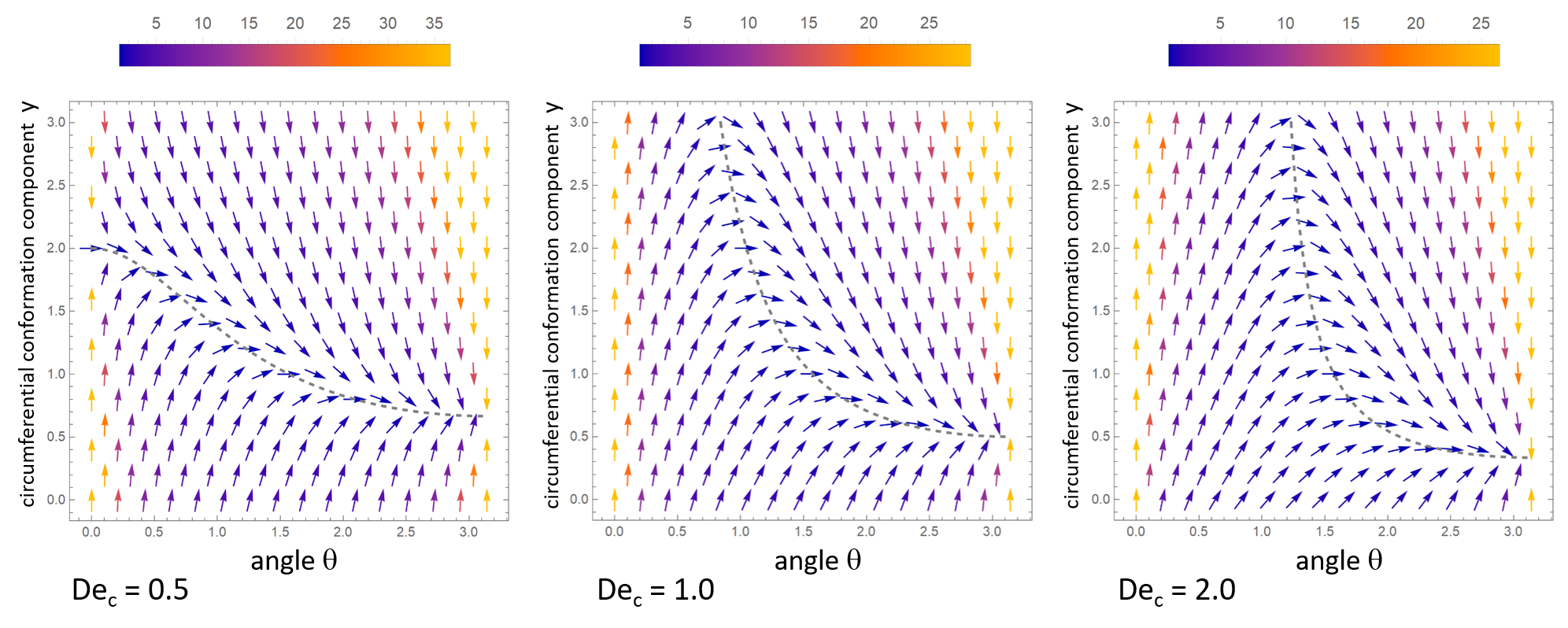} \\
\includegraphics[width=\textwidth]{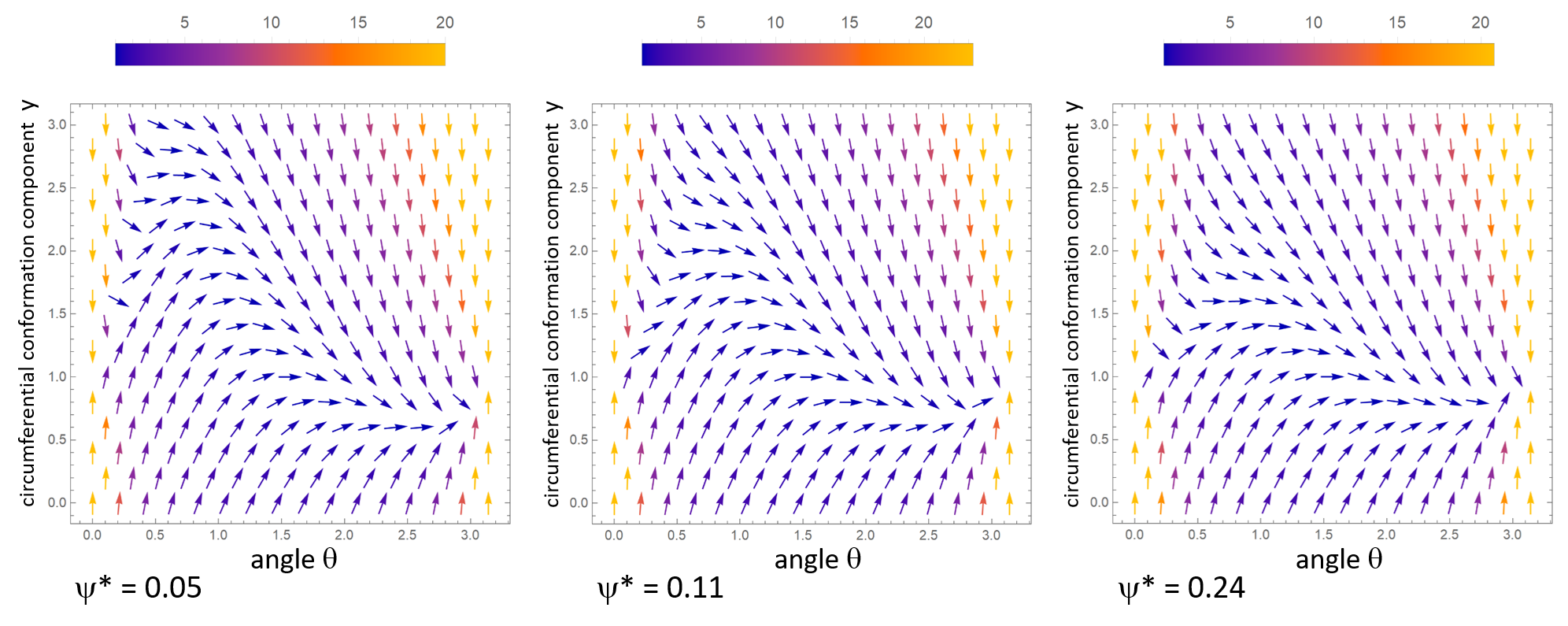}\vspace{-18pt}
}
\caption{\small Top row: vector fields of the ODE \eqref{eq:conform_phi_theta} for different values of ${\rm De}_c >0$ at $\psi^* = 0$, together with the curve of critical points (maxima) of the right-hand side of \eqref{eq:conform_phi_theta}. Bottom row:
vector fields of the ODE \eqref{eq:conform_phi_theta} for ${\rm De}_c =1$ and different values of $\psi^* > 0$, corresponding to minimal $r^*$ of $1.05$, $1.1$ and $1.2$.}
\label{fig:de_variation}
\end{figure}
Instead, to get a first impression of the influence of especially the Deborah number, the top row of plots in Fig.~\ref{fig:de_variation} shows the vector fields associated with the ODE \eqref{eq:conform_phi_theta} along the streamline $\psi^* = 0$ for different values of ${\rm De}_c >0$. In addition, the dashed curve shows all points $(\theta, y)$, in which the right-hand side of \eqref{eq:conform_phi_theta} vanishes. Hence, a solution which passes through a point on this curve has a horizontal tangent there; in fact, it has a local maximum at this point. Consequently, solutions increase as long as they run below the dashed line, and they decrease while they run above it. Note, however, that this curve
itself is not a solution. The line element plots clearly indicate that solutions which start at some (small) $\theta_0 >0$ with a value of $y_0 \approx 1$, say, reach a maximum which becomes higher for larger values of ${\rm De}_c$. This is also reflected by the dashed curves of critical points of the ODE, which are given as
\begin{equation}\label{maxima-curve}
    y(\theta ;{\rm De}_c) = \frac{1}{1- {\rm De}_c \cos \theta }
    \quad \mbox{for } \theta \in (\theta_0 ({\rm De}_c), \pi ),
\end{equation}
where $\theta_0 ({\rm De}_c)=0$ for ${\rm De}_c \leq 1$ and
$\theta_0 ({\rm De}_c)=\arccos ({\rm De}_c^{-1})$ for ${\rm De}_c > 1$.
In the latter case and for $\theta \in (0, \theta_0)$, every solution
is strictly increasing there for all values of $y>0$.
Equation \eqref{maxima-curve} indicates that ${\rm De}_c =1$ is a critical value for the Deborah number, since the possible maxima are all finite for  ${\rm De}_c < 1$, while arbitrarily large maxima are -- in principle -- possible in case of ${\rm De}_c > 1$.
Concerning the choice of the initial value as $y_0 \approx 1$, recall that the dimensionless orientation-deformation tensor ${\bf L}^*$ equals the identity tensor if the polymer molecules are in a relaxed and unoriented state.
\begin{figure}[t!]    
\centering{
\includegraphics[width=\textwidth]{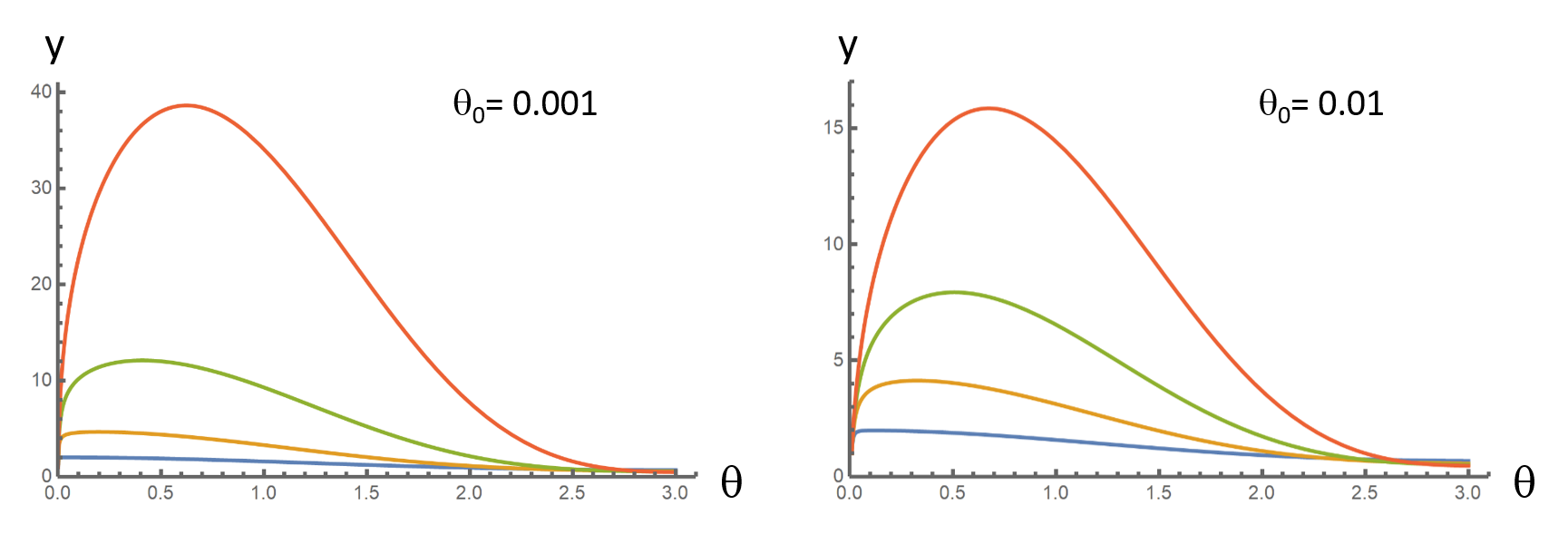}
}
\caption{\small Solutions of \eqref{eq:conform_phi_theta} for different Deborah numbers (${\rm De}_c =0.5$, $0.8$, $1.0$, $1.2$ plotted in blue, orange, green, red, respectively) and  $\psi^* =0$. All solutions start at $y=1$, but at different initial angles: $\theta_0 = 0.001$ (left), $\theta_0 = 0.01$ (right).}
\label{fig:de_variation_solutions}
\end{figure}
\noindent
To come to rigorous conclusions about the influence of the Deborah number, note that in the complete continuum mechanical model from subsection~\ref{subsec:governing-equations}, the forces on the bubble are determined by the pressure distribution and the viscoelastic stresses directly at the bubble surface. While the transport equation \eqref{eq:conform} of the conformation tensor alone does not account
for any back-effect onto the flow field, the limiting case of $\psi^* \to 0+$, corresponding to the streamline running along the bubble surface, nevertheless is the most relevant case. At this point it is important to observe that the right-hand side in \eqref{eq:conform_phi_theta} depends continuously on $\psi^* \geq 0$ and, of course, $r^* (\theta) \to 1$ in the limit as $\psi^* \to 0+$.
The dependence on $\psi^* \geq 0$ is depicted in the bottom row of
Figure~\ref{fig:de_variation} in the critical case ${\rm De}_c =1$: increasing $\psi^*$ corresponds to streamlines passing the bubbles at larger distances and the polymer molecules on more distant streamlines get less stretched in the circumferential direction, experiencing kinematic stretching only at a much weaker intensity.
The solutions of \eqref{eq:conform_phi_theta} for $\psi^* >0$ approach the solutions of the limiting ODE, where the latter reads
\begin{equation}
y'(\theta)=
\frac{2 \cos \theta}{\sin \theta} \, y(\theta)
- \frac{2}{{\rm De}_c \, \sin \theta} (y(\theta) -1).
\label{eq:conform_phi_theta_surface}
\end{equation}
The general solution of \eqref{eq:conform_phi_theta_surface} is
\begin{equation}
y(\theta) = y_p(\theta) + C \frac{\sin^2\theta}{\tan^{2/{\rm De}_c} \big( \theta / 2 \big)}
\end{equation}
with arbitrary $C\in \mathbb{R}$ and the particular solution
\begin{equation}
y_p(\theta)= \frac{1}{2} \sin^2 \theta + \frac{1}{1+ {\rm De}_c} \left(\frac{1-\cos\theta}{2}\right)^2 + \frac{1}{1- {\rm De}_c} \left(\frac{1+ \cos\theta}{2}\right)^2
\end{equation}
in case ${\rm De}_c \neq 1$, and
\begin{equation}
y_p(\theta)= \frac{1}{2} \sin^2 \theta + \frac{1}{2} \left(\frac{1-\cos\theta}{2}\right)^2 + \frac{1}{2} \ln (\tan (\frac{\theta}{2}))
\end{equation}
in case ${\rm De}_c = 1$. 
Now observe that
\begin{equation}
y_p(0+) = \frac{1}{1- {\rm De}_c}
\quad \mbox{as $\theta \to 0+$ in case } {\rm De}_c \neq 1.
\end{equation}
In case ${\rm De}_c >1$, this implies $y(0+) = 1/(1- {\rm De}_c)$
independently of $C\in \mathbb{R}$. On the other hand, in case ${\rm De}_c <1$ the solution has the same limit as $\theta \to 0+$ only if $C=0$,
while for $C\neq 0$ the solution diverges to $\pm \infty$ as $\theta \to 0+$, where the sign is given by ${\rm sgn}(C)$. 
While the behavior differs depending on the value
of the Deborah number, it is not clear how to formulate reasonable initial value problems which allow to directly assess the qualitative behavior of prototypical solutions. This is also supported by the sample solutions shown in Fig.~\ref{fig:de_variation_solutions} (left, middle): the solutions starting with $y_0=1$ but at different $\theta_0>0$ close to zero, strongly depend on $\theta_0$. This is
explained by the fact that the upper pole is an accumulation point
of the velocity field, i.e.\ solutions for smaller $\theta_0$ stay
significantly longer close to this pole.

In order to clarify whether the value of ${\rm De}_c$ impacts the solutions of \eqref{eq:conform_phi_theta_surface} in a way explaining the rise velocity jump, the crucial question is under which conditions the ODE \eqref{eq:conform_phi_theta_surface} has solutions starting at $0< \theta_0 \approx 0$ from values $y_0 \approx 1$ and reach the bubble's equator (i.e., $\theta = \pi /2$) in a significantly extended state, which is still relaxing, such that forces from the contracting polymer molecules will accelerate the bubble.
Evaluating the right-hand side of \eqref{eq:conform_phi_theta_surface} at $\theta = \pi /2$ yields
\begin{equation}\label{eq:9}
y' \left(\frac{\pi}{2}\right)= -\, \frac{1}{2\, {\rm De}_c} 
\big(y(\frac{\pi}{2})-1\big).
\end{equation}
This shows that the polymer molecules passing by the bubble's equator are relaxing ($y' <0$) if and only if they are stretched in the circumferential direction ($y >1$), and the relaxation is the stronger, the larger the extension in the circumferential direction, i.e.\ the larger $y>1$. Remarkably, this relation is independent of ${\rm De}_c$. Note also that the same conclusion holds in the general case of \eqref{eq:conform_phi_theta}. However, we keep focussing on the limiting case $\psi^* \to 0+$, since the solutions of \eqref{eq:conform_phi_theta} converge to those of \eqref{eq:conform_phi_theta_surface} for fixed Deborah number; cf.~Fig.~\ref{fig:de_variation_solutions} (right). 

To answer the question under which conditions the ODE \eqref{eq:conform_phi_theta_surface} allows for solutions which reach a value significantly above $y=1$ at $\theta = \pi/2$, 
we solve \eqref{eq:conform_phi_theta_surface} backwards, starting at $\theta = \pi/2$. We therefore consider $u(\theta)= y(\pi /2 -\theta)-1$ for $\theta \in [0,\pi/2)$ with $u(0)=u_0$, i.e.\ we study the initial value problem
\begin{equation}\label{eq:12}
u'(\theta)=2 \frac{1/{\rm De}_c -  \sin\theta}{\cos\theta} u(\theta) - 2 \frac{\sin\theta}{\cos\theta}, \quad u(0)=u_0.
\end{equation}
For ${\rm De}_c \neq 1$, the initial value problem \eqref{eq:12} has the unique solution
\begin{equation}\label{eq:15}
u(\theta)= \frac{(1+\sin\theta)^2}{4({\rm De}_c^{-1} -1)}
- \frac{(1-\sin\theta)^2}{4({\rm De}_c^{-1} +1)}
+ C_0 \, \cos^2 \theta \Big( \frac{1+\sin\theta}{1-\sin\theta} \Big)^{{\rm De}_c^{-1}}
\end{equation}
with 
\begin{equation}\label{eq:16}
C_0 = u_0 + \frac{1}{4({\rm De}_c^{-1} +1)} -  \frac{1}{4({\rm De}_c^{-1} -1)}.
\end{equation}
To analyze the behavior of the solution $u(\cdot)$ of \eqref{eq:12} as $\theta$ approaches $\pi/2$, we rewrite \eqref{eq:15} as
\begin{equation}\label{eq:15b}
u(\theta)= \frac{(1+\sin\theta)^2}{4({\rm De}_c^{-1} -1)}
- \frac{(1-\sin\theta)^2}{4({\rm De}_c^{-1} +1)}
+ C_0 \, (1+\sin \theta)^2 \Big( \frac{1+\sin\theta}{1-\sin\theta} \Big)^{{\rm De}_c^{-1}-1}.
\end{equation}
Now, as $\theta \to \frac{\pi}{2}-$, all terms converge to a finite value, except for the last factor, which tends to zero in case of ${\rm De}_c >1$, while it diverges to $+ \infty$ in case of ${\rm De}_c <1$. Hence, the qualitative behavior of $u(\cdot)$ for $\theta \to \frac{\pi}{2}-$ is as follows:
\vspace{12pt}\\
{\bf Case ${\rm De}_c <1$. } We must distinguish two subcases:
\begin{equation}
    u(\theta)\to - \infty \mbox{ as } \theta \to \frac{\pi}{2}-
    \; \mbox{ if  }
    u_0 <  \frac{1}{4({\rm De}_c^{-1} -1)} - \frac{1}{4({\rm De}_c^{-1} +1)}
\end{equation}
and
\begin{equation}
    u(\theta)\to + \infty \mbox{ as } \theta \to \frac{\pi}{2}-
    \; \mbox{ if  }
    u_0 >  \frac{1}{4({\rm De}_c^{-1} -1)} - \frac{1}{4({\rm De}_c^{-1} +1)}.
\end{equation}
\begin{figure}[t!]    
\centering{
\includegraphics[width=0.8 \textwidth]{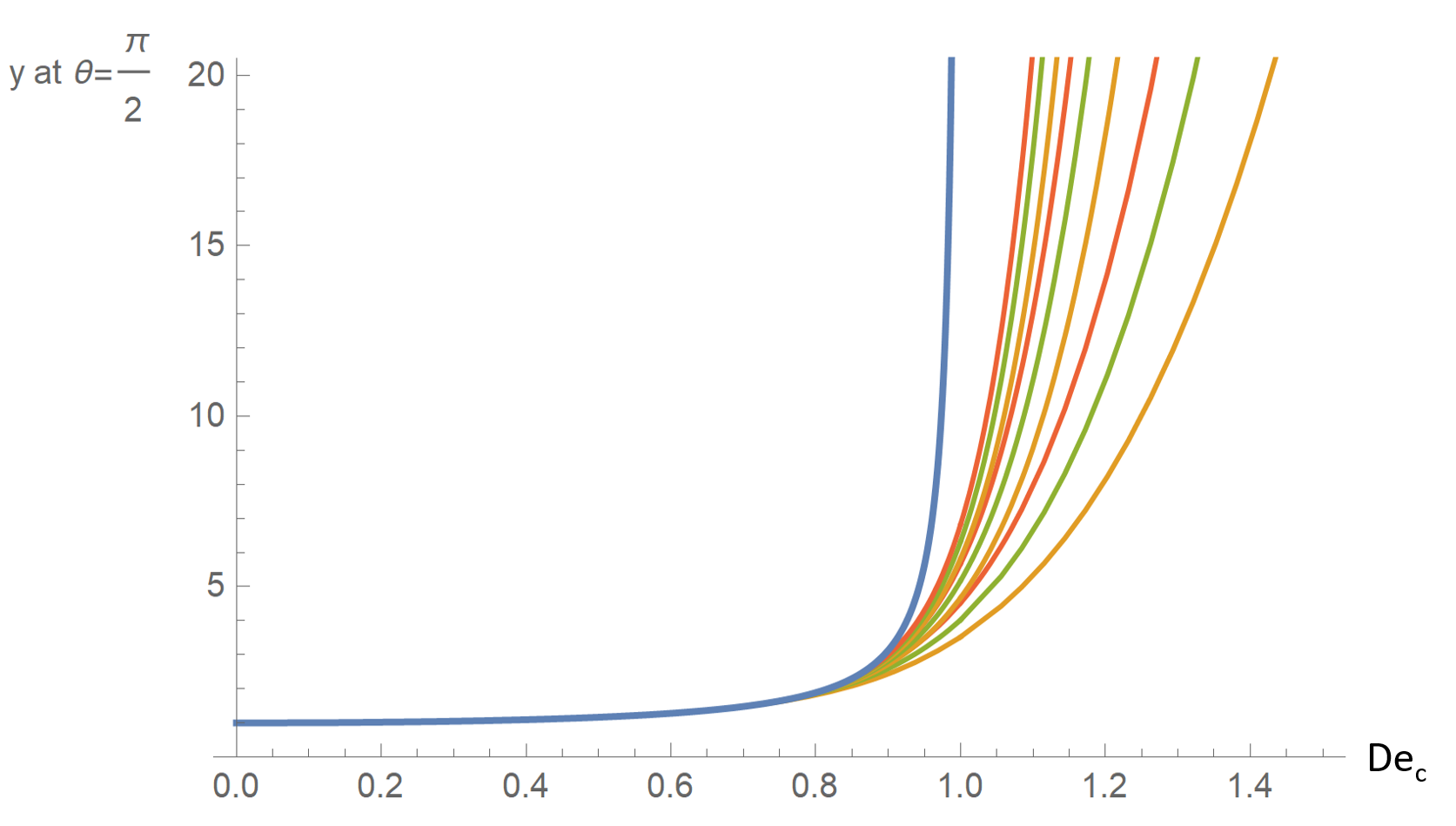}
}
\caption{\small Blue curve: Maximum reachable circumferential conformation component at the bubble equator as a function of the Deborah number in case ${\rm De}_c <1$. Other curves: circumferential conformation component at the bubble equator as a function of the Deborah number for initial values of $y_0=1$ (orange), $y_0=3$ (green) and $y_0= 5$ (red)
and initial polar angles of $\theta_0 =0.01$ (lowest), $\theta_0 =0.001$ (middle) and $\theta_0 = 0.0001$ (highest).}
\label{fig:ymax-plot}
\end{figure}
\noindent
In the first subcase, the limit of $-\infty$ of the backward solution $u(\cdot)$ implies for the original ODE that every $y_0=u_0 + 1$ with such a $u_0$ is reachable at the bubble's equator from an appropriate (possibly large) initial value $y_0$ and initial $\theta_0 >0$ close to zero. In the second subcase, the respective values $y_0=u_0 + 1$ are
not reachable at all. In other words, in case of ${\rm De}_c <1$, there exists a maximal value $y_{\rm max}$, depending on ${\rm De}_c$ according to
\begin{equation}\label{eq:ymax}
    y_{\rm max} ({\rm De}_c) = 1 + \frac{1}{4({\rm De}_c^{-1} -1)} - \frac{1}{4({\rm De}_c^{-1} +1)} = \frac{1- 0.5 \, {\rm De}_c^{2}}{1- {\rm De}_c^{2}},
\end{equation}
which is the maximum reachable circumferential conformation component at this value of the Deborah number. The relation \eqref{eq:ymax} is plotted as the blue curve in Fig.~\ref{fig:ymax-plot}.
For comparison, the other curves shown in this figure visualize the values reachable from a certain set of initial values. \vspace{12pt}\\
{\bf Case ${\rm De}_c >1$: } Then
\begin{equation}
    u(\theta)\to \frac{1}{4({\rm De}_c^{-1} -1)} - \frac{1}{4({\rm De}_c^{-1} +1)} \, < \, 0.
\end{equation}
In this case, arbitrary large values of $y$ are reachable at the bubble equator for appropriate (correspondingly large) initial values $y_0$ at initial $\theta_0 >0$ close to zero. \\

This analysis substantiates and confirms the impression from the vector field plots in Fig.~\ref{fig:de_variation}
and leads us to the following conclusion:
in case of ${\rm De}_c <1$, i.e.\ for fast rel\-ax\-ation of the polymer molecules compared to the convective time scale, the normalized circum\-ferential conformation component $L_{\phi \phi}^*$ cannot attain a value significantly larger than 1 at the equator, starting at a value of about 1 near the upper bubble pole. In turn, in case of
${\rm De}_c >1$, if $L_{\phi \phi}^*$ is just slightly larger than 1 at small polar angle $\theta$, then $L_{\phi \phi}^*$ can reach values significantly above 1 at the bubble's equator. 
Thus, at small Deborah number the major part of the relaxation takes place before the molecules reach the bubble's equator, while, at high Deborah number, a strong relaxation persists until below the equator. In the latter case, the bubble experiences an upward pointing force due to a hoop stress below its equator, which leads to an increased rise velocity. Since this effect is self-amplifying, a velocity jump arises at a critical Deborah number, where the latter is given by ${\rm De}_{\rm crit} =1$ in the idealized scenario considered above.

\section{Discussion and conclusions}
\label{sec:discussconclusions}
The aim of this study is to identify the molecular mechanism behind the formation of sub- or supercritical terminal rise velocities of individual gas bubbles in polymer solutions, providing a physical explanation of the mechanism which leads to a jump discontinuity, i.e.\ to an abrupt increase of the terminal rise velocity for a slight increase in bubble size at a critical volume. 
For this purpose, and in order to allow for a cross-validation of our findings, we combine three different approaches: (i) an experimental study, providing visual information on the bubble shape complemented by PIV-based measurements of the local velocity fields, (ii) a computational approach, employing fully resolved numerical simulations of the continuum physical governing equations employing appropriate rheological models which account for the orientation-deformation state of the polymer molecules via a conformation tensor and (iii) a theoretical investigation, which focusses on understanding the evolution of the most relevant conformation tensor component, which turns out to be the circumferential one. We first summarize the main findings from the different approaches.

Ad (i). While the experiments confirm the occurrence of the rise velocity jump discontinuity at a certain critical bubble volume (cf.\ Fig.~\ref{fig:JumpPIV-P2500}), which depends in particular on the liquid material properties, the results presented in Section~\ref{sec:experimentalresults} go considerably beyond the existing knowledge on integral quantities. Indeed, the PIV data not only display
the negative wake in the velocity field below supercritical bubbles (cf.\ Fig.~\ref{fig:vorticity-P2500-subsupcritical}), but the local velocity information is analyzed regarding Lagrangian transport times of the polymer molecules. This ana\-ly\-sis shows that polymer molecules, starting from the region at the upper pole and traveling along (or close to) the bubble surface in a time interval longer than the relaxation time, cannot reach the equator in a stretched state (cf.\ Fig.~\ref{fig:timescale-P2500-subcritical}(a)), while they evidently reach the lower hemisphere stretched in the supercritical case, in which the transport and stress relaxation times are similar
(cf.\ Fig.~\ref{fig:timescale-P2500-subcritical}(b)).
The crucial point in this experimental finding is that the relaxation time is, hence, comparable to bubble-related hydrodynamic transport times, but clearly different ratios of the relaxation to the convection time scales distinguish the sub- from the supercritical bubbles. This shows that the rise velocity jump discontinuity is related to the interplay of convection and relaxation and their characteristic time scales.

Ad (ii). 
The numerical approach enables simulations which solve the initial-boundary value problem for the buoyancy-driven bubble motion in a polymeric liquid and capture the complete fields of the velocity within the liquid and the gas phase as well as the polymer orientation-deformation tensor as functions of space and time. The simulation results therefore include all the information needed for a physical explanation of the terminal bubble rise velocity, including the genesis of the jump discontinuity. In particular, our numerical approach delivers the correct rise velocities, yields a negative wake in case of supercritical bubbles, and also realistic bubble shapes \cite{Niethammer2019}, although a true cusped shape cannot be accurately reproduced, as with most numerical two-phase flow methods. But the latter is not problematic, since the cusped shape is known to be of minor importance for the jump discontinuity of the rise velocity (cf.\ the survey in Section~\ref{sec:introduction}); we shall briefly come back to this point below.

With fully resolved numerical simulations at hand, it is possible to analyze the local field quantities, in particular the stress acting on the bubble surface. This approach was already employed in \cite{Pillapakkam2007} and in further papers since then, including our work \cite{Niethammer2019}. Up to now, the focus in all contributions was completely on the radial and the polar angular normal stress components. 
The corresponding results, such as the ones shown in Figs.~\ref{fig:tau_TT_sub}--\ref{fig:tau_RR_super} in Appendix~\ref{asec:stresstensor}, show a peculiar behavior:
the polar angular component displays a significant difference between the sub- and the supercritical cases, with a stress boundary layer building 
up in the region around the upper bubble pole, but either relaxing completely along the upper hemisphere, or persisting far beyond the bubble equator and with higher values. The radial normal stress component does not show much difference between the two cases and is localized around the bubble's lower pole.

While this is obviously of interest, it does not help much to explain the underlying mechanism causing such a different behavior.
To this end, the interplay between the local flow and the orientation-deformation state of the polymer molecules needs to be understood. For this purpose, the conformation tensor is the most relevant quantity and, for this reason, the analysis of the local trace of the conformation tensor has indeed already been started in \cite{Pillapakkam2007}. The corresponding results from our present simulations, displayed in Figs.~\ref{fig:tr_C_sub} and \ref{fig:tr_C_super}, show a marked difference between the sub- and the supercritical states, as discussed in Section~\ref{subsec-conformation}. %
But since the trace of the conformation tensor represents the average elongation (length) of the polymer molecules, directional information is not included. To extract information on the polymer orientation, spectral information about the conformation tensor is required. More precisely, as explained in detail in Appendix~\ref{asec:conftens}, the
eigenvector belonging to the largest of the three (positive) eigenvalues gives the direction of the main polymer orientation, with the square root of the eigenvalue measuring the average component of the polymer end-to-end vectors in this direction.
This allows the polymer molecule's average orientation to be visualized
by plotting line elements in the direction of the eigenvectors, colored according to the eigenvalue, as displayed in Figs.~\ref{fig:ev_sub} and \ref{fig:ev_super}. 
The simulations show that, during their passage through the upper pole region, the polymer molecules get oriented mainly in the circumferential direction. 
While this observation about the polymer molecular conformation at the bubble's upper pole region was also already made in \cite{Noh1993}, the new and central finding is that the circumferential stretching and the ratio between the transport time of the polymer molecules along the bubble surface and their relaxation time scale is the key for understanding the rise velocity jump discontinuity. In fact, this molecular orientation-deformation state either relaxes before the molecules pass the equator, or it persists far downstream beyond the bubble's equator. 
This strongly supports the hypothesis that the rise velocity jump discontinuity is related to the interplay of convection and relaxation and their characteristic time scales.
Furthermore, these results underline the crucial importance of the circumferential conformation tensor component.

\begin{figure}[b!]    
\centering{
\includegraphics[width=0.7 \textwidth]{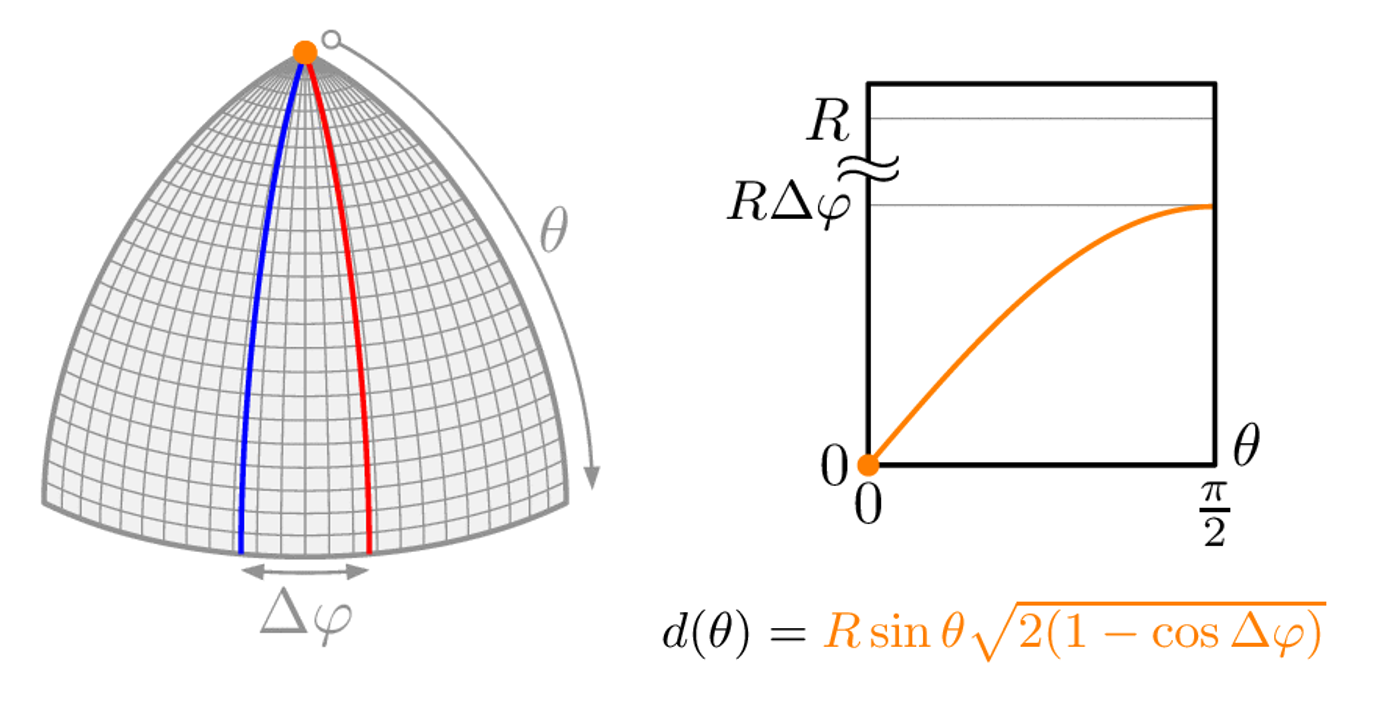}
}
\caption{\small Illustration of the kinematic effect due to diverging (upper pole region) and converging (lower pole region, not shown) trajectories (red and blue) and the evolution of the circumferential distance (orange) $d(\theta)$ of particles moving along neighboring longitudinal trajectories.}
\label{fig:kinematic_effect}
\end{figure}
The reason for the polymer molecules to get oriented and stretched in the circumferential direction near the upper pole is a purely kinematic one. As illustrated in Fig.~\ref{fig:kinematic_effect}, trajectories going down along the upper bubble hemisphere, starting from initially nearby positions with a slight offset in the azimuthal angle,  diverge until their maximum distance at the equator, before they converge again. This qualitative behavior, especially the divergence on the upper hemisphere and the resulting orientation and stretching of the polymer molecules in the circumferential direction, is independent of the bubble volume.
What is influenced by the bubble volume and, hence, by the (transient) rise velocity, is how far the stretched polymer molecules travel along the bubble contour before relaxation is complete. This, of course, decides on where the elastic energy stored in the elongated molecules will be returned (partly) to the fluid, and hence to the bubble: above or below the equator. 

Ad (iii). With a velocity field given, the transport and reorientation-deformation of individual polymer molecules is described by the transport-evolution equation \eqref{eq:conform}. Moreover, under realistic symmetry assumptions, the transport-evolution equation for the circumferential component decouples from the remaining system, allowing for a thorough theoretical analysis in this case. Motivated by both experimental and numerical results on the velocity field close to the bubble's upper hemisphere, we employ the Hadamard-Rybczynski flow field around a fluid sphere translating in a liquid as the velocity field for use with the transport-evolution equation. The resulting differential equation is written as an ODE in the polar angle $\theta$ as the independent variable. From a backward solution analysis, it is rigorously shown that ${\rm De}_c =1$ is a critical Deborah number: For ${\rm De}_c <1$, the maximum values of $C_{\phi \phi}$ are bounded and moderate in value, while arbitrary large values for $C_{\phi \phi}$ are possible in the opposite case ${\rm De}_c >1$. Quantitatively, if ${\rm De}_c < 0.9$, say, $C_{\phi \phi}$ at the bubble's equator is bounded to values around 2, while values of 10 or higher are reached there in case of ${\rm De}_c >1$. This ODE analysis in fact shows that for ${\rm De}_c <1$, polymer relaxation is fast and completed above the bubble's equator, while relaxation is slower than convection and, thus, persists below the equator in case of ${\rm De}_c >1$.

\begin{figure}[t!]    
\centering{
\includegraphics[width=0.96 \textwidth]{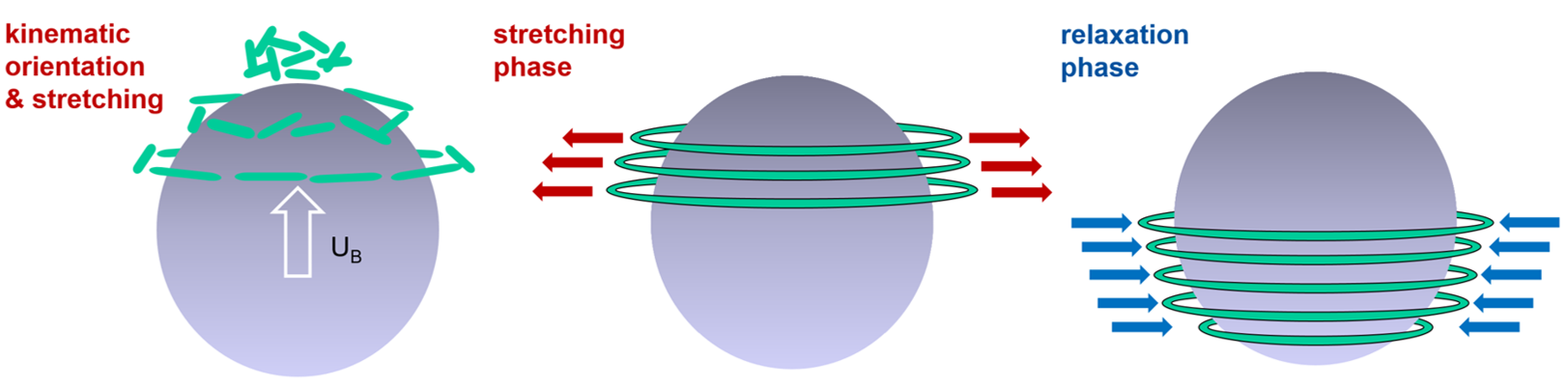}
}
\caption{\small Illustration of the kinematic orientation and stretching of the polymer molecules and the persistence of the relaxation phase for a supercritical bubble, leading to a resulting upward force induced by the hoop stress.}
\label{fig:hoop-stress}
\end{figure}

It is worth noting that this ODE analysis does not contain any back-coupling of the polymer dynamics to the flow fields. The latter is accounted for in the numerical simulations and, there, causes a self-amplification.
To explain this feedback mechanism, we consider the transient
rise of a bubble:
If the transport of the polymer molecule is fast enough such that the relaxation persists until slightly below the equator, the resulting hoop stress accelerates the bubble. Then the subsequent polymer molecules unload their elastic energy a bit further below the bubble's equator, where the force component in the rise direction will be larger due to the different normal direction of the bubble surface. This further accelerates the bubble, and so on. This process amplifies itself until, if this state is reached, the relaxation persists even downstream from the bubble. Then, with increasing distance of the relaxing polymer molecules from the bubble, the upward pushing force vanishes. The mechanism as just explained is illustrated by the schematics in Fig.~\ref{fig:hoop-stress}.

In order to conclude our investigation, let us first recall that the rise velocity jump discontinuity is usually explained as a consequence of either the appearance of a negative wake below the rising bubble, or the formation of a cusp at the bubble's trailing edge, or both features together. The published work on this topic, however, does not give a clear picture.
Even more, none of the existing papers provides a root cause for the existence of a critical bubble volume at which the rise velocity undergoes a true jump, i.e.\ an abrupt increase for only a tiny--almost infinitesimal--change in the bubble volume.
Some exception is the work by Fraggedakis et al.\ \cite{Fraggedakis2016}, which studies the rise velocity jump phenomenon by means of a dynamical systems approach. Employing an FEM-based ALE-type numerical method for axisymmetric two-phase flow in combination with a continuation method, they are able to detect a  curve in the bubble diameter-rise velocity-plane, corresponding to a family of quasi-steady state solutions under the variation of the bubble diameter as a parameter. With this approach they show the existence of multiple steady states of different stability properties: in some range of parameters, three possible rise velocities for the same bubble diameter are found. Two stable ones, corresponding to the slow and the fast bubble, and an unstable rise at an intermediate velocity. These three solutions are placed on an S-shaped curve, where the turning points indicate a change in the stability behavior.
The authors conclude that an increase of the bubble diameter on the arc with low rise velocity ends at such a turning point, and that a further increase of the bubble diameter then leads to the higher rise velocity, i.e.\ to the point on the highest arc of the solution curve, since this is again a stable solution. In contrast to the literature prior to \cite{Fraggedakis2016}, this study provides a dynamical systems theoretical justification for the appearance of a jump. But it still does not provide a physical mechanism that explains the occurrence of arcs of solutions with different rise velocities. 
With the present findings, this mechanism is now identified to be the self-amplifying bubble acceleration due to a vertical hoop stress component.

To confirm and test the proposed mechanism, it is important that it can explain the occurrence of the two stable, hypothetical solutions at different rise velocities for a certain parameter regime, as well as another hypothetical solution which is unstable. This is indeed possible: If a subcritical bubble, with a volume not too far below the critical volume, would by accelerated to an artificial, higher rise velocity by an external force, the relaxation of the polymer molecules, being oriented and stretched in the azimuthal direction, would take place below the bubble's equator, thus further accelerating the bubble. Then--in this thought experiment--if the artificial external force would fade, the bubble still could keep a higher velocity if the relaxation still continues below the equator. This proves the possible existence of a second stable solution with a larger rise velocity, although it will not occur in real experiments. But even the third, intermediate and unstable solution can be explained along the same line: if the bubble would be brought to an intermediate rise velocity at which the relaxation-induced effects of de- and acceleration are just balanced, the bubble could--in principle--also hold its velocity. Of course, any disturbance slightly changing the bubble velocity will induce either a fall back to the lower or an acceleration to the faster rise mode. Here, again, the self-amplifying character of the mechanism is necessary.
Consequently, the mechanism which we propose as the root cause for the bubble rise velocity jump discontinuity, also explains the stability behavior of the dynamical system composed of a buoyancy-driven bubble rising in a viscoelastic liquid.

Furthermore, the formation of a ring of circumferentially oriented and 
stretched polymer molecules at the upper pole region can also help to explain the other characteristic features, i.e.\ the cusped shape and the negative wake of supercritical bubbles.
Indeed, since the supercritical case is correlated with the presence of significant hoop stress below the bubble's equator, this hoop stress
not only transmits an upward force to the bubble, but also pushes the bubble surface towards the axis of symmetry. The fraction of the horizontal force component increases, the more the surface normal turns into a horizontal direction. This is another type of self-amplification, which can lead to a surface normal field that is almost horizontal, uniformly in all azimuthal directions. This produces the shape of a cusp and, thus, shows that the formation of a cusped shape is another consequence of the unloading of circumferentially stretched polymer molecules below the bubble equator. In particular,
the cusped shape is not the cause of the velocity jump discontinuity, but rather another result of the same underlying mechanism.

To see the relation with the negative wake of supercritical bubbles, note that, as visible in Fig.~\ref{fig:ev_super}~(e), the ring of circumferentially oriented, stretched polymer molecules persists even until below the bubble, where it still contracts and, thus, pushes the liquid towards the axis of symmetry. A part of the liquid then moves upwards, the rest moves downwards, creating a negative wake.
Note that this explanation requires the found hoop stress generated by the circumferential orientation of the polymer molecules.
\begin{figure}[t!]    
\centering{
\includegraphics[width=0.68 \textwidth]{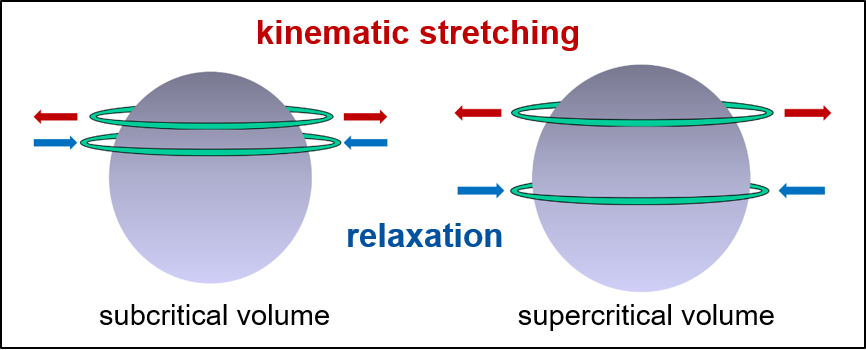}
}
\caption{\small Illustration of the different evolution of the polymer deformation state for sub- and supercritical bubble volume.}
\label{fig:mechanism_comparison}
\end{figure}
The experimental, numerical and theoretical investigations presented above show that the states of steady motion of sub- and a supercritical bubbles in a viscoelastic polymer solution differ concerning the hoop stress set free by the relaxation of polymer molecules oriented and elongated in the flow along the bubble contour. Starting from a stagnation state at the upper pole of the bubble, the molecules are stretched in the polar and azimuthal directions of a biaxial elongational flow along the bubble. The molecules tend to relax in order to maximize the entropy of their state of conformation. The related relaxation time competes with the transport time down the bubble contour. Subcritical bubbles exhibit slow molecule transport, so that the stress relaxation takes place along the upstream hemisphere of the bubble. In contrast, the polymer molecules are transported fast along supercritical bubbles, so that the stress relaxation takes place along the downstream part of the bubble contour. In the two cases, the hoop stress set free hinders or enhances the motion, respectively. The two different scenarios are illustrated in Fig.~\ref{fig:mechanism_comparison}. Let us finally note that the relevance of the circumferential orientation of the polymer molecules found also explains the need for 3D, or at least axisymmetric, simulations to capture the rise velocity jump discontinuity.

\section*{Acknowledgements}
This work was partly supported from the German Research Foundation (DFG) -- Project-ID 265191195 -- SFB 1194. Funding was furthermore provided by the Austrian Science Fund (FWF) under project number P 17624-N07. This support is gratefully acknowledged.
Support of the PIV experiments by Prof. J. Woisetschl{\"a}ger at TU Graz is gratefully acknowledged.
Calculations for this research were conducted on the Lichtenberg high performance computer of the Technische Universit{\"a}t Darmstadt. 

\bibliographystyle{unsrturl}
\bibliography{RisingBubble.bbl}

\appendix
\section{On the conformation tensor}
\label{asec:conftens}
The orientation-deformation state of a polymer molecule is characterized by its end-to-end vector ${\bf r}$, where
${\bf r}/||{\bf r}||$ gives the orientation, while $||{\bf r}||$ is the length of an individual molecule.
The local macroscopic, continuum mechanical state of the polymer solution is determined not by a single polymer molecule but rather by ensembles of molecules with a certain distribution of the individual ${\bf r}$ vectors.
Hence, a stochastic description is required, where ${\bf r}={\bf r}(\omega)$ is a random vector of such
orientation-deformation states. We consider un-polar fluids in which a flip of the polymer molecule orientation, i.e.\ a change from ${\bf r}$ to $- {\bf r}$, is not detectable or at least both micro-states have the same probability (density).
Hence the random vector ${\bf r}$ has zero expectation (mean value):
\begin{equation}\label{eq:A1}
{\rm E}\, [{\bf r} ] := \int_{\Omega} {\bf r}(\omega) {\rm d} P(\omega) =0.
\end{equation}
If the distribution of ${\bf r}$ has the probability density $\psi = \psi ({\bf r})$, then
\begin{equation}\label{eq:A2}
{\rm E}\, [{\bf r} ] = \int_{\mathbb R} {\bf r} \, \psi ({\bf r}) \, {\rm d} {\bf r}.
\end{equation}
Of course, the distribution is a local quantity, i.e.\ $\psi = \psi (t, {\bf x}, {\bf r})$, but this additional dependence on $(t, {\bf x})$ is suppressed here for better readability.

Evidently, the expectation ${\rm E}\, [{\bf r} ]$ (being zero) does not give any information about possible preferred orientations of ${\bf r}$. In order to extract such information, in a statistical sense, from the distribution of ${\bf r}$, let ${\bf n}$ denote a unit vector, thus prescribing a certain direction in the polymer orientation space $\mathbb{R}^3$. The (scalar) random variable
\begin{equation}\label{eq:A3}
R_n = {\bf n}^{\sf T} {\bf r} = \sum_{i=1}^3 {\bf n}_i {\bf r}_i
\end{equation}
gives the length of the projection of the random vector ${\bf r}(\omega)$ onto the one-dimensional subspace (the line)
spanned by ${\bf n}$. While ${\rm E}\, [R_n ]=0$ again, the variance of $R_n$ quantifies to which extend the random vector ${\bf r}$ is aligned with this line. Indeed, the larger this variance is, the larger is the probability that
${\bf r}$ is within a given angle around the direction $(\pm)\, {\bf n}$. Hence, one way to visualize the preferred
orientation of ${\bf r}$ in a way which allows for comparing different directions is
the {\it orientation surface}
\begin{equation}\label{eq:A5}
\Sigma ({\bf r}) := \{ \alpha ({\bf n}) \, {\bf n} \, : \,  {\bf n}\in \mathbb{R}^3 \mbox{ with } || {\bf n}||=1,\,
\alpha ({\bf n}) = {\rm Var}\, [{\bf n}^{\sf T} {\bf r} ]^{1/2}  \}.
\end{equation}
In order to compute this set, notice that, for any given fixed (non-random) vector ${\bf n}$, it holds that
\begin{equation}\label{eq:A6}
{\rm Var}\, [ {\bf n}^{\sf T} {\bf r}] = {\bf n}^{\sf T} \, {\rm Cov}\, [{\bf r}] \, {\bf n},
\end{equation}
where
\begin{equation}\label{eq:A7}
{\rm Cov}\, [ {\bf r}] := {\rm E}\, \big[ ({\bf r} - {\rm E}\, [{\bf r}]) \otimes ({\bf r} - {\rm E}\, [{\bf r}]) \big]
\end{equation}
denotes the (variance-)covariance matrix of the random vector ${\bf r}$.
Since ${\bf r}$ has zero mean, we have
\begin{equation}\label{eq:A8}
{\rm Cov}\, [ {\bf r}] = {\rm E}\, [{\bf r} \otimes {\bf r}]  = \big[ {\rm E}\, [{\bf r}_i \, {\bf r}_j ] \big]_{i,j=1}^3.
\end{equation}
From this identity, it follows that in our setting, the tensor ${\rm Cov}\, [ {\bf r}]$ coincides
with the conformation tensor ${\bf L}={\bf L} ({\bf r})$, i.e.\
${\bf L} ({\bf r})= {\rm Cov}\, [ {\bf r}]$.
Due to \eqref{eq:A6}, the orientation surface from \eqref{eq:A5} is, hence, given as
\begin{equation}\label{eq:A9}
\Sigma ({\bf r}) = \{ \big( {\bf n}^{\sf T}\, {\bf L} ({\bf r})\, {\bf n}   \big)^{1/2} \, {\bf n}
\, : \,  {\bf n}\in \mathbb{R}^3 \mbox{ with } || {\bf n}||=1 \}.
\end{equation}
We employ the spectral representation
\begin{equation}\label{eq:A12}
{\bf L} \, = \, {\bf S} \, {\bf D} \, {\bf S}^{\sf T}
\quad \mbox{ with }
\quad {\bf S}  \, {\bf S}^{\sf T} \, = \, {\bf I}, \quad
{\bf D} \, = \, {\rm diag} ( d_1, d_2, d_3),
\end{equation}
where $d_i >0$ are the eigenvalues of ${\bf L}$; the dependence on ${\bf r}$ is suppressed for better readability.
We now pass to the (local) principal coordinate system (w.r.\ to the spectral decomposition of ${\bf L}$ above), i.e.\ we transform into the orthonormal basis $\{ {\bf u}^1, {\bf u}^2, {\bf u}^3 \}$, given by the eigenvectors of ${\bf L}$ which are in turn the columns of ${\bf S}$. Then a given ${\bf n}$ has representation ${\bf n} = \sum_{i=1}^3 z_i {\bf u}^i$, hence
\begin{equation}\label{eq:A14}
\alpha ({\bf n})^2 = {\bf n}^{\sf T}\, {\bf L} ({\bf r})\, {\bf n} =
\sum_{i,j=1}^3 z_i z_j \langle {\bf u}^i , {\bf L}\, {\bf u}^j \rangle = \sum_{i=1}^3 d_i \, z_i^2.
\end{equation}
It follows that
\begin{equation}\label{eq:A15}
\Sigma ({\bf r}) = \{ \big( \sum_{i=1}^3 d_i \, z_i^2 \big)^{1/2} \sum_{i=1}^3 z_i {\bf u}^i \in \mathbb{R}^3
\, : \, \sum_{i=1}^3 z_i^2=1 \},
\end{equation}
or, in the principal coordinate system,
\begin{equation}\label{eq:A16}
\tilde{\Sigma} ({\bf r}) = \{ \big( \sum_{i=1}^3 d_i \, z_i^2 \big)^{1/2} {\bf z} \in \mathbb{R}^3
\, : \, || {\bf z}|| =1 \}.
\end{equation}
This nicely illustrates the distribution of orientations of the polymer molecules. In particular, the preferred
orientation has its direction given by the eigenvector which is associated to the largest eigenvalue, and the
square root of the eigenvalue represents the deformation state (the degree of stretching).

Finally, observe that $\Sigma ({\bf r})$ is \underline{not} an ellipsoid, but is an inverted ellipsoid,
where the inversion is done w.r.\ to the unit sphere, replacing the point on $\Sigma ({\bf r})$
in direction of ${\bf n}$ by the point on the same line but with reciprocal distance to the origin.

\section{Computation of the tensors $\BT$ and $\OmegaT$}
\label{asec:lddt}
Using the diagonalization $\CT = \QT \dprod \DT \dprod \trans{\QT}$, the tensor $\BT$ can be computed as $\BT = \QT \dprod \tilde{\BT} \dprod \trans{\QT}$, where the elements of the diagonal tensor $\tilde{\BT}$ are given as a function of the tensor $\LT = \QT \dprod \tilde{\LT} \dprod \trans{\QT}$ as $\tilde{b}_{ii} = \tilde{l}_{ii}$. The tensor $\OmegaT$ can be computed as $\OmegaT = \QT \dprod \tilde{\OmegaT} \dprod \trans{\QT}$, where the tensor $\tilde{\OmegaT}$ has zero diagonal entries $\tilde{\omega}_{ii} = 0$, while its off-diagonal elements are given by 
\begin{equation}
\label{diagonalizeCr82}
\tilde{\omega}_{{{ij}, \; {i \neq j}}} = \frac{d_{ii} {\tilde{l}_{{ij}, \; {i \neq j}}} + d_{jj} {\tilde{l}_{{ji}, \; {j \neq i}}}}{d_{jj} - d_{ii}}, \ i, j = 1, 2, 3.
\end{equation}
For a detailed description on the local decomposition of the deformation terms in the convective derivative, we refer to \cite{Niethammer2018}.

\section{Volume averaging definitions}
\label{asec:vadefinitions}
The present section provides the definitions of the volume-averaged quantities in section \ref{sec:vofmodel}. 
Assume that $\PhiT$ is a scalar or tensor quantity such that $\PhiT(t,\cdot)$ is continuous in $\CCV(t)$ and $\PhiT(\cdot, \vec{x})$ is continuously differentiable in $\CCV^\phaseOne \cup \CCV^{\phaseTwo}$. The \textit{volume average} of $\PhiT$ over $\CCV$ is defined as
\begin{align}
    \label{defVA00}
    \overline{\PhiT} := \frac{1}{\vert \Omega \vert} \int_{\Omega} \PhiT(\vec{x}, t) \dV.
\end{align}
Moreover, the intrinsic phase average or phasic average over $\CCV^{\pm}$ is defined as
\begin{align}
    \label{defPA01}
    \overline{\PhiT}^{\pm} := \frac{1}{\vert \CCV^{\pm} \vert} \int_{\CCV^{\pm}} \PhiT(\vec{x}, t) \dV.
\end{align}

\section{Stress tensor components $\tau_{\phi\phi}$, $\tau_{\theta\theta}$ and $\tau_{rr}$}
\label{asec:stresstensor}
\begin{figure}[h!t]    
\centering{
\includegraphics[width=0.95\textwidth]{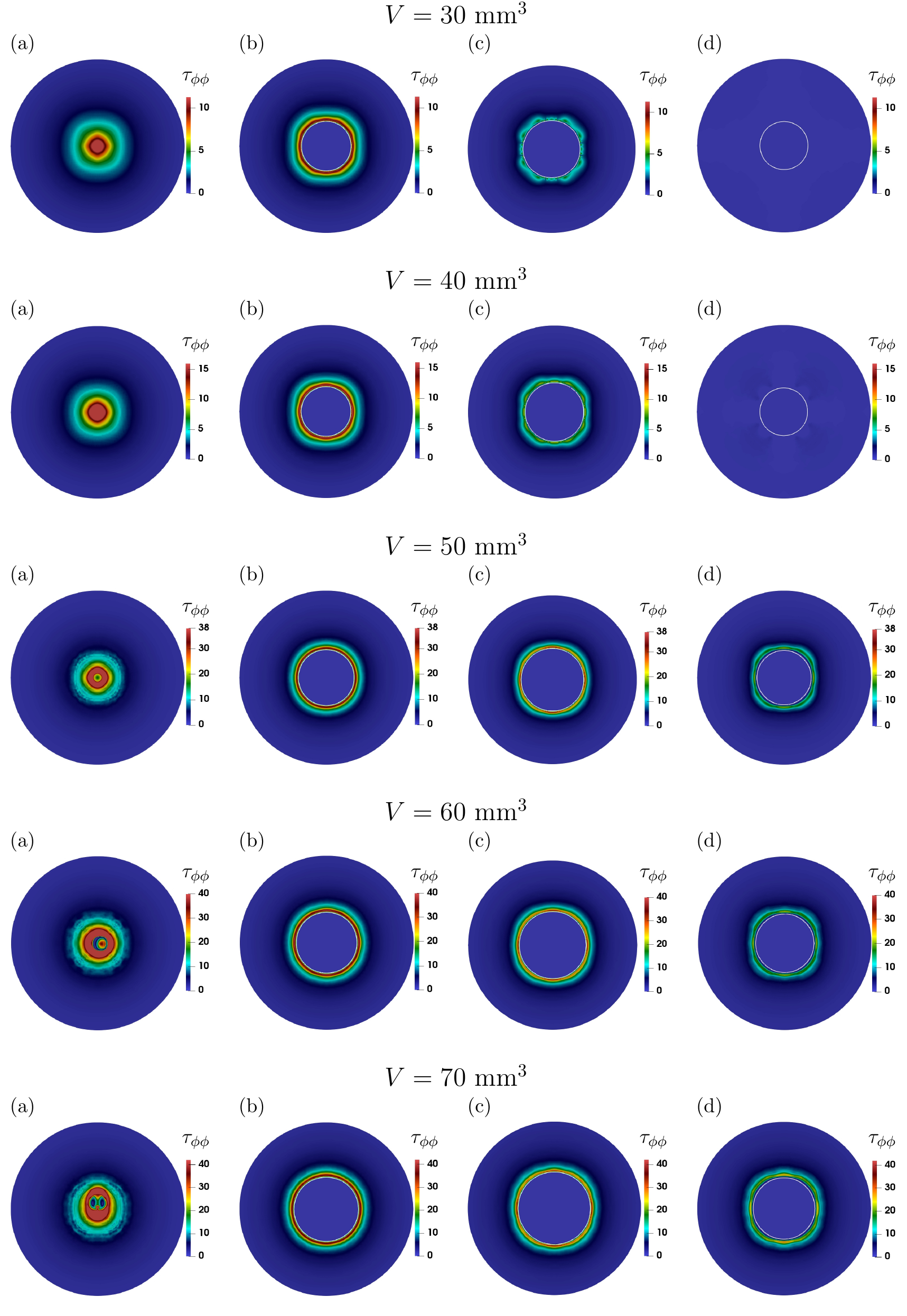}
}
\caption{\small Overview of the stress tensor component ${\tau}_{\phi\phi}$ (in Pa) at different bubble volumes in an aqueous $0.8$~wt.~\% Praestol $2500$ solution. The position of the horizontal cutting planes (a) - (d) corresponds to Figs. \ref{fig:tau_PP_sub} and \ref{fig:tau_PP_super}. The critical bubble volume is $V_\textnormal{crit} = 46$ mm$^3$.}
\label{fig:overview_tau_PP}
\end{figure}

To complement section~\ref{subsec:stressdistribution}, we present an overview of the stress component ${\tau}_{\phi\phi}$ at further sub- and supercritical bubble volumes in Fig. \ref{fig:overview_tau_PP}. Particularly relevant is how the hoop stress is distributed as we approach the critical bubble volume $V_\textnormal{crit} = 46$~mm$^3$. Comparing the two subcritical bubble volumes $V = 30$~mm$^3$ and $V = 40$~mm$^3$, the stress distribution is very similar in the cutting planes (a) to (d) at different heights, with slightly higher stress maxima for $V = 40$~mm$^3$. The supercritical bubble volumes $V = 50$~mm$^3$ to $V = 70$~mm$^3$ show a similar qualitative stress distribution as well. Consequently, the essential difference is between the subcritical and supercritical bubble volume in the planes (c) and (d). For all supercritical volumes, stress relaxation takes place to a large degree in the southern hemisphere of the bubble. For the subcritical volumes, the hoop stress is released almost completely in the northern bubble hemisphere. 

\begin{figure}[t!]    
\centering{
\includegraphics[width=\textwidth]{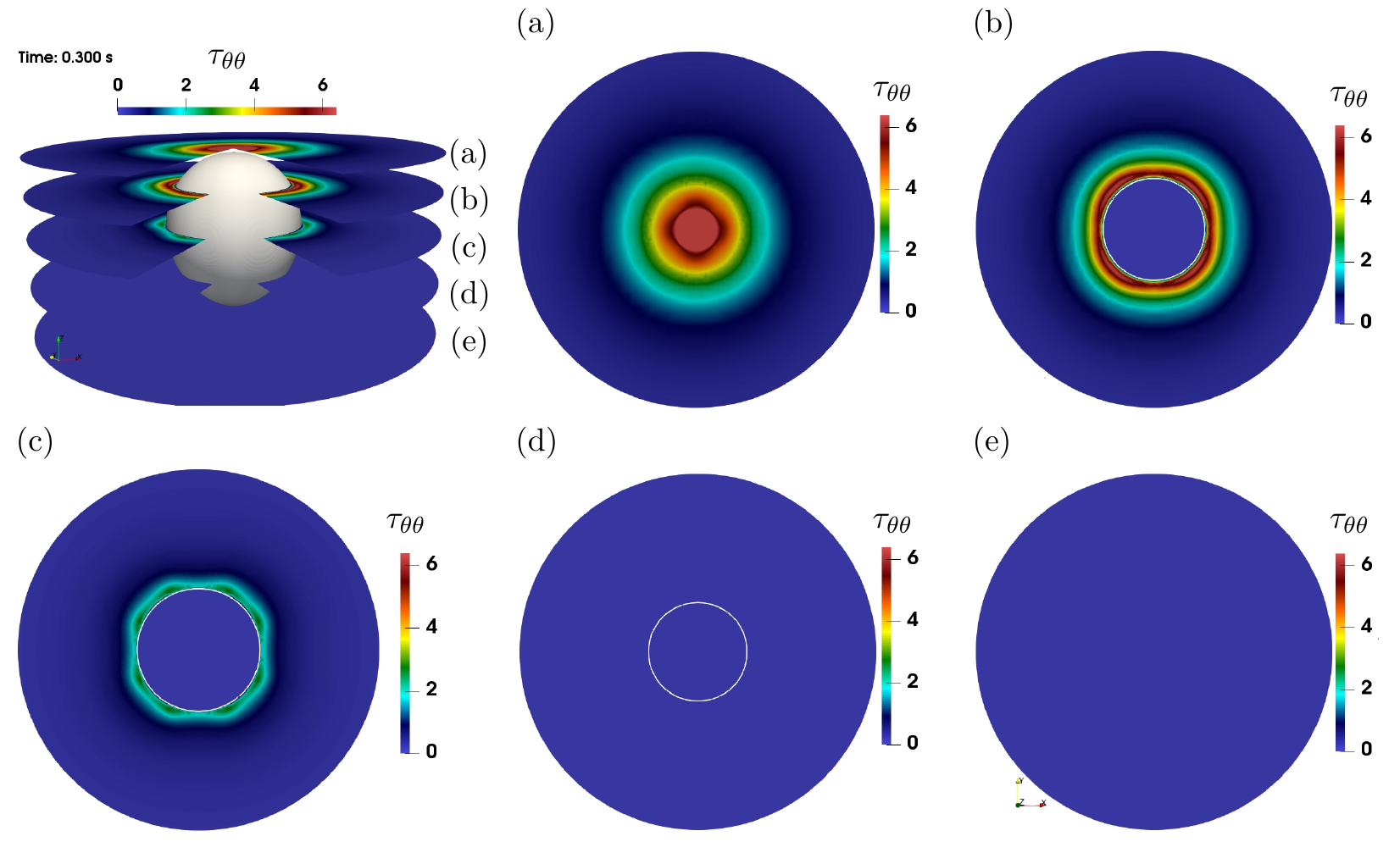}
}
\caption{\small Stress tensor component ${\tau}_{\theta\theta}$ (in Pa) in horizontal cutting planes at different heights (a) - (e). Subcritical bubble volume $V = 30~\textnormal{mm}^3$ in an aqueous $0.8$~wt.~\% Praestol $2500$ solution.}
\label{fig:tau_TT_sub}
\end{figure}
\begin{figure}[h!]    
\centering{
\includegraphics[width=\textwidth]{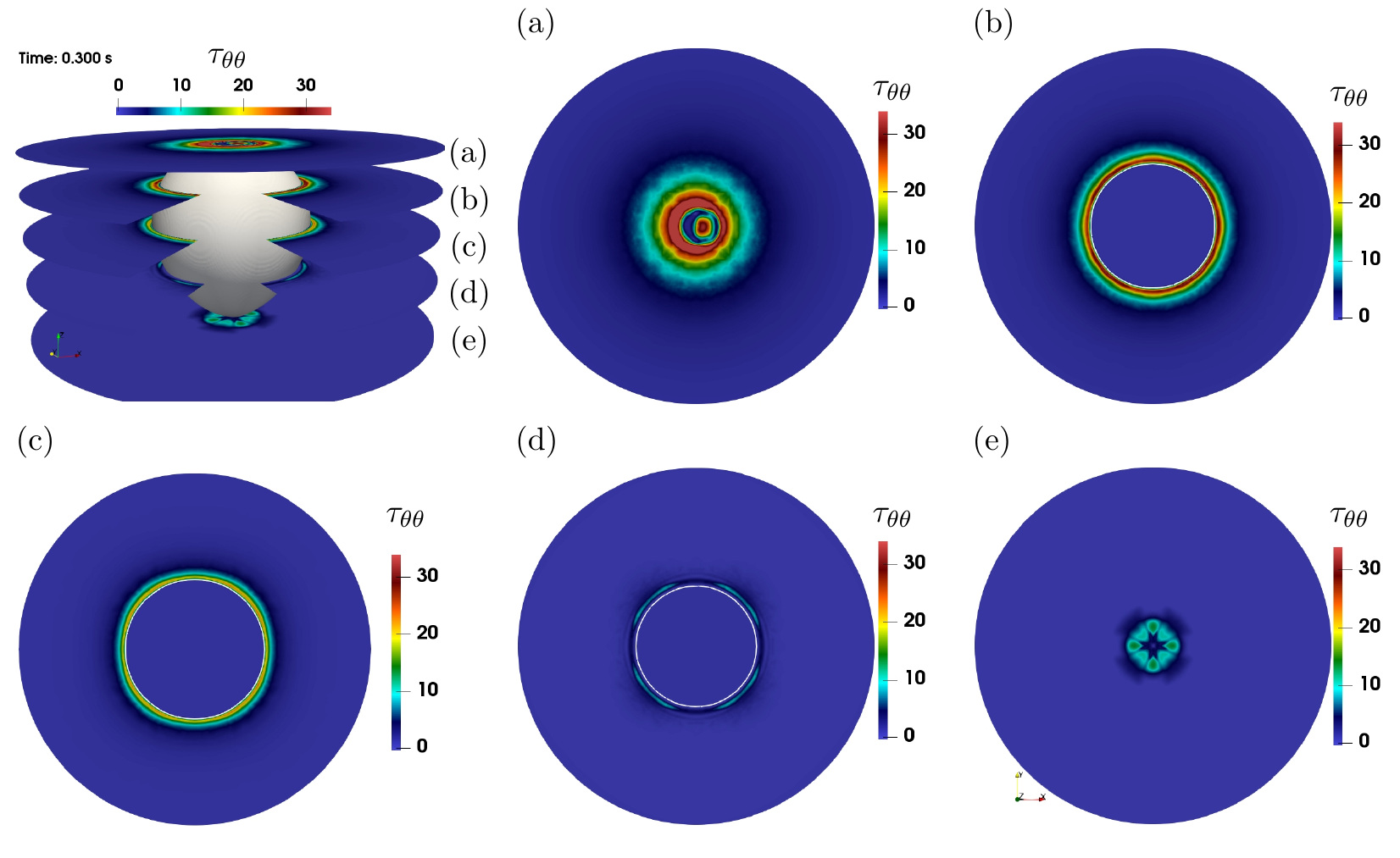}
}
\caption{\small Stress tensor component ${\tau}_{\theta\theta}$ (in Pa) in horizontal cutting planes at different heights (a) - (e). Supercritical bubble volume $V = 60~\textnormal{mm}^3$ in an aqueous $0.8$~wt.~\% Praestol $2500$ solution.}
\label{fig:tau_TT_super}
\end{figure}

Further numerical results are presented for the diagonal components of the stress tensor, \ $\tau_{\theta\theta}$ and $\tau_{rr}$, in horizontal cutting planes. Figs.~\ref{fig:tau_TT_sub} and \ref{fig:tau_TT_super} show $\tau_{\theta\theta}$ in the sub- and supercritical states, respectively. The spatial distribution of $\tau_{\theta\theta}$ is similar to the component $\tau_{\phi\phi}$ shown in Figs.~\ref{fig:tau_PP_sub} and \ref{fig:tau_PP_super}, but with noticeably smaller values.
On the upper hemisphere of the bubble,
the stress component $\tau_{\theta\theta}$ first increases with the polar angle $\theta$, but the relaxation starts already above the equator. In the subcritical state, the relaxation of the stress component $\tau_{\theta\theta}$ is complete at the bubble equator, as visible in  Figs.~\ref{fig:tau_TT_sub}~(a) - (c). On the lower bubble hemisphere and in the bubble wake, $\tau_{\theta\theta}$ is zero. 
In the supercritical state, large values of $\tau_{\theta\theta}$ 
persist up to the bubble equator and the stress relaxation is predominantly located in the lower bubble hemisphere as visible in
Fig.~\ref{fig:tau_TT_super}~(c) and (d).
In contrast to $\tau_{\phi\phi}$, the stress component $\tau_{\theta\theta}$ increases again in the bubble wake (see Fig.~\ref{fig:tau_TT_super}~(e)).

\begin{figure}[b!]    
\centering{
\includegraphics[width=\textwidth]{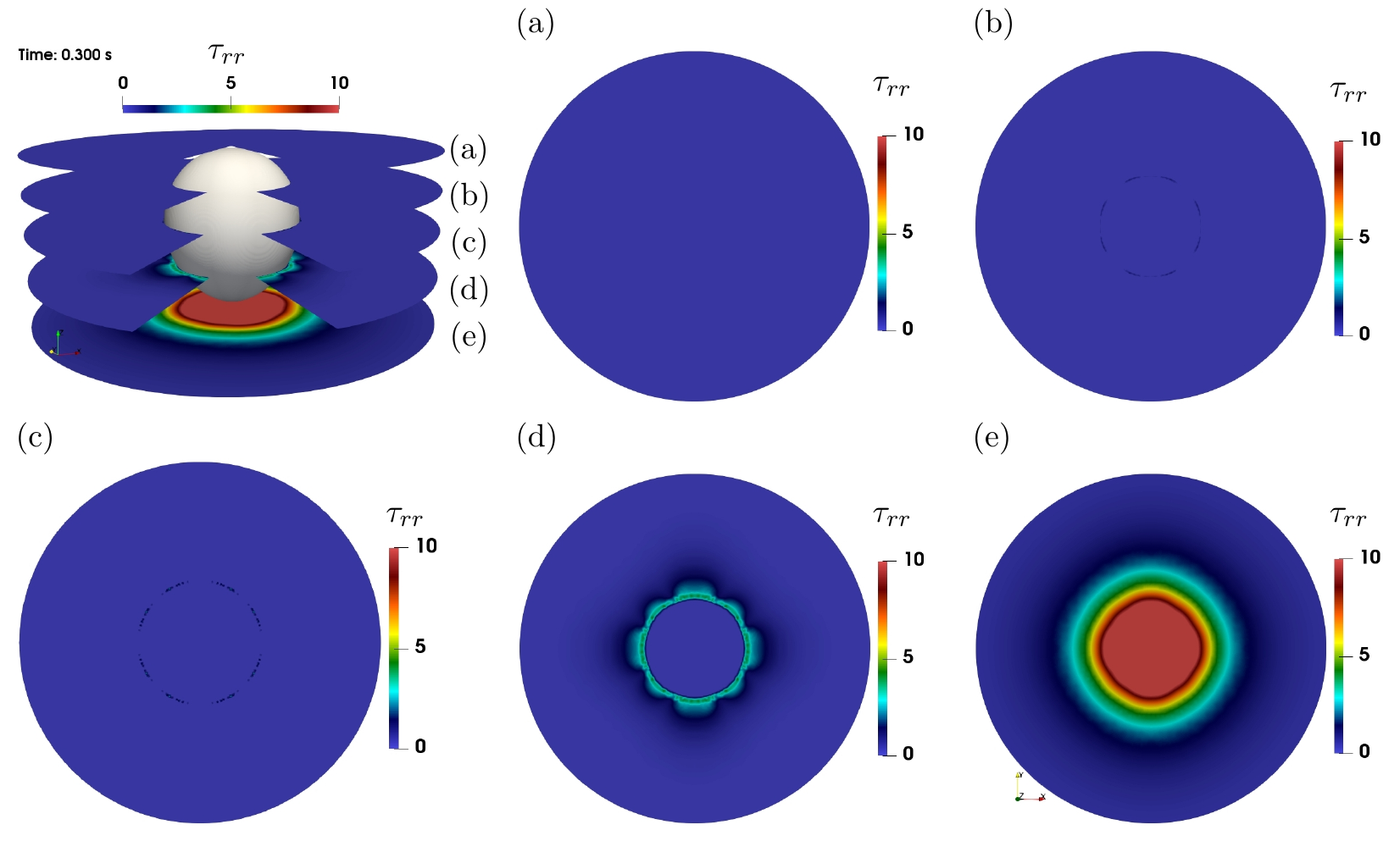}\vspace{-18pt}
}
\caption{\small Stress tensor component ${\tau}_{rr}$ (in Pa) in horizontal cutting planes at different heights (a) - (e). Subcritical bubble volume $V = 30~\textnormal{mm}^3$ in an aqueous $0.8$~wt.~\% Praestol $2500$ solution.}
\label{fig:tau_RR_sub}
\end{figure}

\begin{figure}[htbp!]    
\centering{
\includegraphics[width=\textwidth]{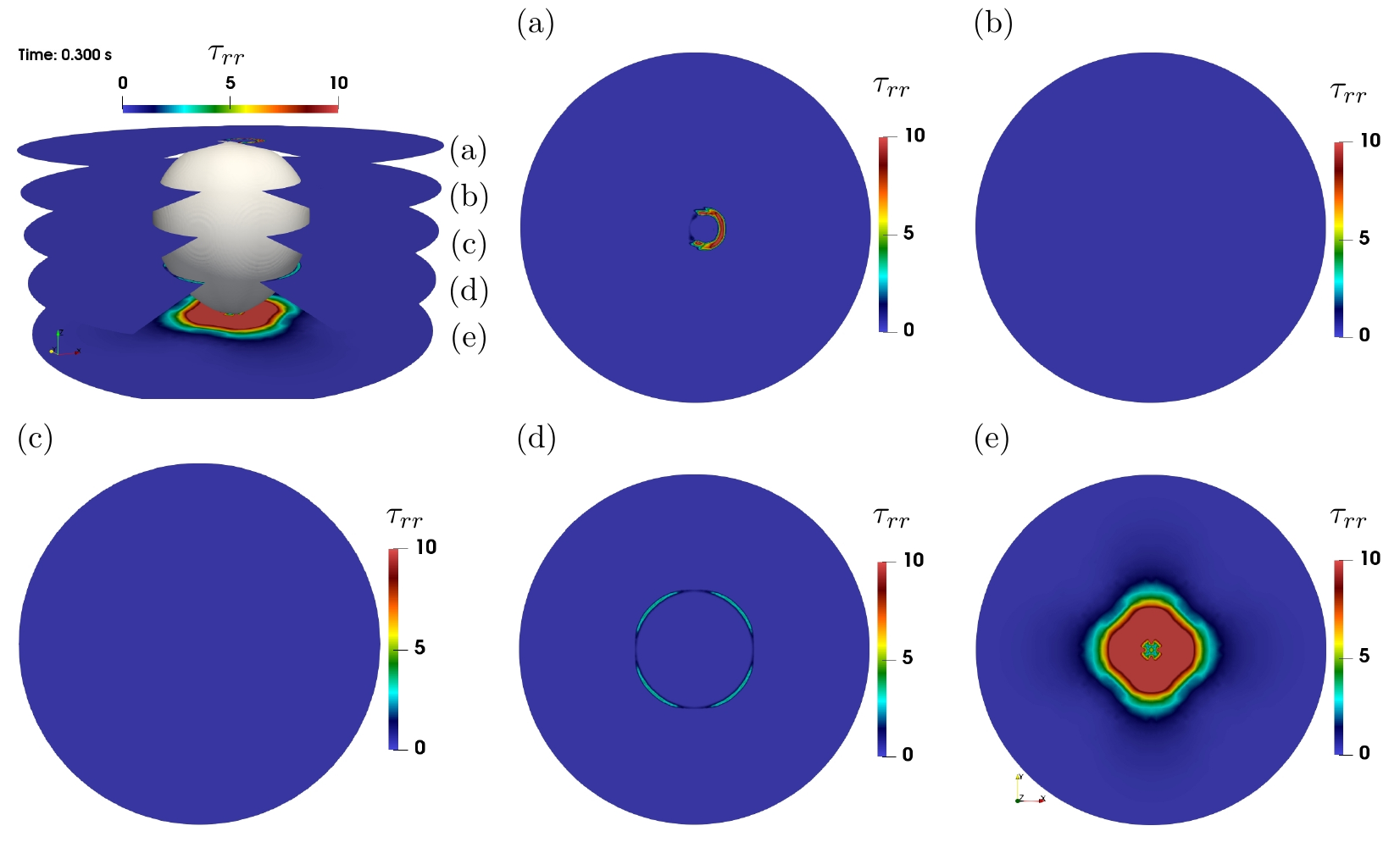}\vspace{-18pt}
}
\caption{\small Stress tensor component ${\tau}_{rr}$ (in Pa) in horizontal cutting planes at different heights (a) - (e). Supercritical bubble volume $V = 60~\textnormal{mm}^3$ in an aqueous $0.8$~wt.~\% Praestol $2500$ solution.}
\label{fig:tau_RR_super}
\end{figure}

Figs.~\ref{fig:tau_RR_sub} and \ref{fig:tau_RR_super} show the stress component $\tau_{rr}$ in the sub- and supercritical states, respectively. For the radial stress component, no qualitative difference between the two cases is visible: only in the region close to the downstream bubble pole, significant values of $\tau_{rr}$ appear.
In particular, the radial stress component is zero on the upper hemisphere, both in the sub- and the supercritical states.
%

\section{Additional data from the conformation tensor analysis}
\label{asec:conformationtensor}
\begin{figure}[htbp!]    
\centering{
\includegraphics[width=\textwidth]{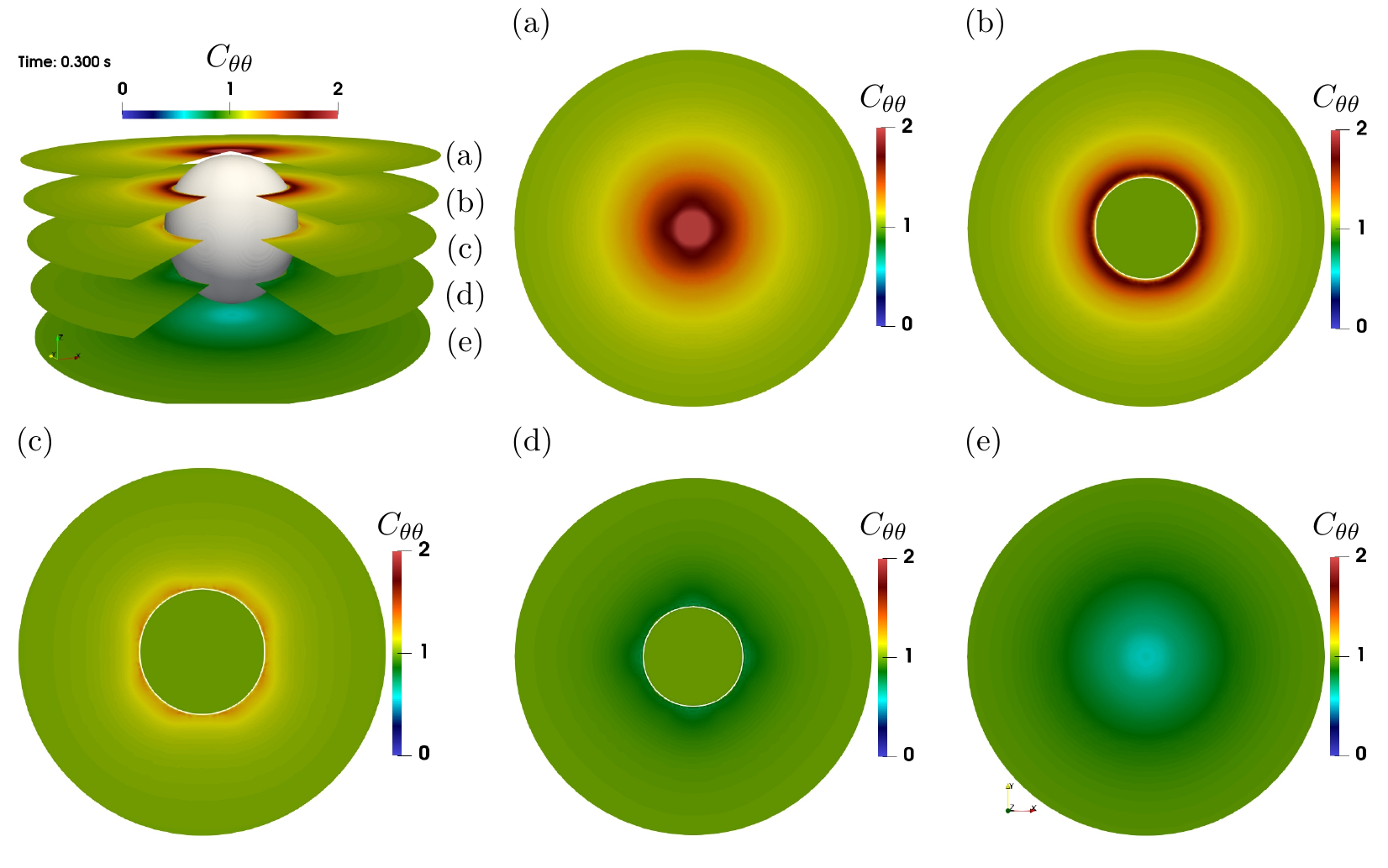}
}
\caption{\small Conformation tensor component ${C}_{\theta\theta}$ in horizontal cutting planes at different heights (a) - (e). Subcritical bubble volume $V = 30~\textnormal{mm}^3$ in an aqueous $0.8$~wt.~\% Praestol $2500$ solution.}
\label{fig:C_TT_sub}
\end{figure}

\begin{figure}[htbp!]    
\centering{
\includegraphics[width=\textwidth]{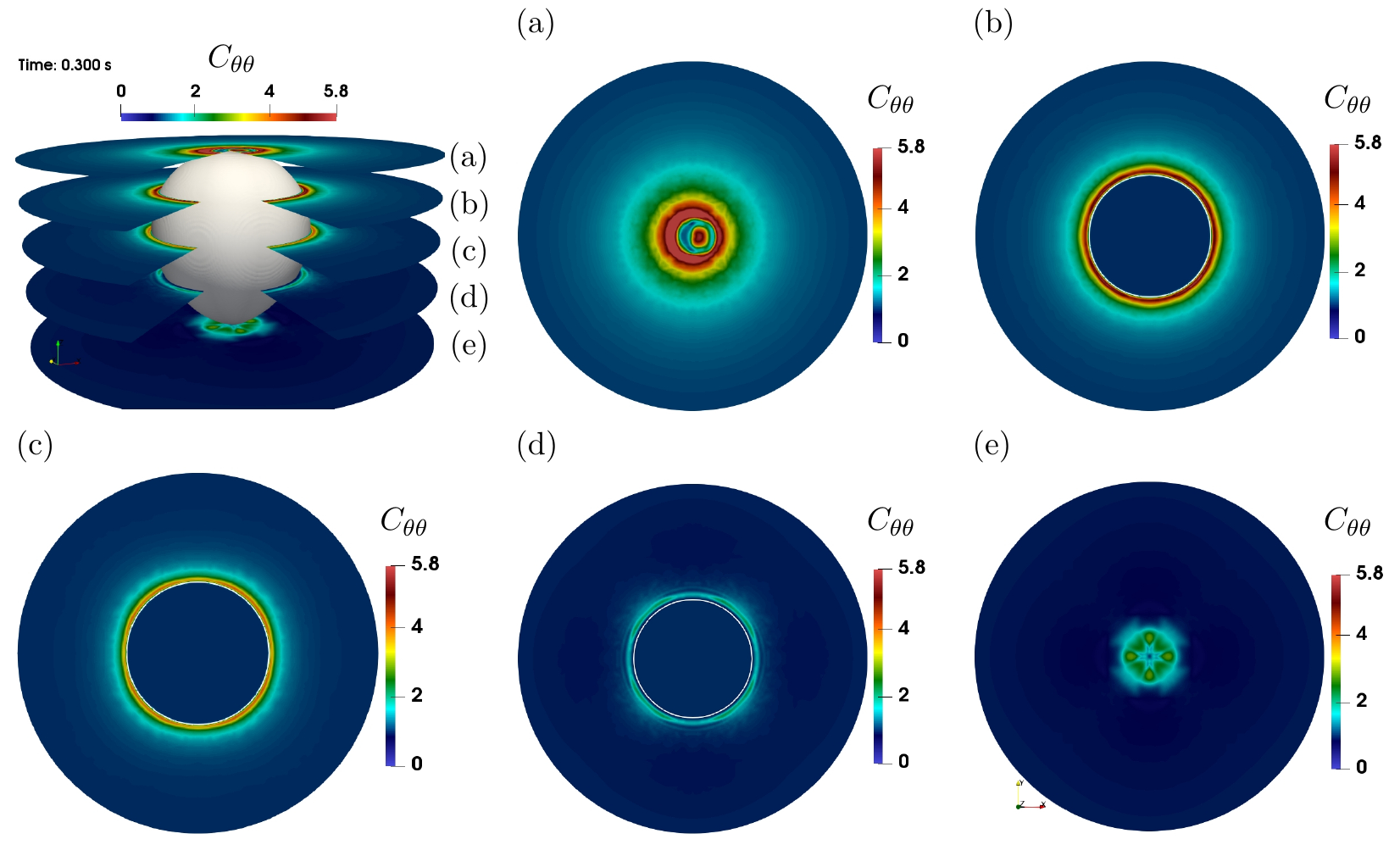}
}
\caption{\small Conformation tensor component ${C}_{\theta\theta}$ in horizontal cutting planes at different heights (a) - (e). Supercritical bubble volume $V = 60~\textnormal{mm}^3$ in an aqueous $0.8$~wt.~\% Praestol $2500$ solution.}
\label{fig:C_TT_super}
\end{figure}

As a supplement to section~\ref{subsec-conformation}, we provide the numerical results for the diagonal components of the conformation tensor, $C_{\theta\theta}$ and $C_{rr}$, in horizontal cutting planes. Figs.~\ref{fig:C_TT_sub} and \ref{fig:C_TT_super} show $C_{\theta\theta}$ in the subcritical and supercritical states, respectively. The component $C_{\theta\theta}$ shows similar behavior around the northern bubble hemisphere as the component $C_{\phi\phi}$ shown in Figs.~\ref{fig:C_PP_sub} and \ref{fig:C_PP_super}. The polymer molecules are stretched in both the latitudinal and the longitudinal directions. In the subcritical state, we observe nearly complete relaxation of the component $C_{\theta\theta}$ towards the equilibrium state upstream from the bubble equator. In the supercritical state, relaxation takes place predominantly below the equator. Unlike $C_{\phi\phi}$, the component $C_{\theta\theta}$ increases again in the bubble wake, as can be seen in Fig.~\ref{fig:C_TT_super} (e). Although the components $C_{\theta\theta}$ and $C_{\phi\phi}$ are similarly distributed, in absolute terms, the component $C_{\phi\phi}$ slightly predominates, and therefore the maximum polymer extension is oriented along the latitudinal direction, as can be seen from the maximum eigenvalues and eigenvectors of the conformation tensor in Figs.~\ref{fig:ev_sub} and \ref{fig:ev_super}.
%
\begin{figure}[t!]    
\centering{
\includegraphics[width=\textwidth]{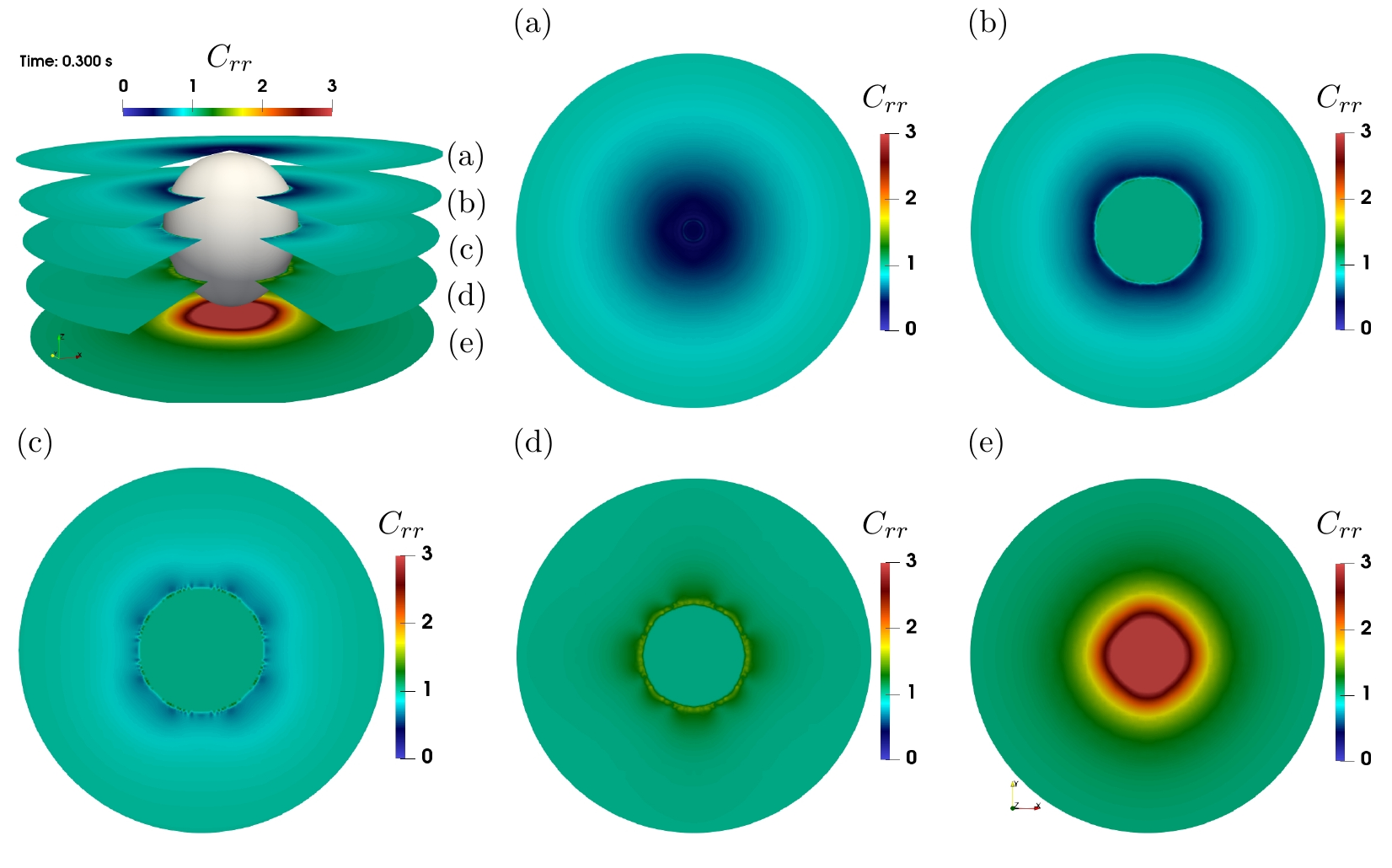}
}\vspace{-6pt}
\caption{\small Conformation tensor component ${C}_{rr}$ in horizontal cutting planes at different heights (a) - (e). Subcritical bubble volume $V = 30~\textnormal{mm}^3$ in an aqueous $0.8$~wt.~\% Praestol $2500$ solution.}
\label{fig:C_RR_sub}
\end{figure}
\begin{figure}[htbp!]    
\centering{
\includegraphics[width=\textwidth]{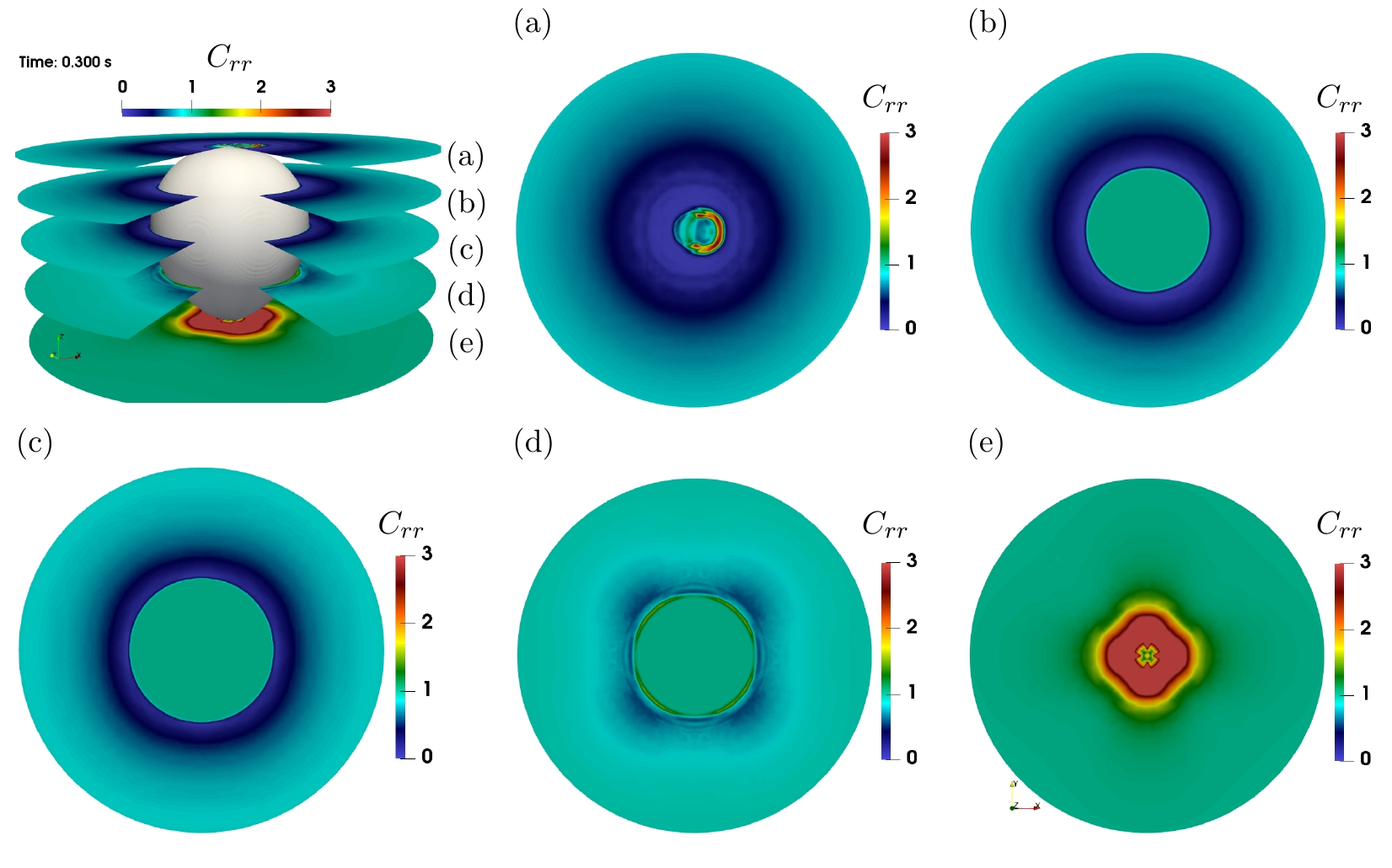}
}\vspace{-6pt}
\caption{\small Conformation tensor component ${C}_{rr}$ in horizontal cutting planes at different heights (a) - (e). Supercritical bubble volume $V = 60~\textnormal{mm}^3$ in an aqueous $0.8$~wt.~\% Praestol $2500$ solution.}
\label{fig:C_RR_super}
\end{figure}
%
Figures~\ref{fig:C_RR_sub} and \ref{fig:C_RR_super} show the component $C_{rr}$ for the subcritical and the supercritical states, respectively. The distribution of $C_{rr}$ clearly differs from the components $C_{\phi\phi}$ and ${C}_{\theta\theta}$. With respect to the conformation tensor equilibrium value of unity, $C_{rr}$ decreases to very small values close to zero in the northern hemisphere. This means that the molecules contract significantly in the radial direction as they flow along the bubble surface. The spatial distribution along the bubble surface, where $C_{rr}$ is less than unity, is wider in the supercritical state. Furthermore, the minimum values of $C_{rr}$ are smaller in the supercritical state, indicating a stronger contraction of the polymer molecules in the radial direction. In the bubble wake, the component $C_{rr}$ increases above the value of three in both the sub- and supercritical states, which corresponds to a substantial elongation of the polymer molecules in the radial direction.

\begin{figure}[b!]    
\centering{
\includegraphics[width=\textwidth]{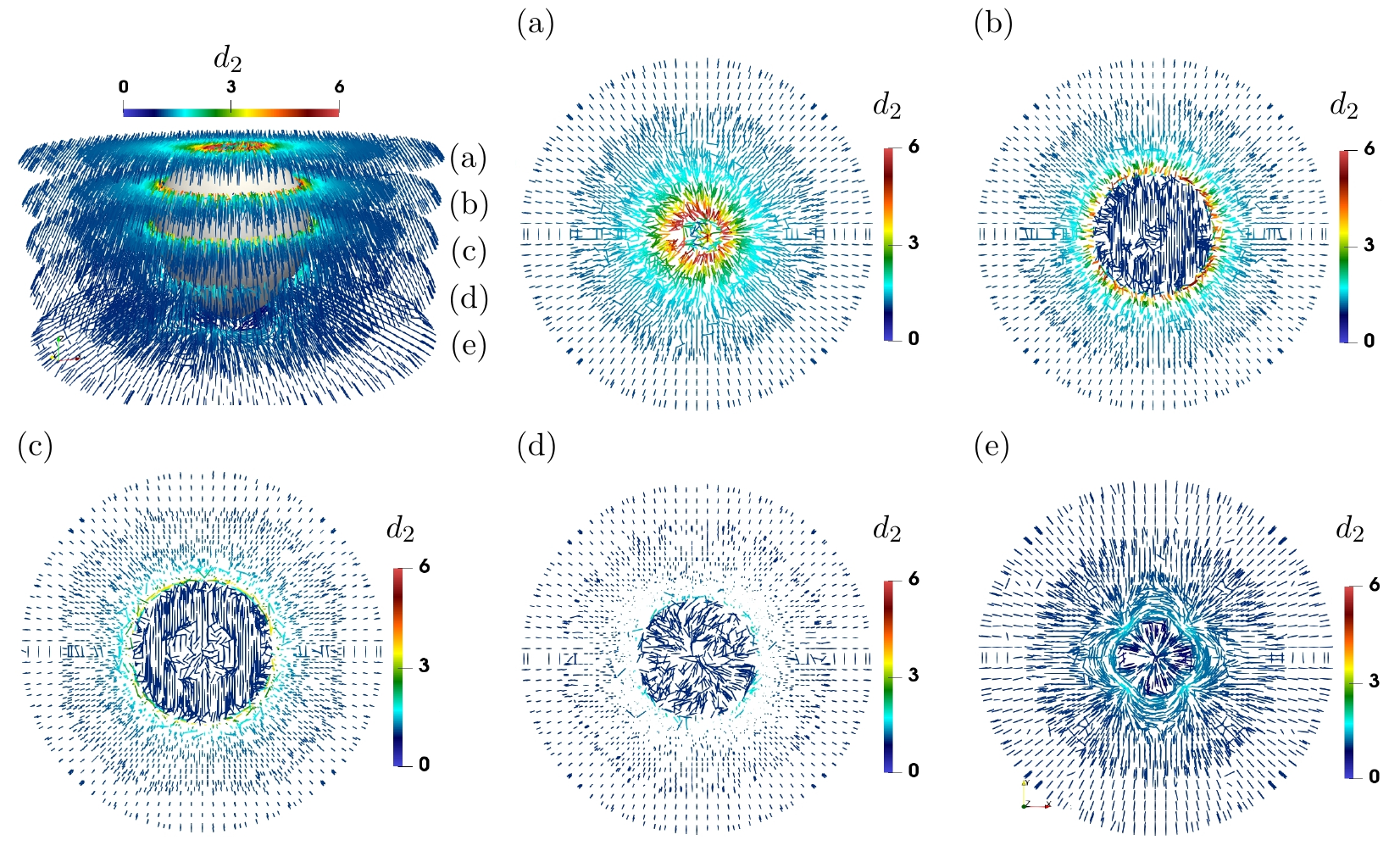}\vspace{-18pt}
}
\caption{\small Second largest eigenvalues ${d_{2}}$ and corresponding eigenvectors $\gvec{v_{2}}$ of the conformation tensor in horizontal cutting planes at different heights (a) - (e). Supercritical bubble volume $V = 60~\textnormal{mm}^3$ in an aqueous $0.8$~wt.~\% Praestol $2500$ solution.}
\label{fig:second_EV_super}
\end{figure}

Fig.~\ref{fig:second_EV_super} displays the second largest eigenvalues ${d_{2}}$ and the corresponding eigenvectors $\gvec{v_{2}}$ of the conformation tensor in horizontal cutting planes for the supercritical bubble volume of $V = 60~\textnormal{mm}^3$. It complements Fig.~\ref{fig:ev_super}, which shows the maximum eigenvalues and eigenvectors. These two figures confirm that, due to the biaxial elongational flow around the bubble, both the principal and the secondary molecular orientations are important. Around the northern bubble hemisphere, the principal molecular orientation is along the $\phi$ direction.~~\begin{figure}[t!]    
\centering{
  \begin{tabular}{@{}cc@{}}
  (a) Leading eigenvalues ${d_{\max}}$ &
  (b) Second largest eigenvalues ${d_{2}}$ \\
\includegraphics[width=0.49 \textwidth]{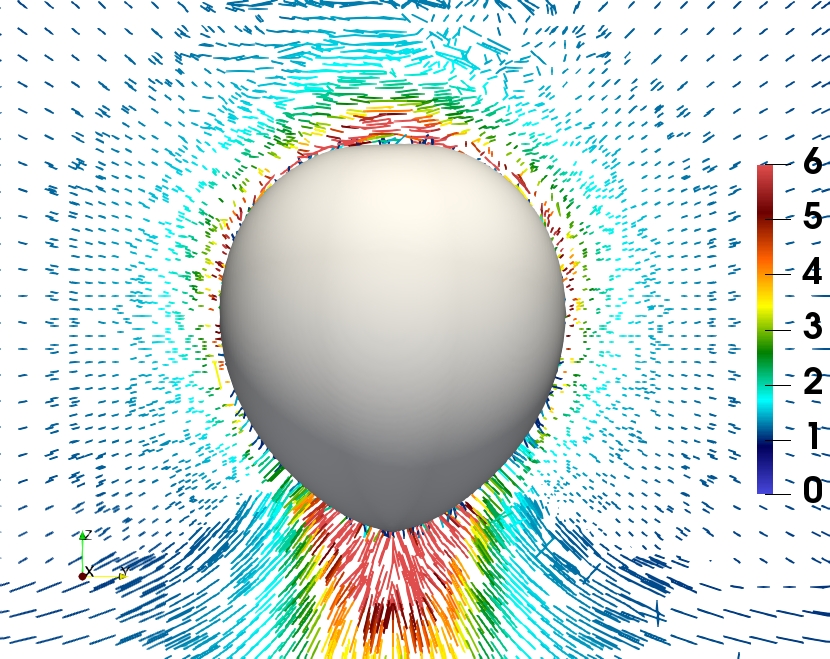} &
\includegraphics[width=0.49 \textwidth]{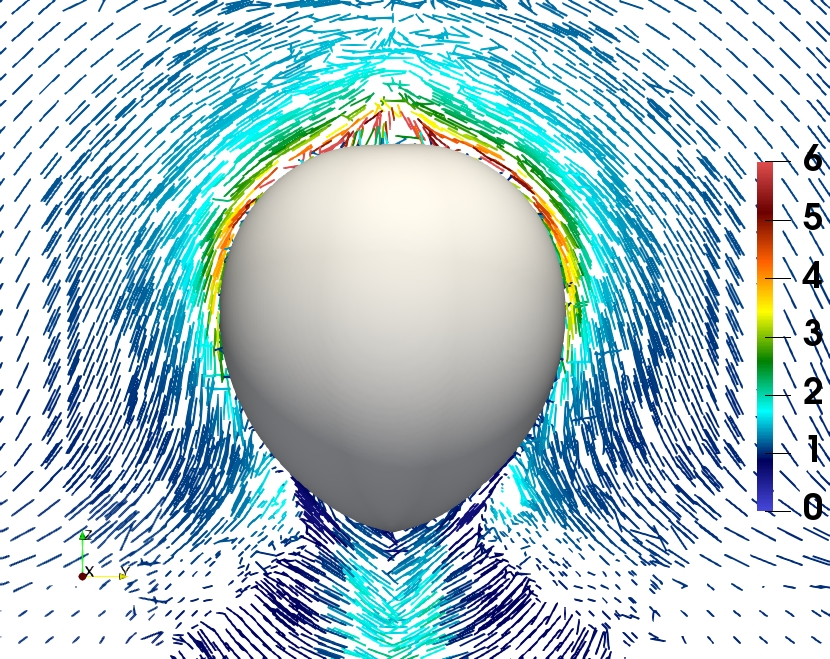}
  \end{tabular}\vspace{-10pt}
}
\caption{\small Leading eigenvalues ${d_{\max}}$ and second largest eigenvalues ${d_{2}}$ with the corresponding eigenvectors $\gvec{v_{\max}}$ and $\gvec{v_{2}}$ of the conformation tensor in vertical cutting planes through the bubble center. Supercritical bubble volume $V = 60~\textnormal{mm}^3$ in an aqueous $0.8$~wt.~\% Praestol $2500$ solution.}
\label{fig:vertical_EV_super}
\end{figure}
 The secondary molecular orientation is along the $\theta$ direction. For this reason, the horizontal planes in Fig.~\ref{fig:second_EV_super} are not ideally suited for visualizing the secondary orientation.
Instead, Fig.~\ref{fig:vertical_EV_super}~(b) shows the secondary molecular orientation in a vertical cutting plane through the bubble center. In the vertical cutting plane, the secondary molecular orientation around the northern bubble hemisphere along the $\theta$ direction is evident. Below the equator, it can be seen from the color scales in Figs.~\ref{fig:vertical_EV_super}~(a) and \ref{fig:vertical_EV_super}~(b) that both the leading and the secondary eigenvalues decrease, indicating molecular relaxation. As molecular contraction proceeds, the primary and secondary molecular orientations undergo changes as well. Fig.~\ref{fig:vertical_EV_super}~(b) shows that, below the equator, the secondary molecular orientation is increasingly influenced by the radial component and is therefore no longer perfectly aligned with the $\theta$ coordinate. With further approach towards the south pole, the radial component becomes the primary molecular orientation, which is shown in Fig.~\ref{fig:vertical_EV_super}~(a). Eventually, near the bubble south pole and in the wake, the leading eigenvalues increase again, indicating molecular stretching in the radial direction. It is interesting to note that, in the bubble wake, only the leading eigenvalues increase, while the secondary eigenvalues remain relatively small. This suggests uniaxial elongational flow in the radial direction in the bubble wake.
\end{document}